\documentclass[%
 aps,
 amsmath,amssymb,
 prb,
 twocolumn,
 superscriptaddress,
]{revtex4-2}

\usepackage{amsmath}
\usepackage{graphicx}
\usepackage{dcolumn}
\usepackage{bm}
\usepackage{diagbox}
\usepackage[]{hyperref}
\usepackage[]{xcolor}
\usepackage{float}

\begin{document}

\title{Moir\'e Superstructures in Marginally-Twisted NbSe$_2$ Bilayers}

\author{J. G. McHugh}
\affiliation{University of Manchester, School of Physics and Astronomy, Manchester M13 9PL, United Kingdom}
\affiliation{National Graphene Institute, University of Manchester, Manchester M13 9PL, United Kingdom}

\author{V.~V.~Enaldiev}
\affiliation{University of Manchester, School of Physics and Astronomy, Manchester M13 9PL, United Kingdom}
\affiliation{National Graphene Institute, University of Manchester, Manchester M13 9PL, United Kingdom}

\author{V.I. Fal'ko}
\affiliation{University of Manchester, School of Physics and Astronomy, Manchester M13 9PL, United Kingdom}
\affiliation{National Graphene Institute, University of Manchester, Manchester M13 9PL, United Kingdom}
\affiliation{Henry Royce Institute, University of Manchester, Manchester M13 9PL, United Kingdom}
\date{\today}
\begin{abstract}
The creation of moir\'e superlattices in twisted bilayers of two-dimensional crystals has been utilised to engineer quantum material properties in graphene and transition metal dichalcogenide (TMD) semiconductors. Here, we examine the structural relaxation and electronic properties in small-angle twisted bilayers of metallic NbSe$_2$\color{black}. Reconstruction appears to be \color{black} particularly strong for misalignment angles $\theta_{P} < 2.9^o$ and $\theta_{AP} < 1.2^o$ for parallel (P) and antiparallel (AP) orientation of monolayers' unit cells, respectively. Multiscale modelling reveals the formation of domains and domain walls with distinct stacking, for which density functional theory (DFT) calculations are used to map the shape of the bilayer Fermi surface \color{black} and the relative phase of the \color{black} charge density wave \color{black} (CDW) order in adjacent layers\color{black}. We find a significant modulation of interlayer coupling across the moir\'e superstructure \color{black} and the existence of preferred interlayer orientations of the CDW phase, necessitating the nucleation of CDW discommensurations at superlattice domain walls. \color{black}
\end{abstract}

\maketitle

\section{Introduction}
\textbf{Introduction} --- The field of twistronics, where a relative twist is applied between adjacent layers in two-dimensional crystalline structures, has recently emerged as a promising method to control electronic \cite{Sung2020,Chen2022} and structural \cite{yoo2019,Weston2020,rosenberger2020,Halbertal2021,McGilly2020,Shabani2021,Edelberg2020,Engelke2022,HETTMD_Moire,Kazmierczak2021} properties. For example, in semiconducting TMDs, with chemical formula MX$_2$ (M = Mo, W; X = S, Se), it has recently been established that the moir\'e pattern in twisted bilayers undergoes significant reconstruction \cite{Weston2020,rosenberger2020,Halbertal2021,Enaldiev2022} at marginal (i.e. sufficiently small) twist angles.

%%%%%%%%%%%%%%%%%%%%%%%%%%%%%%%%%%%%%%%%%%%%%%%%%%%%%%%%%%
\begin{figure}[ht]
  % \centering
    \includegraphics[width=1.0\columnwidth]{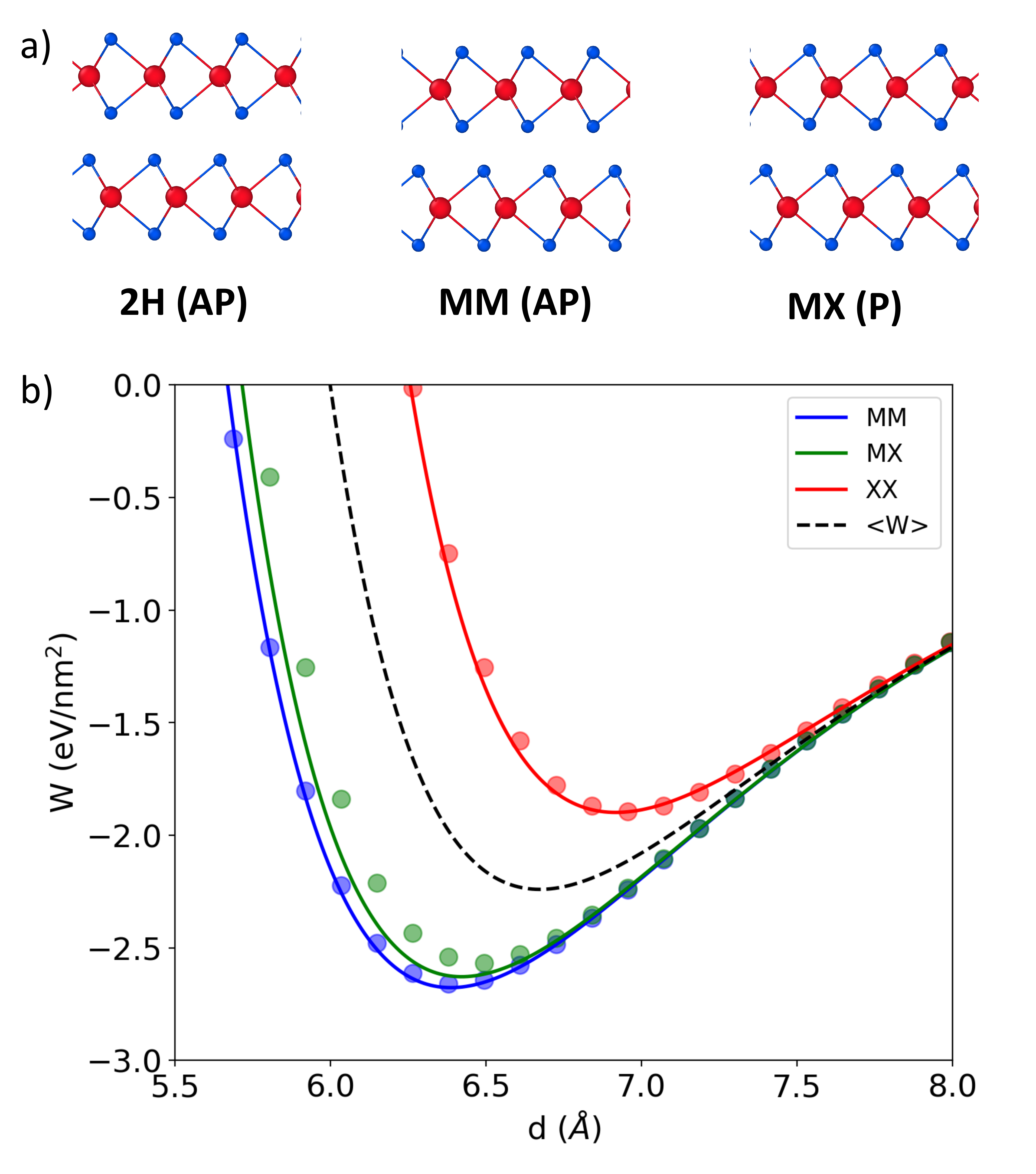}
    \caption{a) Side view of anti-parallel 2H stacking, anti-parallel MM stacking and parallel MX stacking orders of NbSe$_2$ bilayers. \color{black}Additional stacking configurations are shown in Supplemental Information (SI) section S1. \color{black} b) Adhesion energy versus interlayer separation for MM, MX, XX and $\langle$W$\rangle$ (configuration-averaged) stacking configurations.}
    \label{fig:Adhesion}
\end{figure}
%%%%%%%%%%%%%%%%%%%%%%%%%%%%%%%%%%%%%%%%%%%%%%%%%%%%%%%%%%

The niobium (or tantalum) atoms of metallic TMDs host \color{black} one fewer d-electron in the valence shell. The chemical potential then \color{black} lies in the valence band, in contrast to semiconducting TMDs, allowing superconducting and charge-density wave (CDW) phases in monolayer and bulk \cite{He_2018,Fischer2018,Bawden2016,delaBarrera2018,Wickramaratne2020,Johannes2006}. Here, interlayer effects are also found to be important, moderating the critical temperatures of the superconducting \cite{Xi_2015,Noat2015} and CDW \cite{Lin2020,Calandra2009} transitions,  enabling van der Waals Josephson junctions through interlayer twisting \cite{Farrar2021, Yabuki2016} \color{black} and leading to spatial variation of the Gibbs free energy of hydrogen absorption \cite{Zhang_HER}. \color{black} Despite these fascinating properties, there have been no detailed theoretical studies of the structural relaxation of twisted bilayers \color{black} of metallic TMDs, \color{black} and the consequent effect of stacking on electronic properties. Here we address these questions for twisted bilayers of NbSe$_2$ using ab-initio density functional theory (DFT) modelling combined with multiscale analysis of lattice relaxation.

{\bf Ab-initio analysis of adhesion energy} --- DFT calculations were performed as implemented in the Quantum ESPRESSO code \cite{Giannozzi2009, Giannozzi2017}. Core electrons were approximated using Vanderbilt ultrasoft pseudopotentials \cite{Garrity2014} under the generalized gradient approximation (GGA), as parameterised by Perdew, Burke and Ernzerhof \cite{Perdew1996}. A plane-wave cut-off of $E_w = 50$ Ry and charge density cut-off of $E_{\rho} = 600$ Ry were applied. A Monkhorst-Pack \textit{k}-point grid of dimensions $15 \times 15 \times 1$ \cite{Monkhorst1976} and Fermi-Dirac smearing of width $\sigma = 0.01$ eV is applied to aid convergence. Interlayer dispersion is implemented through the optB88-vdW functional \cite{Thonhauser2007, Thonhauser2015, Berland2015,Langreth2009}. Band structure calculations were performed using $E_{w} = 80$ Ry and a  $21 \times 21 \times 1$ \textit{k}-point grid. \color{black} In both band structure \& adhesion parameterisation calculations, we fix the in-plane lattice constant to the bulk value. \color{black}

%%%%%%%%%%%%%%%%%%%%%%%%%%%%%%%%%%%%%%%%%%%%%%%%%%%%%%%%%
\begin{figure}%[ht]
  % \centering
    \includegraphics[width=1.0\columnwidth]{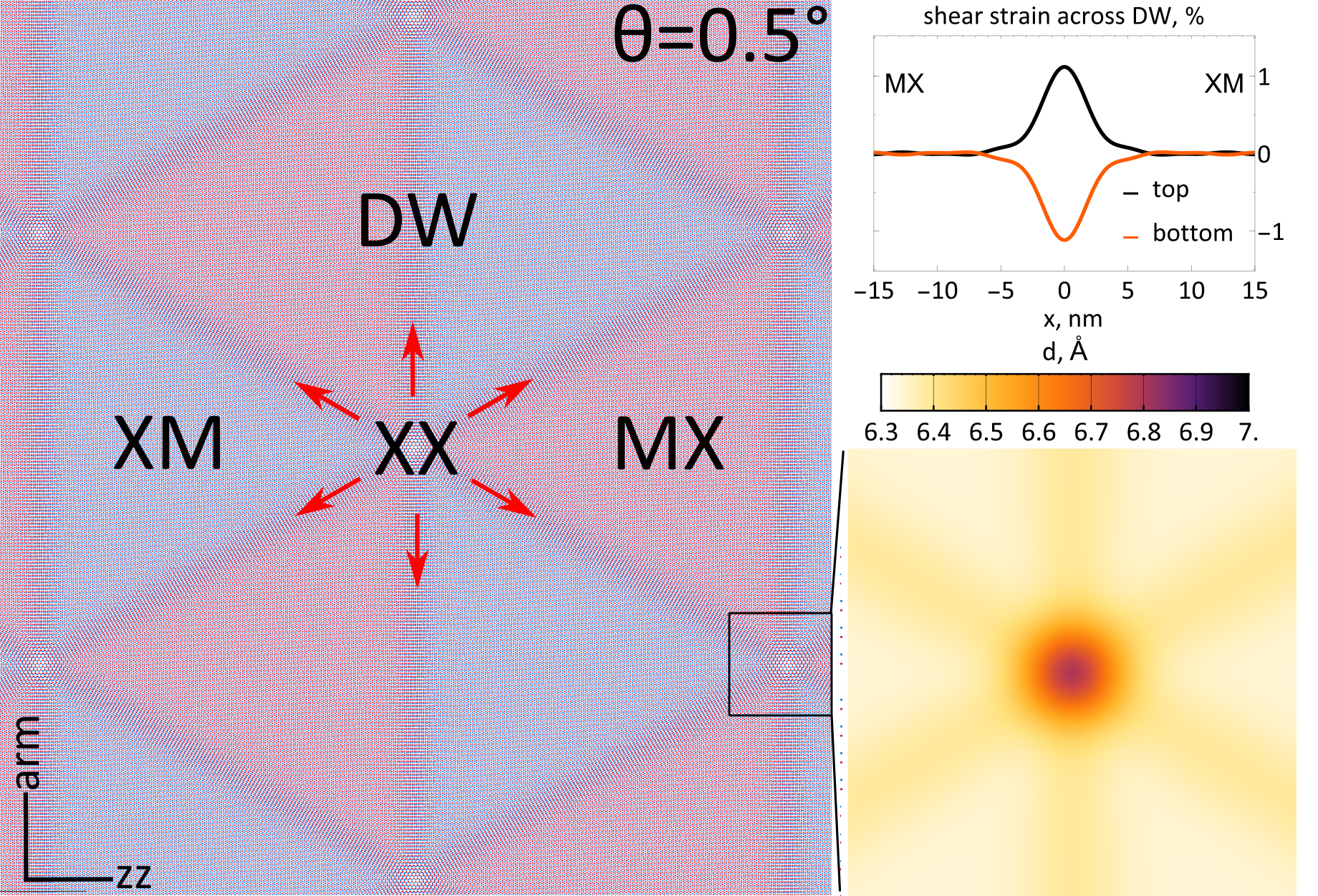}
    \caption{Reconstructed moir\'e superlattice in P-NbSe$_2$ bilayers at $\theta=0.5^{\circ}$. Red arrows show directions of Burgers vectors characterizing shift of atomic register across domain walls. On the upper inset we demonstrate distribution of the only non-zero shear strain across a single domain wall. Lower inset shows map of interlayer distance around XX area given by $d(\bm{r}_0(\bm{r}))$. }
    \label{fig:P_reconst_lat}
\end{figure}

%%%%%%%%%%%%%%%%%%%%%%%%%%%%%%%%%%%%%%%%%%%%%%%%%%%%%%%%%%

%%%%%%%%%%%%%%%%%%%%%%%%%%%%%%%%%%%%%%%%%%%%%%%%%%%%%%%%%
\begin{figure}%[ht]
  % \centering
    \includegraphics[width=1.0\columnwidth]{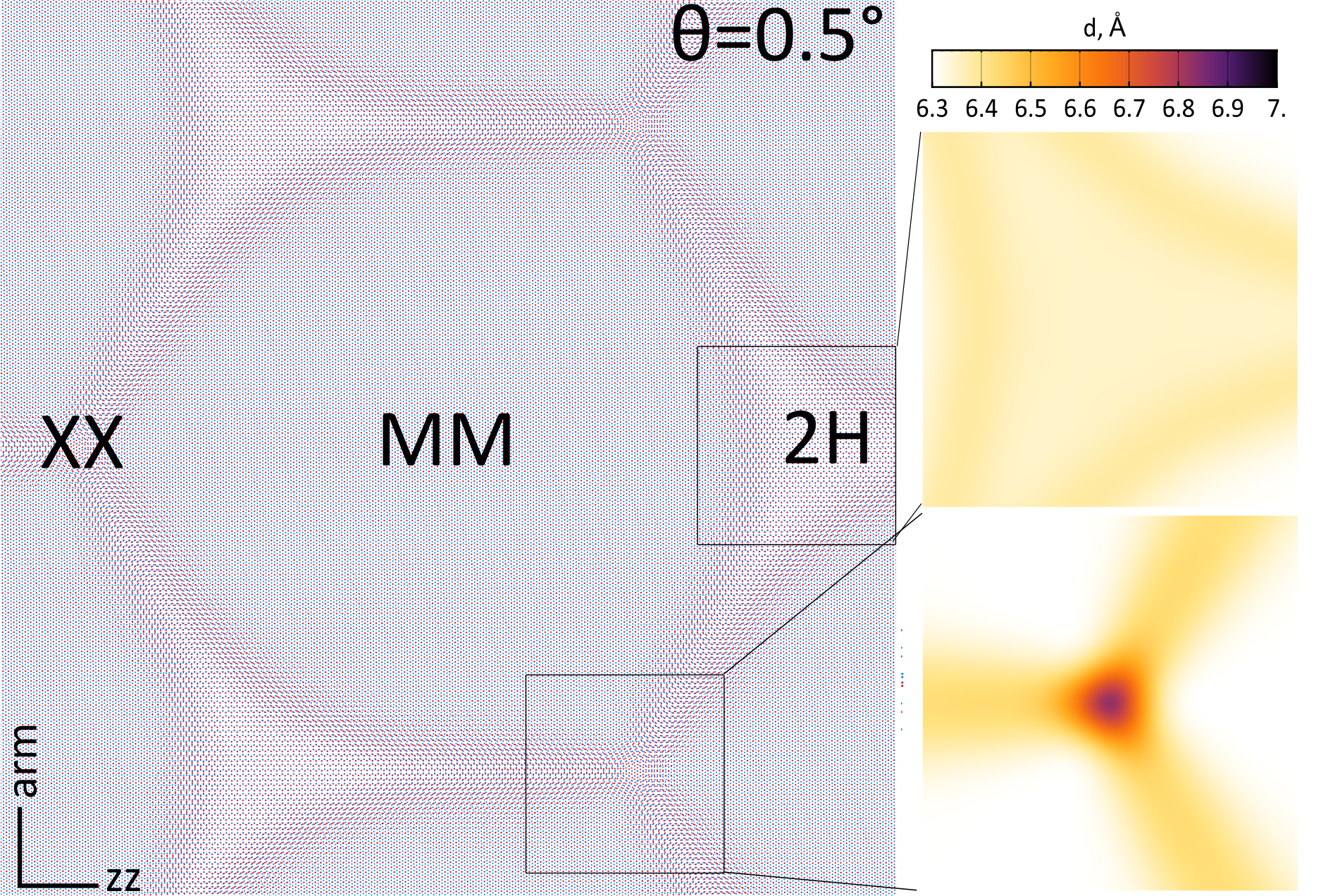}
    \caption{Reconstructed moir\'e superlattice in AP-NbSe$_2$ bilayers at $\theta=0.5^{\circ}$. Upper (lower) insets show interlayer distance map around 2H (XX) stacking areas.}
    \label{fig:AP_reconst_lat}
\end{figure}
%%%%%%%%%%%%%%%%%%%%%%%%%%%%%%%%%%%%%%%%%%%%%%%%%%%%%%%%%%

To assess the preferred stacking order, bilayers of NbSe$_2$ were structurally relaxed in a variety of stacking configurations illustrated by sketches in Fig. \ref{fig:Adhesion}a. Bilayer structures are divided into two classes: anti-parallel (AP), where a 180$^o$ rotation of the top layer restores centrosymmetry, which is absent in a monolayer, and parallel (P) where monolayers have the same orientation and centrosymmetry is absent. Comparison of different stackings shows that the lowest-energy stacking polymorph is the metal-overlapping, MM (also called 2H$_a$ \cite{Meerschaut2001, Kershaw1967}) configuration (see Fig. \ref{fig:Adhesion}b), for which DFT bulk lattice constants, a = 3.45 $\rm{\AA}$ and c = 12.74 $\rm{\AA}$ \cite{Feldman1976}, and elastic moduli \color{black} (see Supplemental Information (SI) section S2), \color{black}  are all found to be in good agreement with available experimental values \cite{Feldman1976, Lv2017, Gardos1990}. 2H-stacking, which is preferred for semiconducting TMDs, is reasonably close in energy. While it has been found with a smaller areal density in CVD-grown NbSe$_2$ bilayers \cite{Wang2017}, it is a metastable state, whereas the MM configuration is  most common and thermodynamically stable. We note that the bulk structure of NbSe$_2$ has frequently been referred to (and sometimes even modelled) as 2H stacking \cite{Valla2004, Nakata2018, Rahn2012, Borisenko2009}, therefore to avoid confusion, below we employ nomenclature of relevant stacking configurations as shown in Fig. \ref{fig:Adhesion}.

In twisted bilayers, moir\'e superlattice reconstruction is fully determined by the stacking-dependent variation of adhesion energy between the constituent layers. Following the approach of Ref. \cite{Enaldiev_PRL} we \color{black} employ an interpolation formula to fit the adhesion energy as a function of local disregistry, $\bm{r}_0$. This is \color{black} parameterised in a coordinate system where XX stacking corresponds to zero displacement, i.e. $\bm{r}_0=(0,0)$, and interlayer distance $d$:
\begin{align}\label{Eq:adhesion}
    W (\bm{r}_0, d) & = f(d)
    + A_1 e^{-(d-d_0)\sqrt{G^2 + \rho_1^{-2} }} \cos\left(\bm{G}_n \bm{r}_0\right) \\
    &+ A_2 e^{-(d-d_0)\sqrt{G^2 + \rho_2^{-2} }} \sin\left(\bm{G}_n \bm{r}_0 + \phi_{P/AP}\right). \nonumber 
\end{align}
Here, $f(d)=C_3/d^{12}+C_2/d^8-C_1/d^4$ is stacking-averaged adhesion energy, characterising long-distance van der Waals attraction between the layers ($\propto -1/d^4$), and short-distance repulsion ($\propto 1/d^{12}$), \color{black} $\bm{G}_n$ are reciprocal lattice vectors and $A_i$, $\rho_i$, are fit parameters. \color{black} The phases, $\phi_{P} = 0$ and $\phi_{AP} = \pi/2$ distinguish P and AP unit cell orientations, reflecting the symmetry of these alignments.

%%%%%%%%%%%%%%%%%%%%%%%%%%%%%%%%%%%%%%%%%%%%%%%%%%%%%%%%%%
\begin{figure*}[ht]
  % \centering
    \includegraphics[width=1.95\columnwidth]{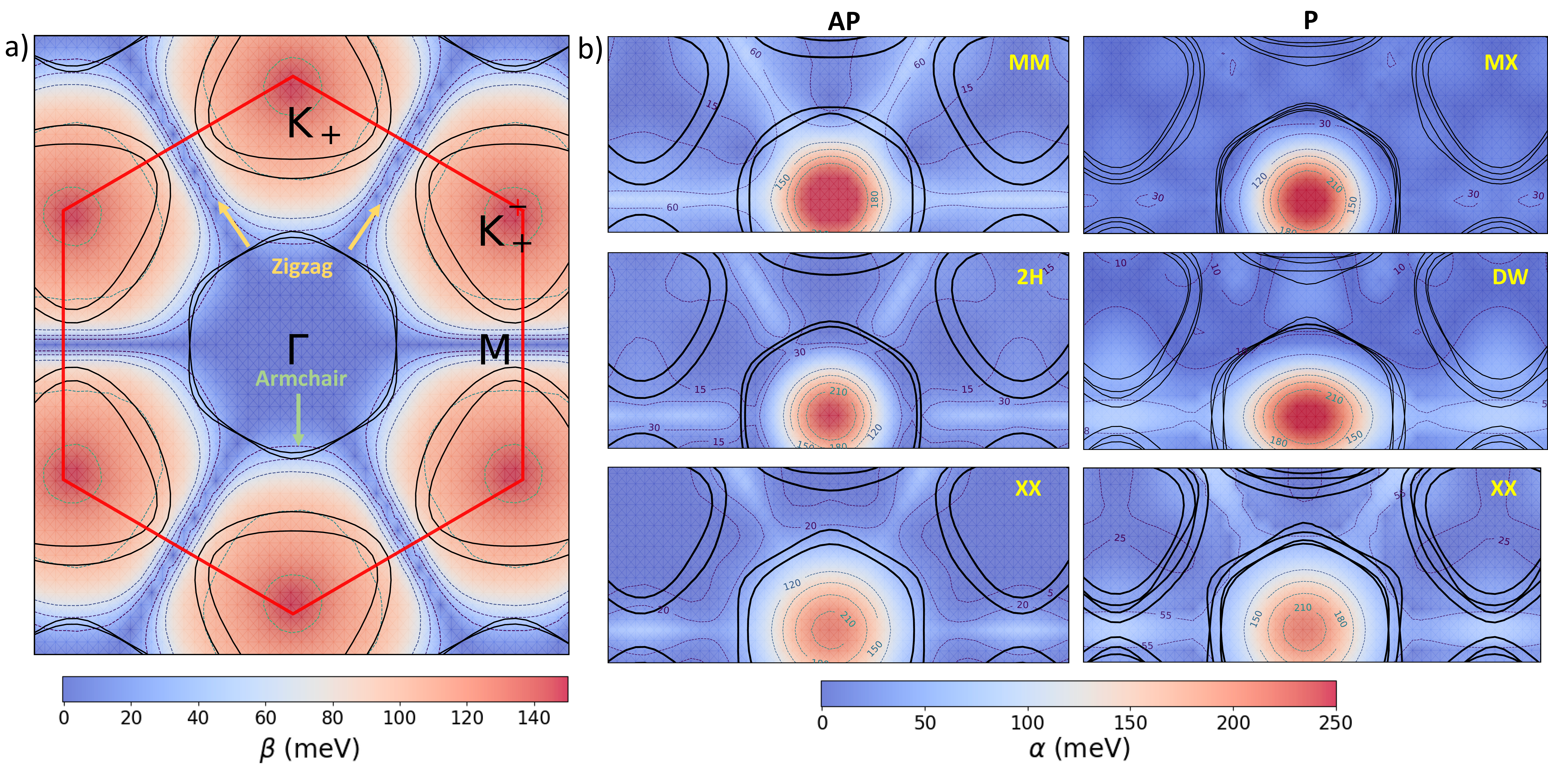}
    \caption{a) Two-dimensional Fermi lines for an NbSe$_2$ monolayer. The colour overlay maps the magnitude of the spin-orbit splitting. The maximum magnitude of band splitting from spin-orbit effects only is $\approx$ 156 meV ($\beta$ = 78 meV). b) Bilayer Fermi lines and magnitude of interlayer-hybridization parameter ($\alpha(\mathbf{k})$) in selected stacking configurations. Splitting around K$_\pm$ pockets is determined by SOC effects, while $\Gamma$ pocket splitting is largely due to interlayer interactions. \color{black} DW corresponds to the P-bilayer "MX-XM" structure in SI S1. \color{black}}
    \label{fig:Fermi}
\end{figure*}
%%%%%%%%%%%%%%%%%%%%%%%%%%%%%%%%%%%%%%%%%%%%%%%%%%%%%%%%%%

Fitting to DFT results leads to optimal interlayer distances $d_0=0.66$\,nm and $d_*=0.65$\,nm for configuration-averaged and ground (MM) stackings, respectively. Expanding Eq. \eqref{Eq:adhesion} around $d_*$ for $\bm{r}_0^{\rm MM}=(0,-2a/\sqrt{3})$, ($W \approx W_{\rm min} + \epsilon_* (d-d_*)^2$, $\epsilon_*\approx 257$\,eV/nm$^4$), \color{black} allows us to estimate the frequency of the layer breathing mode in MM-stacked NbSe$_2$ bilayers as follows:  $\omega_{\rm BM}=\sqrt{4\epsilon_*S_{\rm uc}/\mu}\approx 33.8$\,cm$^{-1}$ ($4.2$\, meV) Here, \color{black} $\mu$ and $S_{\rm uc}$ are mass and area of NbSe$_2$ monolayer unit cell, respectively. The computed value is in a good agreement with the experimental value $\omega_{BM}\approx34$\,cm$^{-1}$ \cite{He_2016} , which validates the DFT-fitted shape of the adhesion energy \eqref{Eq:adhesion} around $d_*$. 

{\bf Lattice reconstruction in twisted NbSe$_2$ bilayers} is modelled using a multiscale approach implemented earlier in Ref. \cite{Enaldiev_PRL} for the analysis of semiconducting TMDs.  This incorporates the microscopic expression for adhesion energy \eqref{Eq:adhesion} and elasticity theory applied to the mesoscale strains across a long-period moir\'e superlattice, which reduces to minimization of total (adhesion and elastic) energy over the moir\'e supercell. In the elastic energy, characterised by elastic moduli $\lambda$ and $\mu$, we take into account only in-plane strains in top (t) and bottom (b) layers, $U=\sum_{l={t,b}}\left[(\lambda/2)\left(u_{ii}^{(l)}\right)^{2} + \mu u_{ij}^{(l)}u_{ji}^{(l)}\right]$, neglecting minor bending energies of layers (see SM in \cite{Enaldiev_PRL}) due to  adjustment of local optimal interlayer distance with corresponding stacking across the moir\'e superlattice. To find the latter, we expand $f(d)$ around its extremum ($f(d_0)+\varepsilon\left(d-d_0\right)^2$), exponential functions in \eqref{Eq:adhesion} up to linear order in $d-d_0$, and, then, minimise the resulting expression as a function of $d$, to obtain an expression for the optimal interlayer distance for every stacking, $d(\bm{r}_0)$. Here, $\bm{r}_0(\bm{r})=\theta \hat{z}\times \bm{r}+\bm{u}_t-\bm{u}_b$, where the first term describes contribution of geometrical twist between the layers, and in-plane displacements, $\bm{u}_{t/b}$, responsible for local deformations in t/b-layer, are found from minimisation to the total energy.

In Fig. \ref{fig:P_reconst_lat} we display the reconstructed moir\'e superlattice resulting from minimisation of total energy in twisted NbSe$_2$ bilayers. Similarly to semiconducting TMDs, at \footnote{Value of critical angles $\theta_{P/AP}^*$ result from equality of gain from formation of domains to the costs of domain wall network.} \mbox{$\theta\leq \theta_{P}^*\approx 2.9^{\circ}$} \color{black} this results in the formation of arrays \color{black} of triangular domains with rhombohedral stacking (XM and MX), separated by a domain wall network. Each domain wall in the network is a partial screw dislocation with dominating shear strain and a Burgers vector length $a/\sqrt{3}$, shown by red arrows for several DWs merging into a DW network node. The magnitude of shear strain reaches 1\% in the middle of domain wall, see Fig. \ref{fig:P_reconst_lat}, inset.  

For AP-NbSe$_2$ bilayers with \mbox{$\theta\leq \theta_{AP}^*\approx 1.2^{\circ}$}, Fig. \ref{fig:AP_reconst_lat}, lattice reconstruction \color{black} leads to the expansion of the lowest energy MM domains, and the formation of a hexagonal superlattice domain wall network. \color{black} The other high-symmetry (2H and XX) stackings occupy corners of the domains, linked by perfect screw dislocations, characterized by a stacking shift of a single translation vector along zigzag axes. \color{black} We note that the critical angles, $\theta_{P,AP}^*$ are slightly higher than those for semiconducting TMDs \cite{Enaldiev_PRL} because of the softer NbSe$_2$ lattice. \color{black}

%%%%%%%%%%%%%%%%%%%%%%%%%%%%%%%%%%%%%%%%%%%%%%%%%%%%%%%%%%
\begin{figure*}[ht]
  % \centering
    \includegraphics[width=1.95\columnwidth]{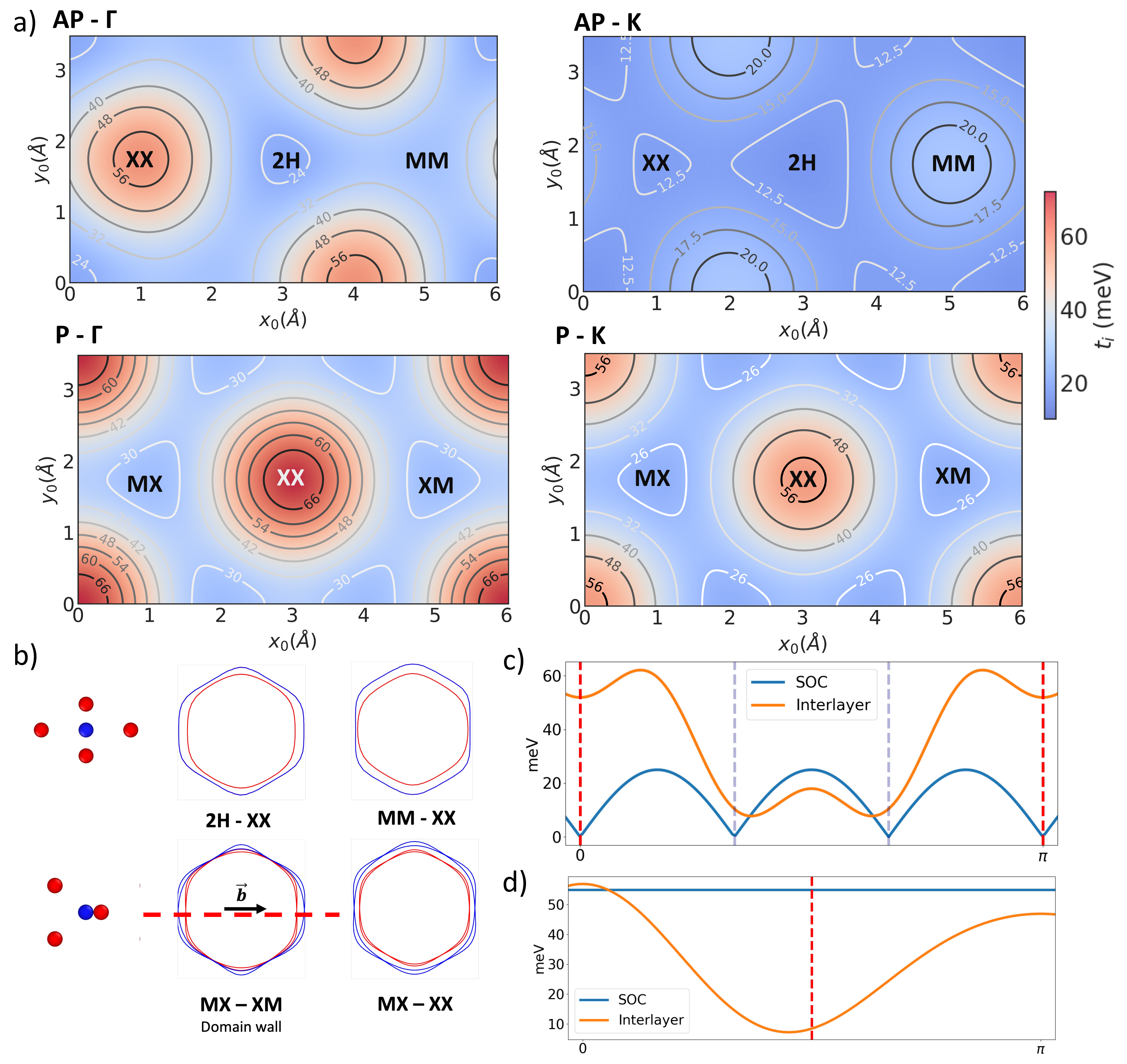}
    \caption{a) DFT-calculated variation of the average interlayer hopping parameter ($\bar{t}_i$)with disregistry, calculated by averaging splitting at the Fermi-level along six distinct crystallographic directions. Energy values are in meV. b) Modulation of the shape of the $\Gamma$-pocket Fermi lines for different intermediate stacking configurations in AP \& P bilayers, \color{black}and corresponding interlayer orientation of Nb atoms. \color{black} c) Variation in the magnitude of the spin-orbit and interlayer contributions vs angular oientation relative to pocket centre, for the $\Gamma$ and d) K pockets at a domain wall in a P-bilayer, extracted from DFT calculations.}
    \label{fig:Hybridization}
\end{figure*}
%%%%%%%%%%%%%%%%%%%%%%%%%%%%%%%%%%%%%%%%%%%%%%%%%%%%%%%%%%

%All unknown parameters in \eqref{Eq:adhesion} were computed using $d$-dependences of adhesion energies between the layers for several stackings shown in Fig. 1 using Full DFT fit values. Full details of the fit and all fit values are listed in the SI.

%Before examining lattice reconstruction of twisted NbSe$_2$ bilayers, we first review the electronic structure in a monolayer,  before considering in detail the effect which a rigid displacement between two adjacent sheets has on the bilayer electronic structure, which will allow us to resolve the electronic properties in different twisted domains and across domain boundaries.

{\bf Electronic structure} --- The Fermi surface of an NbSe$_2$ monolayer is shown in Fig. \ref{fig:Fermi}a. There are three distinct hole pockets across the Brillouin zone: one $\Gamma$-centred pocket and a pair of triangular K$_\pm$ pockets. The dispersion around each pocket ($i = \Gamma/K_\pm$) is
\begin{equation}
    \epsilon_{i}(\bm{p}) = E_{0,i} + \frac{\hbar^2 \bm{p}^2}{2 m^*_i} + C_i |\bm{p}^3| cos(3 \phi) + \beta_i \sigma_z,
\end{equation}
where $\bm{p} = \bm{k} - \bm{K}_i$ is Bloch state momentum relative to the pocket centre, \color{black} $m^*_i$ is effective mass, $C_i$ parameters account for trigonal and hexagonal warping, $\sigma_z$ is a Pauli matrix operating on spin, \color{black} $\phi =$ arctan$(k_y/k_x)$ \cite{Kormnyos2015}, and $E_{0,i}$ is the energy difference between the band edge and the Fermi level. The shape of the metallic band is overall extremely similar to the conduction band in semiconducting TMDs, with similar effective masses $m^*_i$ and a high degree of trigonal warping around the K-pockets (see Table 1) \cite{Bawden2016}. Significant Ising spin-orbit coupling (SOC) is evident across the BZ, with splitting $\propto \beta(k) \sigma_z$. A map of $\beta(k)$ across the entire BZ is displayed in Fig. \ref{fig:Fermi}a, with a maximum value of 78 meV deep in the K$_{\pm}$ pockets. In the $\Gamma$-pocket SOC vanishes along $\Gamma$-M lines, $\beta_\Gamma = \lambda_\Gamma |p|^3 cos (3 \phi)$, while it is approximately constant, $\beta_K = \pm \lambda_K$ in the $K_{\pm}$ pockets \cite{Shaffer2020}.

\begin{table}[h]
\begin{ruledtabular}
\begin{tabular}{ccccc}
Pocket & $E_0$ (eV) & $m^*/m_e$  & $C_{3/6}$ (eV\AA$^3$) & $\lambda_{SOC}$ (meV)\\ \hline
$\Gamma$ & 0.51 [0.58] & -2.52 [-1.92] & 1.29 [0.37] & 35 [35]\\
K  & 0.31 [0.53] & -1.40 [-0.61] & 3.41 [13.50] & 55 [67]
\end{tabular}
\end{ruledtabular}
\caption{\label{Fit_params} Fitted parameters of NbSe$_2$ bands calculated at the Fermi level (brackets: close to centre of pocket). $E_0$ is pocket depth, $m$ is electron effective mass, C$_{3/6}$ accounts for trigonal/hexagonal warping at the K/$\Gamma$ pocket, respectively, and $\lambda_{SOC}$ is the magnitude of SOC. }
\end{table}

Stacking modulation of interlayer hybridization results in variation of the electronic Fermi surface around each pocket. To quantify this, we relate DFT-calculated energy eigenvalues across the entire Brillouin zone (BZ) to a momentum-dependent hybridization parameter $\alpha(k)$, assuming a simple layer-hybridized wavefunction, \color{black} $\Psi(k) = (\psi_1 (k) \pm \psi_2 (k))\sqrt{2}$, with energies, $E_{ML}(k) \pm \alpha(k)$, where $E_{ML}(k)$ are monolayer energy eigenvalues.  \cite{flicker}\color{black} 

Plots of this parameter calculated across the BZ are overlaid with the DFT-calculated Fermi surface in Fig. \ref{fig:Fermi}b. We observe that in centrosymmetric AP configurations, interlayer hybridization leads to a pair of spin-layer locked, layer-hybridized bands which cross the Fermi level in all stacking configurations. In contrast, broken centrosymmetry of the P-bilayers leads to four Fermi lines, which are essentially a pair of layer-hybridized copies of the monolayer bands. DFT calculations in mirror-reflected supercells also demonstrate the presence of a ferroelectric charge transfer, resulting in $\approx$11 meV potential energy drop between the layers (see SI section S3), which is notably smaller than for semiconducting TMDs \cite{Ferreira2021}. \color{black} We note that this charge transfer in principle allows for a layer-dependent potential contribution to interlayer band splitting of P-bilayers, however, explicit incorporation of this term does not have a significant effect on interlayer hybridization due to the small degree of charge transfer between the layers \color{black} (see comparison between Fermi surface fitting with and without an explicit layer-polarization term, SI section S4). 
\color{black} 

In both bilayer orientations, there is a substantial six/three-fold modulation of the hybridization parameter around the $\Gamma$/K$_\pm$ pockets, \color{black} as shown in Fig. \ref{fig:Fermi}b (see also SI section S5). \color{black} This modulation is proportional to the out-of-plane $d_z^2$-orbital component, leading to maxima along the $\Gamma$-M and K-M lines around each pocket. The resulting interlayer hybridization, and the associated shape of the Fermi lines, is moderated by the interlayer coordination of Nb atoms. Consequently, there are also distinct modulations of the interlayer hybridization at domain walls, which is evident as a two-fold "squeezing", of both the $\Gamma$ and K-pockets, along distinguished directions of the Brillouin zone for the intermediate stacking configurations occurring at domain walls (see Fig. \ref{fig:Fermi}b, P-DW). For example, at the domain walls of a P-oriented moir\'e we find that there is significantly stronger hybridization of $\Gamma$-pocket electrons with crystal momenta parallel to the dislocation line compared to the perpendicular direction, while the opposite is the case for the $K$-pocket. \color{black}A comparison of the degree of asymmetric warping, and the corresponding positions of Nb atoms in both layers, is shown in Fig. \ref{fig:Hybridization}b for various intermediate stacking configurations of P and AP bilayers, which demonstrates qualitatively similar distortion between the two bilayers as a function of Nb-interlayer offset. Furthermore we note that, in both pockets, the degree of interlayer coupling can be lifted below or above that of the spin-orbit term, depending on pocket index and crystal momentum. (see Fig. \ref{fig:Hybridization}c,d). \color{black}

%%%%%%%%%%%%%%%%%%%%%%%%%%%%%%%%%%%%%%%%%%%%%%%%%%%%%%%%%%
\begin{figure*}[ht]
  % \centering
    \includegraphics[width=1.95\columnwidth]{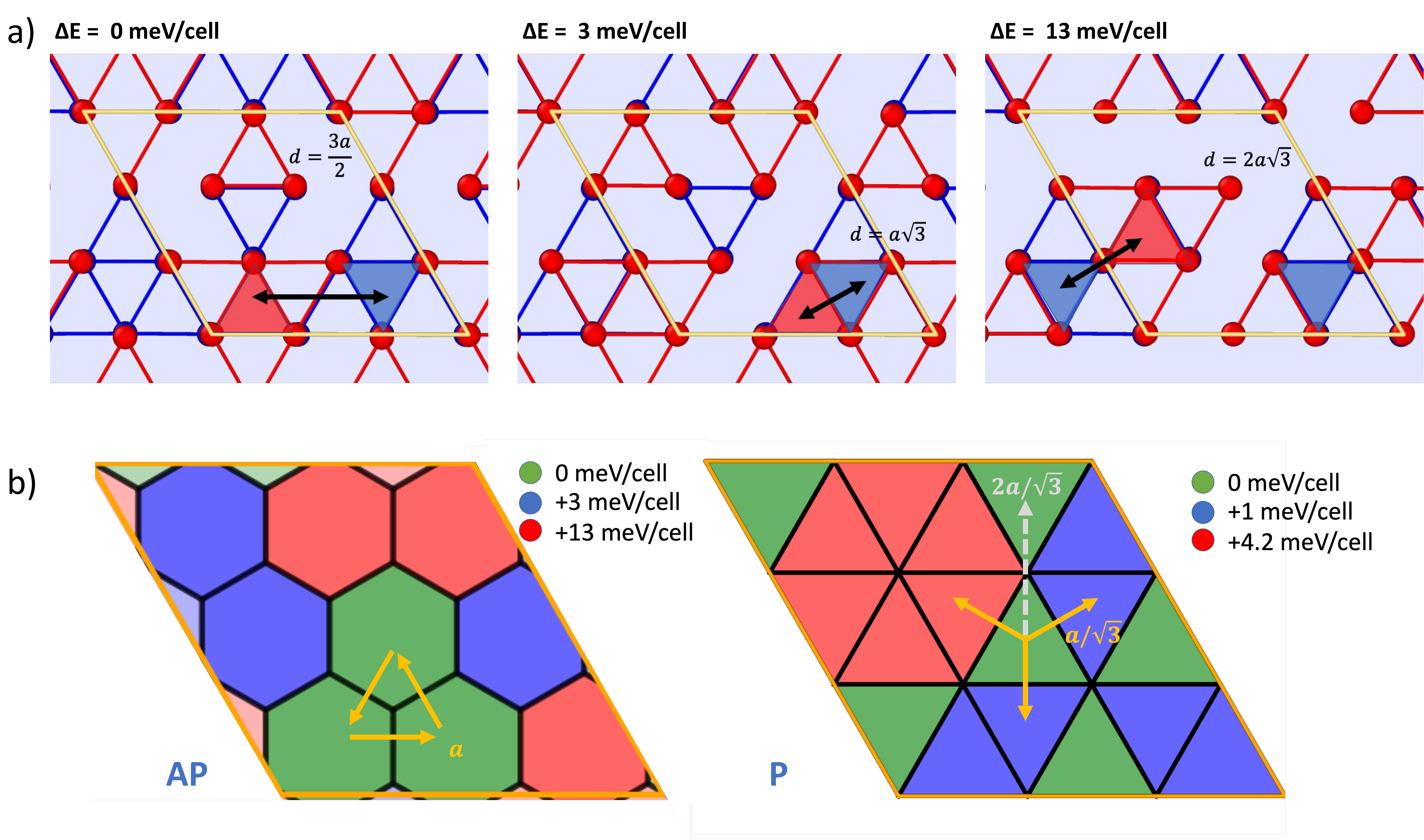}
    \caption{\color{black}a) DFT-relaxed bilayer structures with different interlayer offsets of the CDW order between layers. For clarity, only the Nb sublattice is shown in red (blue) for the top (bottom) layer, and the centre of the CDW distortion is marked as the centre of the "large" triangle in each layer. \color{black} b) Energy of the CDW phase, relative to the most favourable configuration, at different orientations of an AP-bilayer (left) and a P-bilayer (right), in units of meV/unit cell.}
    \label{fig:CDW}
\end{figure*}
%%%%%%%%%%%%%%%%%%%%%%%%%%%%%%%%%%%%%%%%%%%%%%%%%%%%%%%%%%

Using DFT data, we quantify effective interlayer hopping around each pocket by introducing an extra index which acts in layer space. This is encoded in a bilayer Hamiltonian,
\color{black}
\begin{align}
    \mathcal{H}(\mathbf{p}) = & \sum_{i} d^\dagger_{i,\mathbf{p},s,l}
    \big[ \epsilon_i(\mathbf{p}) \sigma_0 \eta_0 + \alpha_i (\mathbf{p}) \sigma_0 \eta_x \\
    & + \beta_i (\mathbf{p}) \sigma_z^a \eta_0 \big] d_{i,\mathbf{p},s,l} \nonumber
\end{align}
\color{black}
where $d^{\dagger}$/$d$ are electron creation/annihilation operators, $i$ is pocket index, $\bm{p}$ is Bloch state momentum relative to the pocket centre, $s$ and $l$ are spin and layer indices. \color{black}$\eta_{x, 0}$ \color{black} are Pauli matrices operating on the layer index, where the parameter \color{black} a = $1, 2$ for a AP, P bilayers respectively, \color{black} and $\alpha_i$ is the strength of hybridization around the corresponding pocket, where modulation is accounted for using appropriate periodic terms in $\alpha_i$. The hybridization term then takes the form,
\color{black}
\begin{equation}
    \alpha_{i}(p) = \bar{t}_i + \sum_n t_{i, n} \cos(n \phi),
\end{equation}
\color{black}
where $\bar{t}_i$ is the average hybridization around pocket. We find that hybridization is approximately constant for momenta $p \approx |p_F|$, with angular variation $\propto \phi$ incorporated via $\bar{t}_{i,n}$ periodic terms. \color{black} Further details of these terms in high-symmetry stackings are given in SI sections S6 and S7.

\color{black} To facilitate simple comparison between different domains in the moir\'e superstructure, we analyse the average interlayer hopping, $\bar{t}_i$ (the first term in Eq. 4), at each pocket as a function of interlayer disregistry. \color{black} We numerically extract this parameter from DFT calculations by averaging \color{black} interlayer splitting around each pocket along the high-symmetry directions ($\Gamma$-M, $\Gamma$-K, K-M, K-$\Gamma$). This procedure was repeated for relaxed bilayers with different interlayer offsets, \color{black} and the extracted value of $\bar{t}_i$ was found to fit to an expansion using only the first star of reciprocal lattice vectors with reasonable accuracy (see SI section S8 and S9). Fig. \ref{fig:Hybridization}a shows this expansion for both pockets in AP- and P-bilayers.

Interestingly, we find distinct variation of interlayer hybridization between the $\Gamma$ and K pockets in AP bilayers. For the former, which is most sensitive to the interlayer distance between Se atoms, it is maximal at XX corners, while interlayer hybridization at the K-pocket is strongest inside the lowest-energy MM domains. Interactions are weakest at the 2H corners for both pockets. An opposite trend was found for the moir\'e of a P-bilayer. In this case, the interlayer separation of chalcogen atoms is rather larger within domains due to trigonal interlayer coordination and there are correspondingly weaker interlayer interactions, which becomes larger at corners and domain walls where the interlayer Se distance is reduced. The overall variation is qualitatively similar at both pockets.

\textbf{CDW Modulation} --- Lastly, we examine the ability of the reconstructed moir\'e superlattice to impact the relative phase of the CDW distortion in adjacent layers, as a function of the interlayer disregistry. CDW order in NbSe$_2$ monolayers induces a $3 \times 3$ \color{black} reconstruction \cite{Ugeda2015, Dai2014, Xi_2015, Arguello2014, SilvaGuilln2016}. In a monolayer, we find two low energy triangular reconstructions of the lattice \cite{Guster2019}, which are characterised by one of two structural distortions, classified as "Hollow" or "Filled", based on the distortion pattern of atoms in the Nb sublattice \cite{Lian2018} (see SI section 10), and that the Hollow phase has lowest overall energy of the two. 

Both monolayer reconstructions are then used to create bilayer cells with different interlayer offsets of the CDW phase between the layers. The difference in DFT total energy between the normal (no reconstruction) and Hollow CDW phase on both layers, $\Delta E = E_{CDW} - E_{Normal}$, is extracted to quantify the realtive strength of the CDW phase between different structures. In agreement with the monolayer case, we find that structures with the Hollow reconstruction on both layers are lowest in energy. We then calculate the relative strength of the CDW distortion at each of the 18 MX/XM (P) and 9 MM (AP) possible stacking configurations of the CDW in adjacent $3 \times 3$ reconstructed layers. Fig. \ref{fig:CDW}a shows these three unique structural configurations in the AP case, where only the Nb-sublattice is shown to improve clarity (see SI section 10 for additional figures of the monolayer CDW reconstructions, and the different interlayer configurartions of these structures considered in this work.)
\color{black}

We observe a notable trend in this energy difference with interlayer disregistry, shown in the bottom left (right) of Fig. \ref{fig:CDW}b for the AP (P) bilayer. While it is possible to reach another low energy mutual orientation of the two CDW  distortions in each layer via a full dislocation in the AP bilayer, this is not possible for P bilayers, where shifting by a partial dislocation always leads to a higher energy configuration. In a moir\'e superlattice, this suggests that the CDW order in adjacent layers will lock-in inside a domain. Glide of the lattice by a partial or full screw dislocation across domain boundaries will rigidly shift the relative orientation of the CDW on each layer, and, \color{black} in order to attain a low-energy interlayer configuration, CDW discommensurations should nucleate at moir\'e dislocation boundaries \cite{Lim2020,McMillan1976}, \color{black} leading to CDW triplet domains in an AP-moir\'e \cite{Goodwin2022} and isolated CDW domains in a P-moir\'e.

\textbf{In conclusion}, multiscale relaxation and electronic structure calculations were performed on twisted bilayer NbSe$_2$. \color{black} Compared to semiconducting TMDs, monolayer NbSe$_2$ is relatively soft, and \color{black} significant relaxation begins at slightly larger twist angles. The resulting domain structure for AP-bilayers has hexagonal MM stacking domains with domain walls featuring XX stacking and seeds of 2H stacking in the alternating corners. The Fermi surface undergoes significant modulation across the moir\'e, with variation of the interlayer coupling at the K points on the order of 10-20 meV (25-55 meV) between different domains for an AP(P)-bilayer.
\color{black} Notably, there is significant anisotropy in the interlayer coupling is observed along domain walls. \color{black} For AP-bilayers the interlayer coupling of metallic bands is strongest inside domains, in contrast to our finding regarding P-bilayers. In P-bilayers, the triangular MX/XM domains have smaller interlayer coupling of electron bands than along the triangular domain wall network. 

\color{black} A further observation arising from this work is that discommensurations \cite{McMillan1976} of the CDW phase should occur due to the rigid displacement of the relative orientation of the CDW in each layer which occurs at superlattice domain walls. CDW discommensurations are known to enhance the superconducting order in TiSe$_2$ \cite{Chen2019,Joe2014,Leridon2020}. \color{black} This suggests a promising application for moir\'e superlattices in metallic TMDs systems as a method to control the CDW, and potentially, other order parameters. \color{black} The potential for distinct modulation of interlayer hopping at each pocket, in addition to control of the CDW phase, along with anisotropic hopping at dislocation boundaries, suggests that marginally-twisted NbSe$_2$ would be an interesting test system to further probe interlayer effects on correlated superconducting and CDW phases of the metallic TMDs \cite{Xi_2015}.

\color{black} Finally, we remark on possible implications of our results for Ising superconductivity in marginally-twisted bilayers. At small angles, the interlayer coupling in both pockets across a significant area of the moir\'e supercell is of a similar magnitude to the MM bilayer, for both twisted AP and P bilayers. In multilayer NbSe$_2$ \cite{Xi_2015} both the critical temperature, $T_c$, and in-plane magnetic field $H_c$ of the superconducting transition show a significant dependence on the number of layers, suggesting the importance of interlayer electronic effects to the magnitude and form of the superconducting gap. If the breaking of spin-layer locking through increased interlayer hopping is the active mechanism in determining the superconducting transition temperature as a function of layer number, this suggests that the significant modulation of interlayer hopping by the moir\'e superstructure could be utilised to engineer superconductivity in specific domains of the superlattice. In an applied field, superconductivity would preferentially persist in regions of the moir\'e with smaller interlayer coupling, which are MX/XM domains (P-bilayer), or domain walls and corners (AP-bilayer). 

\color{black}

{\bf Acknowledgements.} This work was supported by EC-FET European Graphene Flagship Core3 Project, EC-FET Quantum Flagship Project 2D-SIPC, EPSRC grants EP/S030719/1 and EP/V007033/1, and the Lloyd Register Foundation Nanotechnology Grant.

\nocite{*}

%%%%%%%%%%%%%%%%%%%%%%%%%%%%%%%%%%%%%%
%%%%%%%%%%%%%%%%%%%%%%%%%%%%%%%%%%%%%%
%%%%%%%%%%%%%%%%%%%%%%%%%%%%%%%%%%%%%%
%%%%%%%%%%%%%%%%%%%%%%%%%%%%%%%%%%%%%%

\bibliography{references}

%apsrev4-2.bst 2019-01-14 (MD) hand-edited version of apsrev4-1.bst
%Control: key (0)
%Control: author (8) initials jnrlst
%Control: editor formatted (1) identically to author
%Control: production of article title (0) allowed
%Control: page (0) single
%Control: year (1) truncated
%Control: production of eprint (0) enabled
\providecommand{\noopsort}[1]{}\providecommand{\singleletter}[1]{#1}%
\begin{thebibliography}{64}%
\makeatletter
\providecommand \@ifxundefined [1]{%
 \@ifx{#1\undefined}
}%
\providecommand \@ifnum [1]{%
 \ifnum #1\expandafter \@firstoftwo
 \else \expandafter \@secondoftwo
 \fi
}%
\providecommand \@ifx [1]{%
 \ifx #1\expandafter \@firstoftwo
 \else \expandafter \@secondoftwo
 \fi
}%
\providecommand \natexlab [1]{#1}%
\providecommand \enquote  [1]{``#1''}%
\providecommand \bibnamefont  [1]{#1}%
\providecommand \bibfnamefont [1]{#1}%
\providecommand \citenamefont [1]{#1}%
\providecommand \href@noop [0]{\@secondoftwo}%
\providecommand \href [0]{\begingroup \@sanitize@url \@href}%
\providecommand \@href[1]{\@@startlink{#1}\@@href}%
\providecommand \@@href[1]{\endgroup#1\@@endlink}%
\providecommand \@sanitize@url [0]{\catcode `\\12\catcode `\$12\catcode `\&12\catcode `\#12\catcode `\^12\catcode `\_12\catcode `\%12\relax}%
\providecommand \@@startlink[1]{}%
\providecommand \@@endlink[0]{}%
\providecommand \url  [0]{\begingroup\@sanitize@url \@url }%
\providecommand \@url [1]{\endgroup\@href {#1}{\urlprefix }}%
\providecommand \urlprefix  [0]{URL }%
\providecommand \Eprint [0]{\href }%
\providecommand \doibase [0]{https://doi.org/}%
\providecommand \selectlanguage [0]{\@gobble}%
\providecommand \bibinfo  [0]{\@secondoftwo}%
\providecommand \bibfield  [0]{\@secondoftwo}%
\providecommand \translation [1]{[#1]}%
\providecommand \BibitemOpen [0]{}%
\providecommand \bibitemStop [0]{}%
\providecommand \bibitemNoStop [0]{.\EOS\space}%
\providecommand \EOS [0]{\spacefactor3000\relax}%
\providecommand \BibitemShut  [1]{\csname bibitem#1\endcsname}%
\let\auto@bib@innerbib\@empty
%</preamble>
\bibitem [{\citenamefont {Sung}\ \emph {et~al.}(2020)\citenamefont {Sung}, \citenamefont {Zhou}, \citenamefont {Scuri}, \citenamefont {Z{\'o}lyomi}, \citenamefont {Andersen}, \citenamefont {Yoo}, \citenamefont {Wild}, \citenamefont {Joe}, \citenamefont {Gelly}, \citenamefont {Heo}, \citenamefont {Magorrian}, \citenamefont {B{\'e}rub{\'e}}, \citenamefont {Valdivia}, \citenamefont {Taniguchi}, \citenamefont {Watanabe}, \citenamefont {Lukin}, \citenamefont {Kim}, \citenamefont {Fal'ko},\ and\ \citenamefont {Park}}]{Sung2020}%
  \BibitemOpen
  \bibfield  {author} {\bibinfo {author} {\bibfnamefont {J.}~\bibnamefont {Sung}}, \bibinfo {author} {\bibfnamefont {Y.}~\bibnamefont {Zhou}}, \bibinfo {author} {\bibfnamefont {G.}~\bibnamefont {Scuri}}, \bibinfo {author} {\bibfnamefont {V.}~\bibnamefont {Z{\'o}lyomi}}, \bibinfo {author} {\bibfnamefont {T.~I.}\ \bibnamefont {Andersen}}, \bibinfo {author} {\bibfnamefont {H.}~\bibnamefont {Yoo}}, \bibinfo {author} {\bibfnamefont {D.~S.}\ \bibnamefont {Wild}}, \bibinfo {author} {\bibfnamefont {A.~Y.}\ \bibnamefont {Joe}}, \bibinfo {author} {\bibfnamefont {R.~J.}\ \bibnamefont {Gelly}}, \bibinfo {author} {\bibfnamefont {H.}~\bibnamefont {Heo}}, \bibinfo {author} {\bibfnamefont {S.~J.}\ \bibnamefont {Magorrian}}, \bibinfo {author} {\bibfnamefont {D.}~\bibnamefont {B{\'e}rub{\'e}}}, \bibinfo {author} {\bibfnamefont {A.~M.~M.}\ \bibnamefont {Valdivia}}, \bibinfo {author} {\bibfnamefont {T.}~\bibnamefont {Taniguchi}}, \bibinfo {author} {\bibfnamefont {K.}~\bibnamefont {Watanabe}}, \bibinfo {author} {\bibfnamefont
  {M.~D.}\ \bibnamefont {Lukin}}, \bibinfo {author} {\bibfnamefont {P.}~\bibnamefont {Kim}}, \bibinfo {author} {\bibfnamefont {V.~I.}\ \bibnamefont {Fal'ko}},\ and\ \bibinfo {author} {\bibfnamefont {H.}~\bibnamefont {Park}},\ }\bibfield  {title} {\bibinfo {title} {Broken mirror symmetry in excitonic response of reconstructed domains in twisted {MoSe}$_2$/{MoSe}$_2$ bilayers},\ }\href {https://doi.org/10.1038/s41565-020-0728-z} {\bibfield  {journal} {\bibinfo  {journal} {Nature Nanotechnology}\ }\textbf {\bibinfo {volume} {15}},\ \bibinfo {pages} {750} (\bibinfo {year} {2020})}\BibitemShut {NoStop}%
\bibitem [{\citenamefont {Chen}\ \emph {et~al.}(2022)\citenamefont {Chen}, \citenamefont {Lian}, \citenamefont {Huang}, \citenamefont {Su}, \citenamefont {Rashetnia}, \citenamefont {Ma}, \citenamefont {Yan}, \citenamefont {Blei}, \citenamefont {Xiang}, \citenamefont {Taniguchi}, \citenamefont {Watanabe}, \citenamefont {Tongay}, \citenamefont {Smirnov}, \citenamefont {Wang}, \citenamefont {Zhang}, \citenamefont {Cui},\ and\ \citenamefont {Shi}}]{Chen2022}%
  \BibitemOpen
  \bibfield  {author} {\bibinfo {author} {\bibfnamefont {D.}~\bibnamefont {Chen}}, \bibinfo {author} {\bibfnamefont {Z.}~\bibnamefont {Lian}}, \bibinfo {author} {\bibfnamefont {X.}~\bibnamefont {Huang}}, \bibinfo {author} {\bibfnamefont {Y.}~\bibnamefont {Su}}, \bibinfo {author} {\bibfnamefont {M.}~\bibnamefont {Rashetnia}}, \bibinfo {author} {\bibfnamefont {L.}~\bibnamefont {Ma}}, \bibinfo {author} {\bibfnamefont {L.}~\bibnamefont {Yan}}, \bibinfo {author} {\bibfnamefont {M.}~\bibnamefont {Blei}}, \bibinfo {author} {\bibfnamefont {L.}~\bibnamefont {Xiang}}, \bibinfo {author} {\bibfnamefont {T.}~\bibnamefont {Taniguchi}}, \bibinfo {author} {\bibfnamefont {K.}~\bibnamefont {Watanabe}}, \bibinfo {author} {\bibfnamefont {S.}~\bibnamefont {Tongay}}, \bibinfo {author} {\bibfnamefont {D.}~\bibnamefont {Smirnov}}, \bibinfo {author} {\bibfnamefont {Z.}~\bibnamefont {Wang}}, \bibinfo {author} {\bibfnamefont {C.}~\bibnamefont {Zhang}}, \bibinfo {author} {\bibfnamefont {Y.-T.}\ \bibnamefont {Cui}},\ and\ \bibinfo
  {author} {\bibfnamefont {S.-F.}\ \bibnamefont {Shi}},\ }\bibfield  {title} {\bibinfo {title} {{Excitonic Insulator in a Heterojunction Moir{\'{e}} Superlattice}},\ }\href {https://doi.org/10.1038/s41567-022-01703-y} {\bibfield  {journal} {\bibinfo  {journal} {Nature Physics}\ }\textbf {\bibinfo {volume} {18}},\ \bibinfo {pages} {1171} (\bibinfo {year} {2022})}\BibitemShut {NoStop}%
\bibitem [{\citenamefont {Yoo}\ \emph {et~al.}(2019)\citenamefont {Yoo}, \citenamefont {Engelke}, \citenamefont {Carr}, \citenamefont {Fang}, \citenamefont {Zhang}, \citenamefont {Cazeaux}, \citenamefont {Sung}, \citenamefont {Hovden}, \citenamefont {Tsen}, \citenamefont {Taniguchi},\ and\ \citenamefont {et~al}}]{yoo2019}%
  \BibitemOpen
  \bibfield  {author} {\bibinfo {author} {\bibfnamefont {H.}~\bibnamefont {Yoo}}, \bibinfo {author} {\bibfnamefont {R.}~\bibnamefont {Engelke}}, \bibinfo {author} {\bibfnamefont {S.}~\bibnamefont {Carr}}, \bibinfo {author} {\bibfnamefont {S.}~\bibnamefont {Fang}}, \bibinfo {author} {\bibfnamefont {K.}~\bibnamefont {Zhang}}, \bibinfo {author} {\bibfnamefont {P.}~\bibnamefont {Cazeaux}}, \bibinfo {author} {\bibfnamefont {S.~H.}\ \bibnamefont {Sung}}, \bibinfo {author} {\bibfnamefont {R.}~\bibnamefont {Hovden}}, \bibinfo {author} {\bibfnamefont {A.~W.}\ \bibnamefont {Tsen}}, \bibinfo {author} {\bibfnamefont {T.}~\bibnamefont {Taniguchi}},\ and\ \bibinfo {author} {\bibnamefont {et~al}},\ }\bibfield  {title} {\bibinfo {title} {Atomic and electronic reconstruction at the van der {W}aals interface in twisted bilayer graphene},\ }\href {https://doi.org/10.1038/s41563-019-0346-z} {\bibfield  {journal} {\bibinfo  {journal} {Nature materials}\ }\textbf {\bibinfo {volume} {18}},\ \bibinfo {pages} {448} (\bibinfo {year}
  {2019})}\BibitemShut {NoStop}%
\bibitem [{\citenamefont {Weston}\ \emph {et~al.}(2020)\citenamefont {Weston}, \citenamefont {Zou}, \citenamefont {Enaldiev}, \citenamefont {Summerfield}, \citenamefont {Clark}, \citenamefont {Z{\'o}lyomi}, \citenamefont {Graham}, \citenamefont {Yelgel}, \citenamefont {Magorrian}, \citenamefont {Zhou}, \citenamefont {Zultak}, \citenamefont {Hopkinson}, \citenamefont {Barinov}, \citenamefont {Bointon}, \citenamefont {Kretinin}, \citenamefont {Wilson}, \citenamefont {Beton}, \citenamefont {Fal'ko}, \citenamefont {Haigh},\ and\ \citenamefont {Gorbachev}}]{Weston2020}%
  \BibitemOpen
  \bibfield  {author} {\bibinfo {author} {\bibfnamefont {A.}~\bibnamefont {Weston}}, \bibinfo {author} {\bibfnamefont {Y.}~\bibnamefont {Zou}}, \bibinfo {author} {\bibfnamefont {V.}~\bibnamefont {Enaldiev}}, \bibinfo {author} {\bibfnamefont {A.}~\bibnamefont {Summerfield}}, \bibinfo {author} {\bibfnamefont {N.}~\bibnamefont {Clark}}, \bibinfo {author} {\bibfnamefont {V.}~\bibnamefont {Z{\'o}lyomi}}, \bibinfo {author} {\bibfnamefont {A.}~\bibnamefont {Graham}}, \bibinfo {author} {\bibfnamefont {C.}~\bibnamefont {Yelgel}}, \bibinfo {author} {\bibfnamefont {S.}~\bibnamefont {Magorrian}}, \bibinfo {author} {\bibfnamefont {M.}~\bibnamefont {Zhou}}, \bibinfo {author} {\bibfnamefont {J.}~\bibnamefont {Zultak}}, \bibinfo {author} {\bibfnamefont {D.}~\bibnamefont {Hopkinson}}, \bibinfo {author} {\bibfnamefont {A.}~\bibnamefont {Barinov}}, \bibinfo {author} {\bibfnamefont {T.~H.}\ \bibnamefont {Bointon}}, \bibinfo {author} {\bibfnamefont {A.}~\bibnamefont {Kretinin}}, \bibinfo {author} {\bibfnamefont {N.~R.}\
  \bibnamefont {Wilson}}, \bibinfo {author} {\bibfnamefont {P.~H.}\ \bibnamefont {Beton}}, \bibinfo {author} {\bibfnamefont {V.~I.}\ \bibnamefont {Fal'ko}}, \bibinfo {author} {\bibfnamefont {S.~J.}\ \bibnamefont {Haigh}},\ and\ \bibinfo {author} {\bibfnamefont {R.}~\bibnamefont {Gorbachev}},\ }\bibfield  {title} {\bibinfo {title} {Atomic reconstruction in twisted bilayers of transition metal dichalcogenides},\ }\href {https://doi.org/10.1038/s41565-020-0682-9} {\bibfield  {journal} {\bibinfo  {journal} {Nature Nanotechnology}\ }\textbf {\bibinfo {volume} {15}},\ \bibinfo {pages} {592} (\bibinfo {year} {2020})}\BibitemShut {NoStop}%
\bibitem [{\citenamefont {Rosenberger}\ \emph {et~al.}(2020)\citenamefont {Rosenberger}, \citenamefont {Chuang}, \citenamefont {Phillips}, \citenamefont {Oleshko}, \citenamefont {McCreary}, \citenamefont {Sivaram}, \citenamefont {Hellberg},\ and\ \citenamefont {Jonker}}]{rosenberger2020}%
  \BibitemOpen
  \bibfield  {author} {\bibinfo {author} {\bibfnamefont {M.~R.}\ \bibnamefont {Rosenberger}}, \bibinfo {author} {\bibfnamefont {H.-J.}\ \bibnamefont {Chuang}}, \bibinfo {author} {\bibfnamefont {M.}~\bibnamefont {Phillips}}, \bibinfo {author} {\bibfnamefont {V.~P.}\ \bibnamefont {Oleshko}}, \bibinfo {author} {\bibfnamefont {K.~M.}\ \bibnamefont {McCreary}}, \bibinfo {author} {\bibfnamefont {S.~V.}\ \bibnamefont {Sivaram}}, \bibinfo {author} {\bibfnamefont {C.~S.}\ \bibnamefont {Hellberg}},\ and\ \bibinfo {author} {\bibfnamefont {B.~T.}\ \bibnamefont {Jonker}},\ }\bibfield  {title} {\bibinfo {title} {Twist angle-dependent atomic reconstruction and moir{\'e} patterns in transition metal dichalcogenide heterostructures},\ }\href {https://doi.org/10.1021/acsnano.0c00088} {\bibfield  {journal} {\bibinfo  {journal} {ACS Nano}\ }\textbf {\bibinfo {volume} {14}},\ \bibinfo {pages} {4550} (\bibinfo {year} {2020})}\BibitemShut {NoStop}%
\bibitem [{\citenamefont {Halbertal}\ \emph {et~al.}(2021)\citenamefont {Halbertal}, \citenamefont {Finney}, \citenamefont {Sunku}, \citenamefont {Kerelsky}, \citenamefont {Rubio-Verd{\'u}}, \citenamefont {Shabani}, \citenamefont {Xian}, \citenamefont {Carr}, \citenamefont {Chen}, \citenamefont {Zhang}, \citenamefont {Wang}, \citenamefont {Gonzalez-Acevedo}, \citenamefont {McLeod}, \citenamefont {Rhodes}, \citenamefont {Watanabe}, \citenamefont {Taniguchi}, \citenamefont {Kaxiras}, \citenamefont {Dean}, \citenamefont {Hone}, \citenamefont {Pasupathy}, \citenamefont {Kennes}, \citenamefont {Rubio},\ and\ \citenamefont {Basov}}]{Halbertal2021}%
  \BibitemOpen
  \bibfield  {author} {\bibinfo {author} {\bibfnamefont {D.}~\bibnamefont {Halbertal}}, \bibinfo {author} {\bibfnamefont {N.~R.}\ \bibnamefont {Finney}}, \bibinfo {author} {\bibfnamefont {S.~S.}\ \bibnamefont {Sunku}}, \bibinfo {author} {\bibfnamefont {A.}~\bibnamefont {Kerelsky}}, \bibinfo {author} {\bibfnamefont {C.}~\bibnamefont {Rubio-Verd{\'u}}}, \bibinfo {author} {\bibfnamefont {S.}~\bibnamefont {Shabani}}, \bibinfo {author} {\bibfnamefont {L.}~\bibnamefont {Xian}}, \bibinfo {author} {\bibfnamefont {S.}~\bibnamefont {Carr}}, \bibinfo {author} {\bibfnamefont {S.}~\bibnamefont {Chen}}, \bibinfo {author} {\bibfnamefont {C.}~\bibnamefont {Zhang}}, \bibinfo {author} {\bibfnamefont {L.}~\bibnamefont {Wang}}, \bibinfo {author} {\bibfnamefont {D.}~\bibnamefont {Gonzalez-Acevedo}}, \bibinfo {author} {\bibfnamefont {A.~S.}\ \bibnamefont {McLeod}}, \bibinfo {author} {\bibfnamefont {D.}~\bibnamefont {Rhodes}}, \bibinfo {author} {\bibfnamefont {K.}~\bibnamefont {Watanabe}}, \bibinfo {author} {\bibfnamefont
  {T.}~\bibnamefont {Taniguchi}}, \bibinfo {author} {\bibfnamefont {E.}~\bibnamefont {Kaxiras}}, \bibinfo {author} {\bibfnamefont {C.~R.}\ \bibnamefont {Dean}}, \bibinfo {author} {\bibfnamefont {J.~C.}\ \bibnamefont {Hone}}, \bibinfo {author} {\bibfnamefont {A.~N.}\ \bibnamefont {Pasupathy}}, \bibinfo {author} {\bibfnamefont {D.~M.}\ \bibnamefont {Kennes}}, \bibinfo {author} {\bibfnamefont {A.}~\bibnamefont {Rubio}},\ and\ \bibinfo {author} {\bibfnamefont {D.~N.}\ \bibnamefont {Basov}},\ }\bibfield  {title} {\bibinfo {title} {Moir{\'e} metrology of energy landscapes in van der waals heterostructures},\ }\href {https://doi.org/10.1038/s41467-020-20428-1} {\bibfield  {journal} {\bibinfo  {journal} {Nature Communications}\ }\textbf {\bibinfo {volume} {12}},\ \bibinfo {pages} {242} (\bibinfo {year} {2021})}\BibitemShut {NoStop}%
\bibitem [{\citenamefont {McGilly}\ \emph {et~al.}(2020)\citenamefont {McGilly}, \citenamefont {Kerelsky}, \citenamefont {Finney}, \citenamefont {Shapovalov}, \citenamefont {Shih}, \citenamefont {Ghiotto}, \citenamefont {Zeng}, \citenamefont {Moore}, \citenamefont {Wu}, \citenamefont {Bai}, \citenamefont {Watanabe}, \citenamefont {Taniguchi}, \citenamefont {Stengel}, \citenamefont {Zhou}, \citenamefont {Hone}, \citenamefont {Zhu}, \citenamefont {Basov}, \citenamefont {Dean}, \citenamefont {Dreyer},\ and\ \citenamefont {Pasupathy}}]{McGilly2020}%
  \BibitemOpen
  \bibfield  {author} {\bibinfo {author} {\bibfnamefont {L.~J.}\ \bibnamefont {McGilly}}, \bibinfo {author} {\bibfnamefont {A.}~\bibnamefont {Kerelsky}}, \bibinfo {author} {\bibfnamefont {N.~R.}\ \bibnamefont {Finney}}, \bibinfo {author} {\bibfnamefont {K.}~\bibnamefont {Shapovalov}}, \bibinfo {author} {\bibfnamefont {E.-M.}\ \bibnamefont {Shih}}, \bibinfo {author} {\bibfnamefont {A.}~\bibnamefont {Ghiotto}}, \bibinfo {author} {\bibfnamefont {Y.}~\bibnamefont {Zeng}}, \bibinfo {author} {\bibfnamefont {S.~L.}\ \bibnamefont {Moore}}, \bibinfo {author} {\bibfnamefont {W.}~\bibnamefont {Wu}}, \bibinfo {author} {\bibfnamefont {Y.}~\bibnamefont {Bai}}, \bibinfo {author} {\bibfnamefont {K.}~\bibnamefont {Watanabe}}, \bibinfo {author} {\bibfnamefont {T.}~\bibnamefont {Taniguchi}}, \bibinfo {author} {\bibfnamefont {M.}~\bibnamefont {Stengel}}, \bibinfo {author} {\bibfnamefont {L.}~\bibnamefont {Zhou}}, \bibinfo {author} {\bibfnamefont {J.}~\bibnamefont {Hone}}, \bibinfo {author} {\bibfnamefont {X.}~\bibnamefont
  {Zhu}}, \bibinfo {author} {\bibfnamefont {D.~N.}\ \bibnamefont {Basov}}, \bibinfo {author} {\bibfnamefont {C.}~\bibnamefont {Dean}}, \bibinfo {author} {\bibfnamefont {C.~E.}\ \bibnamefont {Dreyer}},\ and\ \bibinfo {author} {\bibfnamefont {A.~N.}\ \bibnamefont {Pasupathy}},\ }\bibfield  {title} {\bibinfo {title} {Visualization of moir\'e superlattices},\ }\href {https://doi.org/10.1038/s41565-020-0708-3} {\bibfield  {journal} {\bibinfo  {journal} {Nature Nanotechnology}\ }\textbf {\bibinfo {volume} {15}},\ \bibinfo {pages} {580} (\bibinfo {year} {2020})}\BibitemShut {NoStop}%
\bibitem [{\citenamefont {Shabani}\ \emph {et~al.}(2021)\citenamefont {Shabani}, \citenamefont {Halbertal}, \citenamefont {Wu}, \citenamefont {Chen}, \citenamefont {Liu}, \citenamefont {Hone}, \citenamefont {Yao}, \citenamefont {Basov}, \citenamefont {Zhu},\ and\ \citenamefont {Pasupathy}}]{Shabani2021}%
  \BibitemOpen
  \bibfield  {author} {\bibinfo {author} {\bibfnamefont {S.}~\bibnamefont {Shabani}}, \bibinfo {author} {\bibfnamefont {D.}~\bibnamefont {Halbertal}}, \bibinfo {author} {\bibfnamefont {W.}~\bibnamefont {Wu}}, \bibinfo {author} {\bibfnamefont {M.}~\bibnamefont {Chen}}, \bibinfo {author} {\bibfnamefont {S.}~\bibnamefont {Liu}}, \bibinfo {author} {\bibfnamefont {J.}~\bibnamefont {Hone}}, \bibinfo {author} {\bibfnamefont {W.}~\bibnamefont {Yao}}, \bibinfo {author} {\bibfnamefont {D.~N.}\ \bibnamefont {Basov}}, \bibinfo {author} {\bibfnamefont {X.}~\bibnamefont {Zhu}},\ and\ \bibinfo {author} {\bibfnamefont {A.~N.}\ \bibnamefont {Pasupathy}},\ }\bibfield  {title} {\bibinfo {title} {Deep moir{\'{e}} potentials in twisted transition metal dichalcogenide bilayers},\ }\href {https://doi.org/10.1038/s41567-021-01174-7} {\bibfield  {journal} {\bibinfo  {journal} {Nature Physics}\ }\textbf {\bibinfo {volume} {17}},\ \bibinfo {pages} {720} (\bibinfo {year} {2021})}\BibitemShut {NoStop}%
\bibitem [{\citenamefont {Edelberg}\ \emph {et~al.}(2020)\citenamefont {Edelberg}, \citenamefont {Kumar}, \citenamefont {Shenoy}, \citenamefont {Ochoa},\ and\ \citenamefont {Pasupathy}}]{Edelberg2020}%
  \BibitemOpen
  \bibfield  {author} {\bibinfo {author} {\bibfnamefont {D.}~\bibnamefont {Edelberg}}, \bibinfo {author} {\bibfnamefont {H.}~\bibnamefont {Kumar}}, \bibinfo {author} {\bibfnamefont {V.}~\bibnamefont {Shenoy}}, \bibinfo {author} {\bibfnamefont {H.}~\bibnamefont {Ochoa}},\ and\ \bibinfo {author} {\bibfnamefont {A.~N.}\ \bibnamefont {Pasupathy}},\ }\bibfield  {title} {\bibinfo {title} {Tunable strain soliton networks confine electrons in van der waals materials},\ }\href {https://doi.org/10.1038/s41567-020-0953-2} {\bibfield  {journal} {\bibinfo  {journal} {Nature Physics}\ }\textbf {\bibinfo {volume} {16}},\ \bibinfo {pages} {1097} (\bibinfo {year} {2020})}\BibitemShut {NoStop}%
\bibitem [{\citenamefont {Engelke}\ \emph {et~al.}(2022)\citenamefont {Engelke}, \citenamefont {Yoo}, \citenamefont {Carr}, \citenamefont {Xu}, \citenamefont {Cazeaux}, \citenamefont {Allen}, \citenamefont {Valdivia}, \citenamefont {Luskin}, \citenamefont {Kaxiras}, \citenamefont {Kim},\ and\ \citenamefont {et~al}}]{Engelke2022}%
  \BibitemOpen
  \bibfield  {author} {\bibinfo {author} {\bibfnamefont {R.}~\bibnamefont {Engelke}}, \bibinfo {author} {\bibfnamefont {H.}~\bibnamefont {Yoo}}, \bibinfo {author} {\bibfnamefont {S.}~\bibnamefont {Carr}}, \bibinfo {author} {\bibfnamefont {K.}~\bibnamefont {Xu}}, \bibinfo {author} {\bibfnamefont {P.}~\bibnamefont {Cazeaux}}, \bibinfo {author} {\bibfnamefont {R.}~\bibnamefont {Allen}}, \bibinfo {author} {\bibfnamefont {A.~M.}\ \bibnamefont {Valdivia}}, \bibinfo {author} {\bibfnamefont {M.}~\bibnamefont {Luskin}}, \bibinfo {author} {\bibfnamefont {E.}~\bibnamefont {Kaxiras}}, \bibinfo {author} {\bibfnamefont {M.}~\bibnamefont {Kim}},\ and\ \bibinfo {author} {\bibnamefont {et~al}},\ }\bibfield  {title} {\bibinfo {title} {Non-abelian topological defects and strain mapping in 2d moir$\backslash$'e materials},\ }\href@noop {} {\bibfield  {journal} {\bibinfo  {journal} {arXiv:2207.05276}\ } (\bibinfo {year} {2022})}\BibitemShut {NoStop}%
\bibitem [{\citenamefont {Van~Winkle}\ \emph {et~al.}(2022)\citenamefont {Van~Winkle}, \citenamefont {Craig}, \citenamefont {Carr}, \citenamefont {Dandu}, \citenamefont {Bustillo}, \citenamefont {Ciston}, \citenamefont {Ophus}, \citenamefont {Taniguchi}, \citenamefont {Watanabe}, \citenamefont {Raja}, \citenamefont {Griffin},\ and\ \citenamefont {Bediako}}]{HETTMD_Moire}%
  \BibitemOpen
  \bibfield  {author} {\bibinfo {author} {\bibfnamefont {M.}~\bibnamefont {Van~Winkle}}, \bibinfo {author} {\bibfnamefont {I.~M.}\ \bibnamefont {Craig}}, \bibinfo {author} {\bibfnamefont {S.}~\bibnamefont {Carr}}, \bibinfo {author} {\bibfnamefont {M.}~\bibnamefont {Dandu}}, \bibinfo {author} {\bibfnamefont {K.~C.}\ \bibnamefont {Bustillo}}, \bibinfo {author} {\bibfnamefont {J.}~\bibnamefont {Ciston}}, \bibinfo {author} {\bibfnamefont {C.}~\bibnamefont {Ophus}}, \bibinfo {author} {\bibfnamefont {T.}~\bibnamefont {Taniguchi}}, \bibinfo {author} {\bibfnamefont {K.}~\bibnamefont {Watanabe}}, \bibinfo {author} {\bibfnamefont {A.}~\bibnamefont {Raja}}, \bibinfo {author} {\bibfnamefont {S.~M.}\ \bibnamefont {Griffin}},\ and\ \bibinfo {author} {\bibfnamefont {D.~K.}\ \bibnamefont {Bediako}},\ }\href {https://doi.org/10.48550/ARXIV.2212.07006} {\bibinfo {title} {Quantitative imaging of intrinsic and extrinsic strain in transition metal dichalcogenide moiré bilayers}} (\bibinfo {year} {2022})\BibitemShut {NoStop}%
\bibitem [{\citenamefont {Kazmierczak}\ \emph {et~al.}(2021)\citenamefont {Kazmierczak}, \citenamefont {Winkle}, \citenamefont {Ophus}, \citenamefont {Bustillo}, \citenamefont {Carr}, \citenamefont {Brown}, \citenamefont {Ciston}, \citenamefont {Taniguchi}, \citenamefont {Watanabe},\ and\ \citenamefont {Bediako}}]{Kazmierczak2021}%
  \BibitemOpen
  \bibfield  {author} {\bibinfo {author} {\bibfnamefont {N.~P.}\ \bibnamefont {Kazmierczak}}, \bibinfo {author} {\bibfnamefont {M.~V.}\ \bibnamefont {Winkle}}, \bibinfo {author} {\bibfnamefont {C.}~\bibnamefont {Ophus}}, \bibinfo {author} {\bibfnamefont {K.~C.}\ \bibnamefont {Bustillo}}, \bibinfo {author} {\bibfnamefont {S.}~\bibnamefont {Carr}}, \bibinfo {author} {\bibfnamefont {H.~G.}\ \bibnamefont {Brown}}, \bibinfo {author} {\bibfnamefont {J.}~\bibnamefont {Ciston}}, \bibinfo {author} {\bibfnamefont {T.}~\bibnamefont {Taniguchi}}, \bibinfo {author} {\bibfnamefont {K.}~\bibnamefont {Watanabe}},\ and\ \bibinfo {author} {\bibfnamefont {D.~K.}\ \bibnamefont {Bediako}},\ }\bibfield  {title} {\bibinfo {title} {Strain fields in twisted bilayer graphene},\ }\href {https://doi.org/10.1038/s41563-021-00973-w} {\bibfield  {journal} {\bibinfo  {journal} {Nature Materials}\ }\textbf {\bibinfo {volume} {20}},\ \bibinfo {pages} {956} (\bibinfo {year} {2021})}\BibitemShut {NoStop}%
\bibitem [{\citenamefont {Enaldiev}\ \emph {et~al.}(2022)\citenamefont {Enaldiev}, \citenamefont {Ferreira}, \citenamefont {McHugh},\ and\ \citenamefont {Fal'ko}}]{Enaldiev2022}%
  \BibitemOpen
  \bibfield  {author} {\bibinfo {author} {\bibfnamefont {V.~V.}\ \bibnamefont {Enaldiev}}, \bibinfo {author} {\bibfnamefont {F.}~\bibnamefont {Ferreira}}, \bibinfo {author} {\bibfnamefont {J.~G.}\ \bibnamefont {McHugh}},\ and\ \bibinfo {author} {\bibfnamefont {V.~I.}\ \bibnamefont {Fal'ko}},\ }\bibfield  {title} {\bibinfo {title} {Self-organized quantum dots in marginally twisted {MoSe}2/{WSe}2 and {MoS}2/{WS}2 bilayers},\ }\bibfield  {journal} {\bibinfo  {journal} {npj 2D Materials and Applications}\ }\textbf {\bibinfo {volume} {6}},\ \href {https://doi.org/10.1038/s41699-022-00346-0} {10.1038/s41699-022-00346-0} (\bibinfo {year} {2022})\BibitemShut {NoStop}%
\bibitem [{\citenamefont {He}\ \emph {et~al.}(2018)\citenamefont {He}, \citenamefont {Zhou}, \citenamefont {He}, \citenamefont {Yuan}, \citenamefont {Zhang},\ and\ \citenamefont {Law}}]{He_2018}%
  \BibitemOpen
  \bibfield  {author} {\bibinfo {author} {\bibfnamefont {W.-Y.}\ \bibnamefont {He}}, \bibinfo {author} {\bibfnamefont {B.~T.}\ \bibnamefont {Zhou}}, \bibinfo {author} {\bibfnamefont {J.~J.}\ \bibnamefont {He}}, \bibinfo {author} {\bibfnamefont {N.~F.~Q.}\ \bibnamefont {Yuan}}, \bibinfo {author} {\bibfnamefont {T.}~\bibnamefont {Zhang}},\ and\ \bibinfo {author} {\bibfnamefont {K.~T.}\ \bibnamefont {Law}},\ }\bibfield  {title} {\bibinfo {title} {Magnetic field driven nodal topological superconductivity in monolayer transition metal dichalcogenides},\ }\bibfield  {journal} {\bibinfo  {journal} {Communications Physics}\ }\textbf {\bibinfo {volume} {1}},\ \href {https://doi.org/10.1038/s42005-018-0041-4} {10.1038/s42005-018-0041-4} (\bibinfo {year} {2018})\BibitemShut {NoStop}%
\bibitem [{\citenamefont {Fischer}\ \emph {et~al.}(2018)\citenamefont {Fischer}, \citenamefont {Sigrist},\ and\ \citenamefont {Agterberg}}]{Fischer2018}%
  \BibitemOpen
  \bibfield  {author} {\bibinfo {author} {\bibfnamefont {M.~H.}\ \bibnamefont {Fischer}}, \bibinfo {author} {\bibfnamefont {M.}~\bibnamefont {Sigrist}},\ and\ \bibinfo {author} {\bibfnamefont {D.~F.}\ \bibnamefont {Agterberg}},\ }\bibfield  {title} {\bibinfo {title} {Superconductivity without inversion and time-reversal symmetries},\ }\bibfield  {journal} {\bibinfo  {journal} {Physical Review Letters}\ }\textbf {\bibinfo {volume} {121}},\ \href {https://doi.org/10.1103/physrevlett.121.157003} {10.1103/physrevlett.121.157003} (\bibinfo {year} {2018})\BibitemShut {NoStop}%
\bibitem [{\citenamefont {Bawden}\ \emph {et~al.}(2016)\citenamefont {Bawden}, \citenamefont {Cooil}, \citenamefont {Mazzola}, \citenamefont {Riley}, \citenamefont {Collins-McIntyre}, \citenamefont {Sunko}, \citenamefont {Hunvik}, \citenamefont {Leandersson}, \citenamefont {Polley}, \citenamefont {Balasubramanian}, \citenamefont {Kim}, \citenamefont {Hoesch}, \citenamefont {Wells}, \citenamefont {Balakrishnan}, \citenamefont {Bahramy},\ and\ \citenamefont {King}}]{Bawden2016}%
  \BibitemOpen
  \bibfield  {author} {\bibinfo {author} {\bibfnamefont {L.}~\bibnamefont {Bawden}}, \bibinfo {author} {\bibfnamefont {S.~P.}\ \bibnamefont {Cooil}}, \bibinfo {author} {\bibfnamefont {F.}~\bibnamefont {Mazzola}}, \bibinfo {author} {\bibfnamefont {J.~M.}\ \bibnamefont {Riley}}, \bibinfo {author} {\bibfnamefont {L.~J.}\ \bibnamefont {Collins-McIntyre}}, \bibinfo {author} {\bibfnamefont {V.}~\bibnamefont {Sunko}}, \bibinfo {author} {\bibfnamefont {K.~W.~B.}\ \bibnamefont {Hunvik}}, \bibinfo {author} {\bibfnamefont {M.}~\bibnamefont {Leandersson}}, \bibinfo {author} {\bibfnamefont {C.~M.}\ \bibnamefont {Polley}}, \bibinfo {author} {\bibfnamefont {T.}~\bibnamefont {Balasubramanian}}, \bibinfo {author} {\bibfnamefont {T.~K.}\ \bibnamefont {Kim}}, \bibinfo {author} {\bibfnamefont {M.}~\bibnamefont {Hoesch}}, \bibinfo {author} {\bibfnamefont {J.~W.}\ \bibnamefont {Wells}}, \bibinfo {author} {\bibfnamefont {G.}~\bibnamefont {Balakrishnan}}, \bibinfo {author} {\bibfnamefont {M.~S.}\ \bibnamefont {Bahramy}},\ and\
  \bibinfo {author} {\bibfnamefont {P.~D.~C.}\ \bibnamefont {King}},\ }\bibfield  {title} {\bibinfo {title} {Spin-valley locking in the normal state of a transition-metal dichalcogenide superconductor},\ }\bibfield  {journal} {\bibinfo  {journal} {Nature Communications}\ }\textbf {\bibinfo {volume} {7}},\ \href {https://doi.org/10.1038/ncomms11711} {10.1038/ncomms11711} (\bibinfo {year} {2016})\BibitemShut {NoStop}%
\bibitem [{\citenamefont {de~la Barrera}\ \emph {et~al.}(2018)\citenamefont {de~la Barrera}, \citenamefont {Sinko}, \citenamefont {Gopalan}, \citenamefont {Sivadas}, \citenamefont {Seyler}, \citenamefont {Watanabe}, \citenamefont {Taniguchi}, \citenamefont {Tsen}, \citenamefont {Xu}, \citenamefont {Xiao},\ and\ \citenamefont {Hunt}}]{delaBarrera2018}%
  \BibitemOpen
  \bibfield  {author} {\bibinfo {author} {\bibfnamefont {S.~C.}\ \bibnamefont {de~la Barrera}}, \bibinfo {author} {\bibfnamefont {M.~R.}\ \bibnamefont {Sinko}}, \bibinfo {author} {\bibfnamefont {D.~P.}\ \bibnamefont {Gopalan}}, \bibinfo {author} {\bibfnamefont {N.}~\bibnamefont {Sivadas}}, \bibinfo {author} {\bibfnamefont {K.~L.}\ \bibnamefont {Seyler}}, \bibinfo {author} {\bibfnamefont {K.}~\bibnamefont {Watanabe}}, \bibinfo {author} {\bibfnamefont {T.}~\bibnamefont {Taniguchi}}, \bibinfo {author} {\bibfnamefont {A.~W.}\ \bibnamefont {Tsen}}, \bibinfo {author} {\bibfnamefont {X.}~\bibnamefont {Xu}}, \bibinfo {author} {\bibfnamefont {D.}~\bibnamefont {Xiao}},\ and\ \bibinfo {author} {\bibfnamefont {B.~M.}\ \bibnamefont {Hunt}},\ }\bibfield  {title} {\bibinfo {title} {Tuning ising superconductivity with layer and spin{-orbit coupling in two-dimensional transition-metal dichalcogenides}},\ }\bibfield  {journal} {\bibinfo  {journal} {Nature Communications}\ }\textbf {\bibinfo {volume} {9}},\ \href
  {https://doi.org/10.1038/s41467-018-03888-4} {10.1038/s41467-018-03888-4} (\bibinfo {year} {2018})\BibitemShut {NoStop}%
\bibitem [{\citenamefont {Wickramaratne}\ \emph {et~al.}(2020)\citenamefont {Wickramaratne}, \citenamefont {Khmelevskyi}, \citenamefont {Agterberg},\ and\ \citenamefont {Mazin}}]{Wickramaratne2020}%
  \BibitemOpen
  \bibfield  {author} {\bibinfo {author} {\bibfnamefont {D.}~\bibnamefont {Wickramaratne}}, \bibinfo {author} {\bibfnamefont {S.}~\bibnamefont {Khmelevskyi}}, \bibinfo {author} {\bibfnamefont {D.~F.}\ \bibnamefont {Agterberg}},\ and\ \bibinfo {author} {\bibfnamefont {I.}~\bibnamefont {Mazin}},\ }\bibfield  {title} {\bibinfo {title} {Ising superconductivity and magnetism in nbse2},\ }\bibfield  {journal} {\bibinfo  {journal} {Physical Review X}\ }\textbf {\bibinfo {volume} {10}},\ \href {https://doi.org/10.1103/physrevx.10.041003} {10.1103/physrevx.10.041003} (\bibinfo {year} {2020})\BibitemShut {NoStop}%
\bibitem [{\citenamefont {Johannes}\ \emph {et~al.}(2006)\citenamefont {Johannes}, \citenamefont {Mazin},\ and\ \citenamefont {Howells}}]{Johannes2006}%
  \BibitemOpen
  \bibfield  {author} {\bibinfo {author} {\bibfnamefont {M.}~\bibnamefont {Johannes}}, \bibinfo {author} {\bibfnamefont {I.}~\bibnamefont {Mazin}},\ and\ \bibinfo {author} {\bibfnamefont {C.}~\bibnamefont {Howells}},\ }\bibfield  {title} {\bibinfo {title} {Fermi-surface nesting and the origin of the charge-density wave in nbse2},\ }\bibfield  {journal} {\bibinfo  {journal} {Physical Review B}\ }\textbf {\bibinfo {volume} {73}},\ \href {https://doi.org/10.1103/physrevb.73.205102} {10.1103/physrevb.73.205102} (\bibinfo {year} {2006})\BibitemShut {NoStop}%
\bibitem [{\citenamefont {Xi}\ \emph {et~al.}(2015)\citenamefont {Xi}, \citenamefont {Wang}, \citenamefont {Zhao}, \citenamefont {Park}, \citenamefont {Law}, \citenamefont {Berger}, \citenamefont {Forr{\'{o}}}, \citenamefont {Shan},\ and\ \citenamefont {Mak}}]{Xi_2015}%
  \BibitemOpen
  \bibfield  {author} {\bibinfo {author} {\bibfnamefont {X.}~\bibnamefont {Xi}}, \bibinfo {author} {\bibfnamefont {Z.}~\bibnamefont {Wang}}, \bibinfo {author} {\bibfnamefont {W.}~\bibnamefont {Zhao}}, \bibinfo {author} {\bibfnamefont {J.-H.}\ \bibnamefont {Park}}, \bibinfo {author} {\bibfnamefont {K.~T.}\ \bibnamefont {Law}}, \bibinfo {author} {\bibfnamefont {H.}~\bibnamefont {Berger}}, \bibinfo {author} {\bibfnamefont {L.}~\bibnamefont {Forr{\'{o}}}}, \bibinfo {author} {\bibfnamefont {J.}~\bibnamefont {Shan}},\ and\ \bibinfo {author} {\bibfnamefont {K.~F.}\ \bibnamefont {Mak}},\ }\bibfield  {title} {\bibinfo {title} {Ising pairing in superconducting {NbSe}2 atomic~layers},\ }\href {https://doi.org/10.1038/nphys3538} {\bibfield  {journal} {\bibinfo  {journal} {Nature Physics}\ }\textbf {\bibinfo {volume} {12}},\ \bibinfo {pages} {139} (\bibinfo {year} {2015})}\BibitemShut {NoStop}%
\bibitem [{\citenamefont {Noat}\ \emph {et~al.}(2015)\citenamefont {Noat}, \citenamefont {Silva-Guill{\'{e}}n}, \citenamefont {Cren}, \citenamefont {Cherkez}, \citenamefont {Brun}, \citenamefont {Pons}, \citenamefont {Debontridder}, \citenamefont {Roditchev}, \citenamefont {Sacks}, \citenamefont {Cario}, \citenamefont {Ordej{\'{o}}n}, \citenamefont {Garc{\'{\i}}a},\ and\ \citenamefont {Canadell}}]{Noat2015}%
  \BibitemOpen
  \bibfield  {author} {\bibinfo {author} {\bibfnamefont {Y.}~\bibnamefont {Noat}}, \bibinfo {author} {\bibfnamefont {J.~A.}\ \bibnamefont {Silva-Guill{\'{e}}n}}, \bibinfo {author} {\bibfnamefont {T.}~\bibnamefont {Cren}}, \bibinfo {author} {\bibfnamefont {V.}~\bibnamefont {Cherkez}}, \bibinfo {author} {\bibfnamefont {C.}~\bibnamefont {Brun}}, \bibinfo {author} {\bibfnamefont {S.}~\bibnamefont {Pons}}, \bibinfo {author} {\bibfnamefont {F.}~\bibnamefont {Debontridder}}, \bibinfo {author} {\bibfnamefont {D.}~\bibnamefont {Roditchev}}, \bibinfo {author} {\bibfnamefont {W.}~\bibnamefont {Sacks}}, \bibinfo {author} {\bibfnamefont {L.}~\bibnamefont {Cario}}, \bibinfo {author} {\bibfnamefont {P.}~\bibnamefont {Ordej{\'{o}}n}}, \bibinfo {author} {\bibfnamefont {A.}~\bibnamefont {Garc{\'{\i}}a}},\ and\ \bibinfo {author} {\bibfnamefont {E.}~\bibnamefont {Canadell}},\ }\bibfield  {title} {\bibinfo {title} {Quasiparticle spectra of nbse2: Two-band superconductivity and the role of tunneling selectivity},\ }\bibfield
  {journal} {\bibinfo  {journal} {Physical Review B}\ }\textbf {\bibinfo {volume} {92}},\ \href {https://doi.org/10.1103/physrevb.92.134510} {10.1103/physrevb.92.134510} (\bibinfo {year} {2015})\BibitemShut {NoStop}%
\bibitem [{\citenamefont {Lin}\ \emph {et~al.}(2020)\citenamefont {Lin}, \citenamefont {Li}, \citenamefont {Wen}, \citenamefont {Berger}, \citenamefont {Forr{\'{o}}}, \citenamefont {Zhou}, \citenamefont {Jia}, \citenamefont {Taniguchi}, \citenamefont {Watanabe}, \citenamefont {Xi},\ and\ \citenamefont {Bahramy}}]{Lin2020}%
  \BibitemOpen
  \bibfield  {author} {\bibinfo {author} {\bibfnamefont {D.}~\bibnamefont {Lin}}, \bibinfo {author} {\bibfnamefont {S.}~\bibnamefont {Li}}, \bibinfo {author} {\bibfnamefont {J.}~\bibnamefont {Wen}}, \bibinfo {author} {\bibfnamefont {H.}~\bibnamefont {Berger}}, \bibinfo {author} {\bibfnamefont {L.}~\bibnamefont {Forr{\'{o}}}}, \bibinfo {author} {\bibfnamefont {H.}~\bibnamefont {Zhou}}, \bibinfo {author} {\bibfnamefont {S.}~\bibnamefont {Jia}}, \bibinfo {author} {\bibfnamefont {T.}~\bibnamefont {Taniguchi}}, \bibinfo {author} {\bibfnamefont {K.}~\bibnamefont {Watanabe}}, \bibinfo {author} {\bibfnamefont {X.}~\bibnamefont {Xi}},\ and\ \bibinfo {author} {\bibfnamefont {M.~S.}\ \bibnamefont {Bahramy}},\ }\bibfield  {title} {\bibinfo {title} {Patterns and driving forces of dimensionality-dependent charge density waves in 2h-type transition metal dichalcogenides},\ }\bibfield  {journal} {\bibinfo  {journal} {Nature Communications}\ }\textbf {\bibinfo {volume} {11}},\ \href
  {https://doi.org/10.1038/s41467-020-15715-w} {10.1038/s41467-020-15715-w} (\bibinfo {year} {2020})\BibitemShut {NoStop}%
\bibitem [{\citenamefont {Calandra}\ \emph {et~al.}(2009)\citenamefont {Calandra}, \citenamefont {Mazin},\ and\ \citenamefont {Mauri}}]{Calandra2009}%
  \BibitemOpen
  \bibfield  {author} {\bibinfo {author} {\bibfnamefont {M.}~\bibnamefont {Calandra}}, \bibinfo {author} {\bibfnamefont {I.~I.}\ \bibnamefont {Mazin}},\ and\ \bibinfo {author} {\bibfnamefont {F.}~\bibnamefont {Mauri}},\ }\bibfield  {title} {\bibinfo {title} {Effect of dimensionality on the charge-density wave in few-layer 2h-nbse2},\ }\bibfield  {journal} {\bibinfo  {journal} {Physical Review B}\ }\textbf {\bibinfo {volume} {80}},\ \href {https://doi.org/10.1103/physrevb.80.241108} {10.1103/physrevb.80.241108} (\bibinfo {year} {2009})\BibitemShut {NoStop}%
\bibitem [{\citenamefont {Farrar}\ \emph {et~al.}(2021)\citenamefont {Farrar}, \citenamefont {Nevill}, \citenamefont {Lim}, \citenamefont {Balakrishnan}, \citenamefont {Dale},\ and\ \citenamefont {Bending}}]{Farrar2021}%
  \BibitemOpen
  \bibfield  {author} {\bibinfo {author} {\bibfnamefont {L.~S.}\ \bibnamefont {Farrar}}, \bibinfo {author} {\bibfnamefont {A.}~\bibnamefont {Nevill}}, \bibinfo {author} {\bibfnamefont {Z.~J.}\ \bibnamefont {Lim}}, \bibinfo {author} {\bibfnamefont {G.}~\bibnamefont {Balakrishnan}}, \bibinfo {author} {\bibfnamefont {S.}~\bibnamefont {Dale}},\ and\ \bibinfo {author} {\bibfnamefont {S.~J.}\ \bibnamefont {Bending}},\ }\bibfield  {title} {\bibinfo {title} {Superconducting quantum interference in twisted van der waals heterostructures},\ }\href {https://doi.org/10.1021/acs.nanolett.1c00152} {\bibfield  {journal} {\bibinfo  {journal} {Nano Letters}\ }\textbf {\bibinfo {volume} {21}},\ \bibinfo {pages} {6725} (\bibinfo {year} {2021})}\BibitemShut {NoStop}%
\bibitem [{\citenamefont {Yabuki}\ \emph {et~al.}(2016)\citenamefont {Yabuki}, \citenamefont {Moriya}, \citenamefont {Arai}, \citenamefont {Sata}, \citenamefont {Morikawa}, \citenamefont {Masubuchi},\ and\ \citenamefont {Machida}}]{Yabuki2016}%
  \BibitemOpen
  \bibfield  {author} {\bibinfo {author} {\bibfnamefont {N.}~\bibnamefont {Yabuki}}, \bibinfo {author} {\bibfnamefont {R.}~\bibnamefont {Moriya}}, \bibinfo {author} {\bibfnamefont {M.}~\bibnamefont {Arai}}, \bibinfo {author} {\bibfnamefont {Y.}~\bibnamefont {Sata}}, \bibinfo {author} {\bibfnamefont {S.}~\bibnamefont {Morikawa}}, \bibinfo {author} {\bibfnamefont {S.}~\bibnamefont {Masubuchi}},\ and\ \bibinfo {author} {\bibfnamefont {T.}~\bibnamefont {Machida}},\ }\bibfield  {title} {\bibinfo {title} {Supercurrent in van der waals josephson junction},\ }\bibfield  {journal} {\bibinfo  {journal} {Nature Communications}\ }\textbf {\bibinfo {volume} {7}},\ \href {https://doi.org/10.1038/ncomms10616} {10.1038/ncomms10616} (\bibinfo {year} {2016})\BibitemShut {NoStop}%
\bibitem [{\citenamefont {Zhang}\ \emph {et~al.}(2021)\citenamefont {Zhang}, \citenamefont {Felser},\ and\ \citenamefont {Fu}}]{Zhang_HER}%
  \BibitemOpen
  \bibfield  {author} {\bibinfo {author} {\bibfnamefont {Y.}~\bibnamefont {Zhang}}, \bibinfo {author} {\bibfnamefont {C.}~\bibnamefont {Felser}},\ and\ \bibinfo {author} {\bibfnamefont {L.}~\bibnamefont {Fu}},\ }\href {https://doi.org/10.48550/ARXIV.2111.03058} {\bibinfo {title} {Moiré metal for catalysis}} (\bibinfo {year} {2021})\BibitemShut {NoStop}%
\bibitem [{\citenamefont {Giannozzi}\ \emph {et~al.}(2009)\citenamefont {Giannozzi}, \citenamefont {Baroni}, \citenamefont {Bonini}, \citenamefont {Calandra}, \citenamefont {Car}, \citenamefont {Cavazzoni}, \citenamefont {Ceresoli}, \citenamefont {Chiarotti}, \citenamefont {Cococcioni}, \citenamefont {Dabo}, \citenamefont {Corso}, \citenamefont {de~Gironcoli}, \citenamefont {Fabris}, \citenamefont {Fratesi}, \citenamefont {Gebauer}, \citenamefont {Gerstmann}, \citenamefont {Gougoussis}, \citenamefont {Kokalj}, \citenamefont {Lazzeri}, \citenamefont {Martin-Samos}, \citenamefont {Marzari}, \citenamefont {Mauri}, \citenamefont {Mazzarello}, \citenamefont {Paolini}, \citenamefont {Pasquarello}, \citenamefont {Paulatto}, \citenamefont {Sbraccia}, \citenamefont {Scandolo}, \citenamefont {Sclauzero}, \citenamefont {Seitsonen}, \citenamefont {Smogunov}, \citenamefont {Umari},\ and\ \citenamefont {Wentzcovitch}}]{Giannozzi2009}%
  \BibitemOpen
  \bibfield  {author} {\bibinfo {author} {\bibfnamefont {P.}~\bibnamefont {Giannozzi}}, \bibinfo {author} {\bibfnamefont {S.}~\bibnamefont {Baroni}}, \bibinfo {author} {\bibfnamefont {N.}~\bibnamefont {Bonini}}, \bibinfo {author} {\bibfnamefont {M.}~\bibnamefont {Calandra}}, \bibinfo {author} {\bibfnamefont {R.}~\bibnamefont {Car}}, \bibinfo {author} {\bibfnamefont {C.}~\bibnamefont {Cavazzoni}}, \bibinfo {author} {\bibfnamefont {D.}~\bibnamefont {Ceresoli}}, \bibinfo {author} {\bibfnamefont {G.~L.}\ \bibnamefont {Chiarotti}}, \bibinfo {author} {\bibfnamefont {M.}~\bibnamefont {Cococcioni}}, \bibinfo {author} {\bibfnamefont {I.}~\bibnamefont {Dabo}}, \bibinfo {author} {\bibfnamefont {A.~D.}\ \bibnamefont {Corso}}, \bibinfo {author} {\bibfnamefont {S.}~\bibnamefont {de~Gironcoli}}, \bibinfo {author} {\bibfnamefont {S.}~\bibnamefont {Fabris}}, \bibinfo {author} {\bibfnamefont {G.}~\bibnamefont {Fratesi}}, \bibinfo {author} {\bibfnamefont {R.}~\bibnamefont {Gebauer}}, \bibinfo {author} {\bibfnamefont
  {U.}~\bibnamefont {Gerstmann}}, \bibinfo {author} {\bibfnamefont {C.}~\bibnamefont {Gougoussis}}, \bibinfo {author} {\bibfnamefont {A.}~\bibnamefont {Kokalj}}, \bibinfo {author} {\bibfnamefont {M.}~\bibnamefont {Lazzeri}}, \bibinfo {author} {\bibfnamefont {L.}~\bibnamefont {Martin-Samos}}, \bibinfo {author} {\bibfnamefont {N.}~\bibnamefont {Marzari}}, \bibinfo {author} {\bibfnamefont {F.}~\bibnamefont {Mauri}}, \bibinfo {author} {\bibfnamefont {R.}~\bibnamefont {Mazzarello}}, \bibinfo {author} {\bibfnamefont {S.}~\bibnamefont {Paolini}}, \bibinfo {author} {\bibfnamefont {A.}~\bibnamefont {Pasquarello}}, \bibinfo {author} {\bibfnamefont {L.}~\bibnamefont {Paulatto}}, \bibinfo {author} {\bibfnamefont {C.}~\bibnamefont {Sbraccia}}, \bibinfo {author} {\bibfnamefont {S.}~\bibnamefont {Scandolo}}, \bibinfo {author} {\bibfnamefont {G.}~\bibnamefont {Sclauzero}}, \bibinfo {author} {\bibfnamefont {A.~P.}\ \bibnamefont {Seitsonen}}, \bibinfo {author} {\bibfnamefont {A.}~\bibnamefont {Smogunov}}, \bibinfo {author}
  {\bibfnamefont {P.}~\bibnamefont {Umari}},\ and\ \bibinfo {author} {\bibfnamefont {R.~M.}\ \bibnamefont {Wentzcovitch}},\ }\bibfield  {title} {\bibinfo {title} {{Quantum ESPRESSO}: a modular and open-source software project for quantum simulations of materials},\ }\href {https://doi.org/10.1088/0953-8984/21/39/395502} {\bibfield  {journal} {\bibinfo  {journal} {Journal of Physics: Condensed Matter}\ }\textbf {\bibinfo {volume} {21}},\ \bibinfo {pages} {395502} (\bibinfo {year} {2009})}\BibitemShut {NoStop}%
\bibitem [{\citenamefont {Giannozzi}\ \emph {et~al.}(2017)\citenamefont {Giannozzi}, \citenamefont {Andreussi}, \citenamefont {Brumme}, \citenamefont {Bunau}, \citenamefont {Nardelli}, \citenamefont {Calandra}, \citenamefont {Car}, \citenamefont {Cavazzoni}, \citenamefont {Ceresoli}, \citenamefont {Cococcioni}, \citenamefont {Colonna}, \citenamefont {Carnimeo}, \citenamefont {Corso}, \citenamefont {de~Gironcoli}, \citenamefont {Delugas}, \citenamefont {DiStasio}, \citenamefont {Ferretti}, \citenamefont {Floris}, \citenamefont {Fratesi}, \citenamefont {Fugallo}, \citenamefont {Gebauer}, \citenamefont {Gerstmann}, \citenamefont {Giustino}, \citenamefont {Gorni}, \citenamefont {Jia}, \citenamefont {Kawamura}, \citenamefont {Ko}, \citenamefont {Kokalj}, \citenamefont {K\"{u}{\c{c}}\"{u}kbenli}, \citenamefont {Lazzeri}, \citenamefont {Marsili}, \citenamefont {Marzari}, \citenamefont {Mauri}, \citenamefont {Nguyen}, \citenamefont {Nguyen}, \citenamefont {de-la Roza}, \citenamefont {Paulatto}, \citenamefont
  {Ponc{\'{e}}}, \citenamefont {Rocca}, \citenamefont {Sabatini}, \citenamefont {Santra}, \citenamefont {Schlipf}, \citenamefont {Seitsonen}, \citenamefont {Smogunov}, \citenamefont {Timrov}, \citenamefont {Thonhauser}, \citenamefont {Umari}, \citenamefont {Vast}, \citenamefont {Wu},\ and\ \citenamefont {Baroni}}]{Giannozzi2017}%
  \BibitemOpen
  \bibfield  {author} {\bibinfo {author} {\bibfnamefont {P.}~\bibnamefont {Giannozzi}}, \bibinfo {author} {\bibfnamefont {O.}~\bibnamefont {Andreussi}}, \bibinfo {author} {\bibfnamefont {T.}~\bibnamefont {Brumme}}, \bibinfo {author} {\bibfnamefont {O.}~\bibnamefont {Bunau}}, \bibinfo {author} {\bibfnamefont {M.~B.}\ \bibnamefont {Nardelli}}, \bibinfo {author} {\bibfnamefont {M.}~\bibnamefont {Calandra}}, \bibinfo {author} {\bibfnamefont {R.}~\bibnamefont {Car}}, \bibinfo {author} {\bibfnamefont {C.}~\bibnamefont {Cavazzoni}}, \bibinfo {author} {\bibfnamefont {D.}~\bibnamefont {Ceresoli}}, \bibinfo {author} {\bibfnamefont {M.}~\bibnamefont {Cococcioni}}, \bibinfo {author} {\bibfnamefont {N.}~\bibnamefont {Colonna}}, \bibinfo {author} {\bibfnamefont {I.}~\bibnamefont {Carnimeo}}, \bibinfo {author} {\bibfnamefont {A.~D.}\ \bibnamefont {Corso}}, \bibinfo {author} {\bibfnamefont {S.}~\bibnamefont {de~Gironcoli}}, \bibinfo {author} {\bibfnamefont {P.}~\bibnamefont {Delugas}}, \bibinfo {author} {\bibfnamefont
  {R.~A.}\ \bibnamefont {DiStasio}}, \bibinfo {author} {\bibfnamefont {A.}~\bibnamefont {Ferretti}}, \bibinfo {author} {\bibfnamefont {A.}~\bibnamefont {Floris}}, \bibinfo {author} {\bibfnamefont {G.}~\bibnamefont {Fratesi}}, \bibinfo {author} {\bibfnamefont {G.}~\bibnamefont {Fugallo}}, \bibinfo {author} {\bibfnamefont {R.}~\bibnamefont {Gebauer}}, \bibinfo {author} {\bibfnamefont {U.}~\bibnamefont {Gerstmann}}, \bibinfo {author} {\bibfnamefont {F.}~\bibnamefont {Giustino}}, \bibinfo {author} {\bibfnamefont {T.}~\bibnamefont {Gorni}}, \bibinfo {author} {\bibfnamefont {J.}~\bibnamefont {Jia}}, \bibinfo {author} {\bibfnamefont {M.}~\bibnamefont {Kawamura}}, \bibinfo {author} {\bibfnamefont {H.-Y.}\ \bibnamefont {Ko}}, \bibinfo {author} {\bibfnamefont {A.}~\bibnamefont {Kokalj}}, \bibinfo {author} {\bibfnamefont {E.}~\bibnamefont {K\"{u}{\c{c}}\"{u}kbenli}}, \bibinfo {author} {\bibfnamefont {M.}~\bibnamefont {Lazzeri}}, \bibinfo {author} {\bibfnamefont {M.}~\bibnamefont {Marsili}}, \bibinfo {author}
  {\bibfnamefont {N.}~\bibnamefont {Marzari}}, \bibinfo {author} {\bibfnamefont {F.}~\bibnamefont {Mauri}}, \bibinfo {author} {\bibfnamefont {N.~L.}\ \bibnamefont {Nguyen}}, \bibinfo {author} {\bibfnamefont {H.-V.}\ \bibnamefont {Nguyen}}, \bibinfo {author} {\bibfnamefont {A.~O.}\ \bibnamefont {de-la Roza}}, \bibinfo {author} {\bibfnamefont {L.}~\bibnamefont {Paulatto}}, \bibinfo {author} {\bibfnamefont {S.}~\bibnamefont {Ponc{\'{e}}}}, \bibinfo {author} {\bibfnamefont {D.}~\bibnamefont {Rocca}}, \bibinfo {author} {\bibfnamefont {R.}~\bibnamefont {Sabatini}}, \bibinfo {author} {\bibfnamefont {B.}~\bibnamefont {Santra}}, \bibinfo {author} {\bibfnamefont {M.}~\bibnamefont {Schlipf}}, \bibinfo {author} {\bibfnamefont {A.~P.}\ \bibnamefont {Seitsonen}}, \bibinfo {author} {\bibfnamefont {A.}~\bibnamefont {Smogunov}}, \bibinfo {author} {\bibfnamefont {I.}~\bibnamefont {Timrov}}, \bibinfo {author} {\bibfnamefont {T.}~\bibnamefont {Thonhauser}}, \bibinfo {author} {\bibfnamefont {P.}~\bibnamefont {Umari}}, \bibinfo
  {author} {\bibfnamefont {N.}~\bibnamefont {Vast}}, \bibinfo {author} {\bibfnamefont {X.}~\bibnamefont {Wu}},\ and\ \bibinfo {author} {\bibfnamefont {S.}~\bibnamefont {Baroni}},\ }\bibfield  {title} {\bibinfo {title} {Advanced capabilities for materials modelling with {Quantum ESPRESSO}},\ }\href {https://doi.org/10.1088/1361-648x/aa8f79} {\bibfield  {journal} {\bibinfo  {journal} {Journal of Physics: Condensed Matter}\ }\textbf {\bibinfo {volume} {29}},\ \bibinfo {pages} {465901} (\bibinfo {year} {2017})}\BibitemShut {NoStop}%
\bibitem [{\citenamefont {Garrity}\ \emph {et~al.}(2014)\citenamefont {Garrity}, \citenamefont {Bennett}, \citenamefont {Rabe},\ and\ \citenamefont {Vanderbilt}}]{Garrity2014}%
  \BibitemOpen
  \bibfield  {author} {\bibinfo {author} {\bibfnamefont {K.~F.}\ \bibnamefont {Garrity}}, \bibinfo {author} {\bibfnamefont {J.~W.}\ \bibnamefont {Bennett}}, \bibinfo {author} {\bibfnamefont {K.~M.}\ \bibnamefont {Rabe}},\ and\ \bibinfo {author} {\bibfnamefont {D.}~\bibnamefont {Vanderbilt}},\ }\bibfield  {title} {\bibinfo {title} {Pseudopotentials for high-throughput {DFT} calculations},\ }\href {https://doi.org/10.1016/j.commatsci.2013.08.053} {\bibfield  {journal} {\bibinfo  {journal} {Computational Materials Science}\ }\textbf {\bibinfo {volume} {81}},\ \bibinfo {pages} {446} (\bibinfo {year} {2014})}\BibitemShut {NoStop}%
\bibitem [{\citenamefont {Perdew}\ \emph {et~al.}(1996)\citenamefont {Perdew}, \citenamefont {Burke},\ and\ \citenamefont {Ernzerhof}}]{Perdew1996}%
  \BibitemOpen
  \bibfield  {author} {\bibinfo {author} {\bibfnamefont {J.~P.}\ \bibnamefont {Perdew}}, \bibinfo {author} {\bibfnamefont {K.}~\bibnamefont {Burke}},\ and\ \bibinfo {author} {\bibfnamefont {M.}~\bibnamefont {Ernzerhof}},\ }\bibfield  {title} {\bibinfo {title} {Generalized gradient approximation made simple},\ }\href {https://doi.org/10.1103/physrevlett.77.3865} {\bibfield  {journal} {\bibinfo  {journal} {Physical Review Letters}\ }\textbf {\bibinfo {volume} {77}},\ \bibinfo {pages} {3865} (\bibinfo {year} {1996})}\BibitemShut {NoStop}%
\bibitem [{\citenamefont {Monkhorst}\ and\ \citenamefont {Pack}(1976)}]{Monkhorst1976}%
  \BibitemOpen
  \bibfield  {author} {\bibinfo {author} {\bibfnamefont {H.~J.}\ \bibnamefont {Monkhorst}}\ and\ \bibinfo {author} {\bibfnamefont {J.~D.}\ \bibnamefont {Pack}},\ }\bibfield  {title} {\bibinfo {title} {Special points for brillouin-zone integrations},\ }\href {https://doi.org/10.1103/physrevb.13.5188} {\bibfield  {journal} {\bibinfo  {journal} {Physical Review B}\ }\textbf {\bibinfo {volume} {13}},\ \bibinfo {pages} {5188} (\bibinfo {year} {1976})}\BibitemShut {NoStop}%
\bibitem [{\citenamefont {Thonhauser}\ \emph {et~al.}(2007)\citenamefont {Thonhauser}, \citenamefont {Cooper}, \citenamefont {Li}, \citenamefont {Puzder}, \citenamefont {Hyldgaard},\ and\ \citenamefont {Langreth}}]{Thonhauser2007}%
  \BibitemOpen
  \bibfield  {author} {\bibinfo {author} {\bibfnamefont {T.}~\bibnamefont {Thonhauser}}, \bibinfo {author} {\bibfnamefont {V.~R.}\ \bibnamefont {Cooper}}, \bibinfo {author} {\bibfnamefont {S.}~\bibnamefont {Li}}, \bibinfo {author} {\bibfnamefont {A.}~\bibnamefont {Puzder}}, \bibinfo {author} {\bibfnamefont {P.}~\bibnamefont {Hyldgaard}},\ and\ \bibinfo {author} {\bibfnamefont {D.~C.}\ \bibnamefont {Langreth}},\ }\bibfield  {title} {\bibinfo {title} {Van der waals density functional: Self-consistent potential and the nature of the van der waals bond},\ }\bibfield  {journal} {\bibinfo  {journal} {Physical Review B}\ }\textbf {\bibinfo {volume} {76}},\ \href {https://doi.org/10.1103/physrevb.76.125112} {10.1103/physrevb.76.125112} (\bibinfo {year} {2007})\BibitemShut {NoStop}%
\bibitem [{\citenamefont {Thonhauser}\ \emph {et~al.}(2015)\citenamefont {Thonhauser}, \citenamefont {Zuluaga}, \citenamefont {Arter}, \citenamefont {Berland}, \citenamefont {Schr\"{o}der},\ and\ \citenamefont {Hyldgaard}}]{Thonhauser2015}%
  \BibitemOpen
  \bibfield  {author} {\bibinfo {author} {\bibfnamefont {T.}~\bibnamefont {Thonhauser}}, \bibinfo {author} {\bibfnamefont {S.}~\bibnamefont {Zuluaga}}, \bibinfo {author} {\bibfnamefont {C.}~\bibnamefont {Arter}}, \bibinfo {author} {\bibfnamefont {K.}~\bibnamefont {Berland}}, \bibinfo {author} {\bibfnamefont {E.}~\bibnamefont {Schr\"{o}der}},\ and\ \bibinfo {author} {\bibfnamefont {P.}~\bibnamefont {Hyldgaard}},\ }\bibfield  {title} {\bibinfo {title} {Spin signature of nonlocal correlation binding in metal-organic frameworks},\ }\bibfield  {journal} {\bibinfo  {journal} {Physical Review Letters}\ }\textbf {\bibinfo {volume} {115}},\ \href {https://doi.org/10.1103/physrevlett.115.136402} {10.1103/physrevlett.115.136402} (\bibinfo {year} {2015})\BibitemShut {NoStop}%
\bibitem [{\citenamefont {Berland}\ \emph {et~al.}(2015)\citenamefont {Berland}, \citenamefont {Cooper}, \citenamefont {Lee}, \citenamefont {Schr\"{o}der}, \citenamefont {Thonhauser}, \citenamefont {Hyldgaard},\ and\ \citenamefont {Lundqvist}}]{Berland2015}%
  \BibitemOpen
  \bibfield  {author} {\bibinfo {author} {\bibfnamefont {K.}~\bibnamefont {Berland}}, \bibinfo {author} {\bibfnamefont {V.~R.}\ \bibnamefont {Cooper}}, \bibinfo {author} {\bibfnamefont {K.}~\bibnamefont {Lee}}, \bibinfo {author} {\bibfnamefont {E.}~\bibnamefont {Schr\"{o}der}}, \bibinfo {author} {\bibfnamefont {T.}~\bibnamefont {Thonhauser}}, \bibinfo {author} {\bibfnamefont {P.}~\bibnamefont {Hyldgaard}},\ and\ \bibinfo {author} {\bibfnamefont {B.~I.}\ \bibnamefont {Lundqvist}},\ }\bibfield  {title} {\bibinfo {title} {van der waals forces in density functional theory: a review of the {vdW}-{DF} method},\ }\href {https://doi.org/10.1088/0034-4885/78/6/066501} {\bibfield  {journal} {\bibinfo  {journal} {Reports on Progress in Physics}\ }\textbf {\bibinfo {volume} {78}},\ \bibinfo {pages} {066501} (\bibinfo {year} {2015})}\BibitemShut {NoStop}%
\bibitem [{\citenamefont {Langreth}\ \emph {et~al.}(2009)\citenamefont {Langreth}, \citenamefont {Lundqvist}, \citenamefont {Chakarova-K\"{a}ck}, \citenamefont {Cooper}, \citenamefont {Dion}, \citenamefont {Hyldgaard}, \citenamefont {Kelkkanen}, \citenamefont {Kleis}, \citenamefont {Kong}, \citenamefont {Li}, \citenamefont {Moses}, \citenamefont {Murray}, \citenamefont {Puzder}, \citenamefont {Rydberg}, \citenamefont {Schr\"{o}der},\ and\ \citenamefont {Thonhauser}}]{Langreth2009}%
  \BibitemOpen
  \bibfield  {author} {\bibinfo {author} {\bibfnamefont {D.~C.}\ \bibnamefont {Langreth}}, \bibinfo {author} {\bibfnamefont {B.~I.}\ \bibnamefont {Lundqvist}}, \bibinfo {author} {\bibfnamefont {S.~D.}\ \bibnamefont {Chakarova-K\"{a}ck}}, \bibinfo {author} {\bibfnamefont {V.~R.}\ \bibnamefont {Cooper}}, \bibinfo {author} {\bibfnamefont {M.}~\bibnamefont {Dion}}, \bibinfo {author} {\bibfnamefont {P.}~\bibnamefont {Hyldgaard}}, \bibinfo {author} {\bibfnamefont {A.}~\bibnamefont {Kelkkanen}}, \bibinfo {author} {\bibfnamefont {J.}~\bibnamefont {Kleis}}, \bibinfo {author} {\bibfnamefont {L.}~\bibnamefont {Kong}}, \bibinfo {author} {\bibfnamefont {S.}~\bibnamefont {Li}}, \bibinfo {author} {\bibfnamefont {P.~G.}\ \bibnamefont {Moses}}, \bibinfo {author} {\bibfnamefont {E.}~\bibnamefont {Murray}}, \bibinfo {author} {\bibfnamefont {A.}~\bibnamefont {Puzder}}, \bibinfo {author} {\bibfnamefont {H.}~\bibnamefont {Rydberg}}, \bibinfo {author} {\bibfnamefont {E.}~\bibnamefont {Schr\"{o}der}},\ and\ \bibinfo {author}
  {\bibfnamefont {T.}~\bibnamefont {Thonhauser}},\ }\bibfield  {title} {\bibinfo {title} {A density functional for sparse matter},\ }\href {https://doi.org/10.1088/0953-8984/21/8/084203} {\bibfield  {journal} {\bibinfo  {journal} {Journal of Physics: Condensed Matter}\ }\textbf {\bibinfo {volume} {21}},\ \bibinfo {pages} {084203} (\bibinfo {year} {2009})}\BibitemShut {NoStop}%
\bibitem [{\citenamefont {Meerschaut}\ and\ \citenamefont {Deudon}(2001)}]{Meerschaut2001}%
  \BibitemOpen
  \bibfield  {author} {\bibinfo {author} {\bibfnamefont {A.}~\bibnamefont {Meerschaut}}\ and\ \bibinfo {author} {\bibfnamefont {C.}~\bibnamefont {Deudon}},\ }\bibfield  {title} {\bibinfo {title} {Crystal structure studies of the 3r-nb1.09s2 and the 2h-{NbSe}2 compounds: correlation between nonstoichiometry and stacking type (= polytypism)},\ }\href {https://doi.org/10.1016/s0025-5408(01)00646-8} {\bibfield  {journal} {\bibinfo  {journal} {Materials Research Bulletin}\ }\textbf {\bibinfo {volume} {36}},\ \bibinfo {pages} {1721} (\bibinfo {year} {2001})}\BibitemShut {NoStop}%
\bibitem [{\citenamefont {Kershaw}\ \emph {et~al.}(1967)\citenamefont {Kershaw}, \citenamefont {Vlasse},\ and\ \citenamefont {Wold}}]{Kershaw1967}%
  \BibitemOpen
  \bibfield  {author} {\bibinfo {author} {\bibfnamefont {R.}~\bibnamefont {Kershaw}}, \bibinfo {author} {\bibfnamefont {M.}~\bibnamefont {Vlasse}},\ and\ \bibinfo {author} {\bibfnamefont {A.}~\bibnamefont {Wold}},\ }\bibfield  {title} {\bibinfo {title} {The preparation of and electrical properties of niobium selenide and tungsten selenide},\ }\href {https://doi.org/10.1021/ic50054a043} {\bibfield  {journal} {\bibinfo  {journal} {Inorganic Chemistry}\ }\textbf {\bibinfo {volume} {6}},\ \bibinfo {pages} {1599} (\bibinfo {year} {1967})}\BibitemShut {NoStop}%
\bibitem [{\citenamefont {Feldman}(1976)}]{Feldman1976}%
  \BibitemOpen
  \bibfield  {author} {\bibinfo {author} {\bibfnamefont {J.}~\bibnamefont {Feldman}},\ }\bibfield  {title} {\bibinfo {title} {Elastic constants of 2h-{MoS}2 and 2h-{NbSe}2 extracted from measured dispersion curves and linear compressibilities},\ }\href {https://doi.org/10.1016/0022-3697(76)90143-8} {\bibfield  {journal} {\bibinfo  {journal} {Journal of Physics and Chemistry of Solids}\ }\textbf {\bibinfo {volume} {37}},\ \bibinfo {pages} {1141} (\bibinfo {year} {1976})}\BibitemShut {NoStop}%
\bibitem [{\citenamefont {Lv}\ \emph {et~al.}(2017)\citenamefont {Lv}, \citenamefont {Wei}, \citenamefont {Sun}, \citenamefont {Huang},\ and\ \citenamefont {Dai}}]{Lv2017}%
  \BibitemOpen
  \bibfield  {author} {\bibinfo {author} {\bibfnamefont {X.}~\bibnamefont {Lv}}, \bibinfo {author} {\bibfnamefont {W.}~\bibnamefont {Wei}}, \bibinfo {author} {\bibfnamefont {Q.}~\bibnamefont {Sun}}, \bibinfo {author} {\bibfnamefont {B.}~\bibnamefont {Huang}},\ and\ \bibinfo {author} {\bibfnamefont {Y.}~\bibnamefont {Dai}},\ }\bibfield  {title} {\bibinfo {title} {A first-principles study of nbse2 monolayer as anode materials for rechargeable lithium-ion and sodium-ion batteries},\ }\href {https://doi.org/10.1088/1361-6463/aa6eca} {\bibfield  {journal} {\bibinfo  {journal} {Journal of Physics D: Applied Physics}\ }\textbf {\bibinfo {volume} {50}},\ \bibinfo {pages} {235501} (\bibinfo {year} {2017})}\BibitemShut {NoStop}%
\bibitem [{\citenamefont {Gardos}(1990)}]{Gardos1990}%
  \BibitemOpen
  \bibfield  {author} {\bibinfo {author} {\bibfnamefont {M.}~\bibnamefont {Gardos}},\ }\bibfield  {title} {\bibinfo {title} {On the elastic constants of thin solid lubricant films},\ }in\ \href {https://doi.org/10.1016/s0167-8922(08)70236-0} {\emph {\bibinfo {booktitle} {Tribology Series}}}\ (\bibinfo  {publisher} {Elsevier},\ \bibinfo {year} {1990})\ pp.\ \bibinfo {pages} {3--13}\BibitemShut {NoStop}%
\bibitem [{\citenamefont {Wang}\ \emph {et~al.}(2017)\citenamefont {Wang}, \citenamefont {Huang}, \citenamefont {Lin}, \citenamefont {Cui}, \citenamefont {Chen}, \citenamefont {Zhu}, \citenamefont {Liu}, \citenamefont {Zeng}, \citenamefont {Zhou}, \citenamefont {Yu}, \citenamefont {Wang}, \citenamefont {He}, \citenamefont {Tsang}, \citenamefont {Gao}, \citenamefont {Suenaga}, \citenamefont {Ma}, \citenamefont {Yang}, \citenamefont {Lu}, \citenamefont {Yu}, \citenamefont {Teo}, \citenamefont {Liu},\ and\ \citenamefont {Liu}}]{Wang2017}%
  \BibitemOpen
  \bibfield  {author} {\bibinfo {author} {\bibfnamefont {H.}~\bibnamefont {Wang}}, \bibinfo {author} {\bibfnamefont {X.}~\bibnamefont {Huang}}, \bibinfo {author} {\bibfnamefont {J.}~\bibnamefont {Lin}}, \bibinfo {author} {\bibfnamefont {J.}~\bibnamefont {Cui}}, \bibinfo {author} {\bibfnamefont {Y.}~\bibnamefont {Chen}}, \bibinfo {author} {\bibfnamefont {C.}~\bibnamefont {Zhu}}, \bibinfo {author} {\bibfnamefont {F.}~\bibnamefont {Liu}}, \bibinfo {author} {\bibfnamefont {Q.}~\bibnamefont {Zeng}}, \bibinfo {author} {\bibfnamefont {J.}~\bibnamefont {Zhou}}, \bibinfo {author} {\bibfnamefont {P.}~\bibnamefont {Yu}}, \bibinfo {author} {\bibfnamefont {X.}~\bibnamefont {Wang}}, \bibinfo {author} {\bibfnamefont {H.}~\bibnamefont {He}}, \bibinfo {author} {\bibfnamefont {S.~H.}\ \bibnamefont {Tsang}}, \bibinfo {author} {\bibfnamefont {W.}~\bibnamefont {Gao}}, \bibinfo {author} {\bibfnamefont {K.}~\bibnamefont {Suenaga}}, \bibinfo {author} {\bibfnamefont {F.}~\bibnamefont {Ma}}, \bibinfo {author} {\bibfnamefont
  {C.}~\bibnamefont {Yang}}, \bibinfo {author} {\bibfnamefont {L.}~\bibnamefont {Lu}}, \bibinfo {author} {\bibfnamefont {T.}~\bibnamefont {Yu}}, \bibinfo {author} {\bibfnamefont {E.~H.~T.}\ \bibnamefont {Teo}}, \bibinfo {author} {\bibfnamefont {G.}~\bibnamefont {Liu}},\ and\ \bibinfo {author} {\bibfnamefont {Z.}~\bibnamefont {Liu}},\ }\bibfield  {title} {\bibinfo {title} {High-quality monolayer superconductor {NbSe}2 grown by chemical vapour deposition},\ }\bibfield  {journal} {\bibinfo  {journal} {Nature Communications}\ }\textbf {\bibinfo {volume} {8}},\ \href {https://doi.org/10.1038/s41467-017-00427-5} {10.1038/s41467-017-00427-5} (\bibinfo {year} {2017})\BibitemShut {NoStop}%
\bibitem [{\citenamefont {Valla}\ \emph {et~al.}(2004)\citenamefont {Valla}, \citenamefont {Fedorov}, \citenamefont {Johnson}, \citenamefont {Glans}, \citenamefont {McGuinness}, \citenamefont {Smith}, \citenamefont {Andrei},\ and\ \citenamefont {Berger}}]{Valla2004}%
  \BibitemOpen
  \bibfield  {author} {\bibinfo {author} {\bibfnamefont {T.}~\bibnamefont {Valla}}, \bibinfo {author} {\bibfnamefont {A.~V.}\ \bibnamefont {Fedorov}}, \bibinfo {author} {\bibfnamefont {P.~D.}\ \bibnamefont {Johnson}}, \bibinfo {author} {\bibfnamefont {P.-A.}\ \bibnamefont {Glans}}, \bibinfo {author} {\bibfnamefont {C.}~\bibnamefont {McGuinness}}, \bibinfo {author} {\bibfnamefont {K.~E.}\ \bibnamefont {Smith}}, \bibinfo {author} {\bibfnamefont {E.~Y.}\ \bibnamefont {Andrei}},\ and\ \bibinfo {author} {\bibfnamefont {H.}~\bibnamefont {Berger}},\ }\bibfield  {title} {\bibinfo {title} {Quasiparticle spectra, charge-density waves, superconductivity, and electron-phonon coupling in 2h-nbse2},\ }\bibfield  {journal} {\bibinfo  {journal} {Physical Review Letters}\ }\textbf {\bibinfo {volume} {92}},\ \href {https://doi.org/10.1103/physrevlett.92.086401} {10.1103/physrevlett.92.086401} (\bibinfo {year} {2004})\BibitemShut {NoStop}%
\bibitem [{\citenamefont {Nakata}\ \emph {et~al.}(2018)\citenamefont {Nakata}, \citenamefont {Sugawara}, \citenamefont {Ichinokura}, \citenamefont {Okada}, \citenamefont {Hitosugi}, \citenamefont {Koretsune}, \citenamefont {Ueno}, \citenamefont {Hasegawa}, \citenamefont {Takahashi},\ and\ \citenamefont {Sato}}]{Nakata2018}%
  \BibitemOpen
  \bibfield  {author} {\bibinfo {author} {\bibfnamefont {Y.}~\bibnamefont {Nakata}}, \bibinfo {author} {\bibfnamefont {K.}~\bibnamefont {Sugawara}}, \bibinfo {author} {\bibfnamefont {S.}~\bibnamefont {Ichinokura}}, \bibinfo {author} {\bibfnamefont {Y.}~\bibnamefont {Okada}}, \bibinfo {author} {\bibfnamefont {T.}~\bibnamefont {Hitosugi}}, \bibinfo {author} {\bibfnamefont {T.}~\bibnamefont {Koretsune}}, \bibinfo {author} {\bibfnamefont {K.}~\bibnamefont {Ueno}}, \bibinfo {author} {\bibfnamefont {S.}~\bibnamefont {Hasegawa}}, \bibinfo {author} {\bibfnamefont {T.}~\bibnamefont {Takahashi}},\ and\ \bibinfo {author} {\bibfnamefont {T.}~\bibnamefont {Sato}},\ }\bibfield  {title} {\bibinfo {title} {Anisotropic band splitting in monolayer {NbSe}2: implications for superconductivity and charge density wave},\ }\bibfield  {journal} {\bibinfo  {journal} {npj 2D Materials and Applications}\ }\textbf {\bibinfo {volume} {2}},\ \href {https://doi.org/10.1038/s41699-018-0057-3} {10.1038/s41699-018-0057-3} (\bibinfo {year}
  {2018})\BibitemShut {NoStop}%
\bibitem [{\citenamefont {Rahn}\ \emph {et~al.}(2012)\citenamefont {Rahn}, \citenamefont {Hellmann}, \citenamefont {Kall\"{a}ne}, \citenamefont {Sohrt}, \citenamefont {Kim}, \citenamefont {Kipp},\ and\ \citenamefont {Rossnagel}}]{Rahn2012}%
  \BibitemOpen
  \bibfield  {author} {\bibinfo {author} {\bibfnamefont {D.~J.}\ \bibnamefont {Rahn}}, \bibinfo {author} {\bibfnamefont {S.}~\bibnamefont {Hellmann}}, \bibinfo {author} {\bibfnamefont {M.}~\bibnamefont {Kall\"{a}ne}}, \bibinfo {author} {\bibfnamefont {C.}~\bibnamefont {Sohrt}}, \bibinfo {author} {\bibfnamefont {T.~K.}\ \bibnamefont {Kim}}, \bibinfo {author} {\bibfnamefont {L.}~\bibnamefont {Kipp}},\ and\ \bibinfo {author} {\bibfnamefont {K.}~\bibnamefont {Rossnagel}},\ }\bibfield  {title} {\bibinfo {title} {Gaps and kinks in the electronic structure of the superconductor 2h-nbse2 from angle-resolved photoemission at 1 k},\ }\bibfield  {journal} {\bibinfo  {journal} {Physical Review B}\ }\textbf {\bibinfo {volume} {85}},\ \href {https://doi.org/10.1103/physrevb.85.224532} {10.1103/physrevb.85.224532} (\bibinfo {year} {2012})\BibitemShut {NoStop}%
\bibitem [{\citenamefont {Borisenko}\ \emph {et~al.}(2009)\citenamefont {Borisenko}, \citenamefont {Kordyuk}, \citenamefont {Zabolotnyy}, \citenamefont {Inosov}, \citenamefont {Evtushinsky}, \citenamefont {B\"{u}chner}, \citenamefont {Yaresko}, \citenamefont {Varykhalov}, \citenamefont {Follath}, \citenamefont {Eberhardt}, \citenamefont {Patthey},\ and\ \citenamefont {Berger}}]{Borisenko2009}%
  \BibitemOpen
  \bibfield  {author} {\bibinfo {author} {\bibfnamefont {S.~V.}\ \bibnamefont {Borisenko}}, \bibinfo {author} {\bibfnamefont {A.~A.}\ \bibnamefont {Kordyuk}}, \bibinfo {author} {\bibfnamefont {V.~B.}\ \bibnamefont {Zabolotnyy}}, \bibinfo {author} {\bibfnamefont {D.~S.}\ \bibnamefont {Inosov}}, \bibinfo {author} {\bibfnamefont {D.}~\bibnamefont {Evtushinsky}}, \bibinfo {author} {\bibfnamefont {B.}~\bibnamefont {B\"{u}chner}}, \bibinfo {author} {\bibfnamefont {A.~N.}\ \bibnamefont {Yaresko}}, \bibinfo {author} {\bibfnamefont {A.}~\bibnamefont {Varykhalov}}, \bibinfo {author} {\bibfnamefont {R.}~\bibnamefont {Follath}}, \bibinfo {author} {\bibfnamefont {W.}~\bibnamefont {Eberhardt}}, \bibinfo {author} {\bibfnamefont {L.}~\bibnamefont {Patthey}},\ and\ \bibinfo {author} {\bibfnamefont {H.}~\bibnamefont {Berger}},\ }\bibfield  {title} {\bibinfo {title} {Two energy gaps and fermi-surface arcs in 2h-nbse2},\ }\bibfield  {journal} {\bibinfo  {journal} {Physical Review Letters}\ }\textbf {\bibinfo {volume} {102}},\
  \href {https://doi.org/10.1103/physrevlett.102.166402} {10.1103/physrevlett.102.166402} (\bibinfo {year} {2009})\BibitemShut {NoStop}%
\bibitem [{\citenamefont {Enaldiev}\ \emph {et~al.}(2020)\citenamefont {Enaldiev}, \citenamefont {Z\'olyomi}, \citenamefont {Yelgel}, \citenamefont {Magorrian},\ and\ \citenamefont {Fal'ko}}]{Enaldiev_PRL}%
  \BibitemOpen
  \bibfield  {author} {\bibinfo {author} {\bibfnamefont {V.~V.}\ \bibnamefont {Enaldiev}}, \bibinfo {author} {\bibfnamefont {V.}~\bibnamefont {Z\'olyomi}}, \bibinfo {author} {\bibfnamefont {C.}~\bibnamefont {Yelgel}}, \bibinfo {author} {\bibfnamefont {S.~J.}\ \bibnamefont {Magorrian}},\ and\ \bibinfo {author} {\bibfnamefont {V.~I.}\ \bibnamefont {Fal'ko}},\ }\bibfield  {title} {\bibinfo {title} {Stacking domains and dislocation networks in marginally twisted bilayers of transition metal dichalcogenides},\ }\href {https://doi.org/10.1103/PhysRevLett.124.206101} {\bibfield  {journal} {\bibinfo  {journal} {Phys. Rev. Lett.}\ }\textbf {\bibinfo {volume} {124}},\ \bibinfo {pages} {206101} (\bibinfo {year} {2020})}\BibitemShut {NoStop}%
\bibitem [{\citenamefont {He}\ \emph {et~al.}(2016)\citenamefont {He}, \citenamefont {van Baren}, \citenamefont {Yan}, \citenamefont {Xi}, \citenamefont {Ye}, \citenamefont {Ye}, \citenamefont {Lu}, \citenamefont {Leong},\ and\ \citenamefont {Lui}}]{He_2016}%
  \BibitemOpen
  \bibfield  {author} {\bibinfo {author} {\bibfnamefont {R.}~\bibnamefont {He}}, \bibinfo {author} {\bibfnamefont {J.}~\bibnamefont {van Baren}}, \bibinfo {author} {\bibfnamefont {J.-A.}\ \bibnamefont {Yan}}, \bibinfo {author} {\bibfnamefont {X.}~\bibnamefont {Xi}}, \bibinfo {author} {\bibfnamefont {Z.}~\bibnamefont {Ye}}, \bibinfo {author} {\bibfnamefont {G.}~\bibnamefont {Ye}}, \bibinfo {author} {\bibfnamefont {I.-H.}\ \bibnamefont {Lu}}, \bibinfo {author} {\bibfnamefont {S.~M.}\ \bibnamefont {Leong}},\ and\ \bibinfo {author} {\bibfnamefont {C.~H.}\ \bibnamefont {Lui}},\ }\bibfield  {title} {\bibinfo {title} {Interlayer breathing and shear modes in nbse2 atomic layers},\ }\href {https://doi.org/10.1088/2053-1583/3/3/031008} {\bibfield  {journal} {\bibinfo  {journal} {2D Materials}\ }\textbf {\bibinfo {volume} {3}},\ \bibinfo {pages} {031008} (\bibinfo {year} {2016})}\BibitemShut {NoStop}%
\bibitem [{Note1()}]{Note1}%
  \BibitemOpen
  \bibinfo {note} {Value of critical angles $\theta _{P/AP}^*$ result from equality of gain from formation of domains to the costs of domain wall network.}\BibitemShut {Stop}%
\bibitem [{\citenamefont {Korm{\'{a}}nyos}\ \emph {et~al.}(2015)\citenamefont {Korm{\'{a}}nyos}, \citenamefont {Burkard}, \citenamefont {Gmitra}, \citenamefont {Fabian}, \citenamefont {Z{\'{o}}lyomi}, \citenamefont {Drummond},\ and\ \citenamefont {Fal'ko}}]{Kormnyos2015}%
  \BibitemOpen
  \bibfield  {author} {\bibinfo {author} {\bibfnamefont {A.}~\bibnamefont {Korm{\'{a}}nyos}}, \bibinfo {author} {\bibfnamefont {G.}~\bibnamefont {Burkard}}, \bibinfo {author} {\bibfnamefont {M.}~\bibnamefont {Gmitra}}, \bibinfo {author} {\bibfnamefont {J.}~\bibnamefont {Fabian}}, \bibinfo {author} {\bibfnamefont {V.}~\bibnamefont {Z{\'{o}}lyomi}}, \bibinfo {author} {\bibfnamefont {N.~D.}\ \bibnamefont {Drummond}},\ and\ \bibinfo {author} {\bibfnamefont {V.}~\bibnamefont {Fal'ko}},\ }\bibfield  {title} {\bibinfo {title} {Corrigendum: k.p theory for two-dimensional transition metal dichalcogenide semiconductors},\ }\href {https://doi.org/10.1088/2053-1583/2/4/049501} {\bibfield  {journal} {\bibinfo  {journal} {2D Materials}\ }\textbf {\bibinfo {volume} {2}},\ \bibinfo {pages} {049501} (\bibinfo {year} {2015})}\BibitemShut {NoStop}%
\bibitem [{\citenamefont {Shaffer}\ \emph {et~al.}(2020)\citenamefont {Shaffer}, \citenamefont {Kang}, \citenamefont {Burnell},\ and\ \citenamefont {Fernandes}}]{Shaffer2020}%
  \BibitemOpen
  \bibfield  {author} {\bibinfo {author} {\bibfnamefont {D.}~\bibnamefont {Shaffer}}, \bibinfo {author} {\bibfnamefont {J.}~\bibnamefont {Kang}}, \bibinfo {author} {\bibfnamefont {F.~J.}\ \bibnamefont {Burnell}},\ and\ \bibinfo {author} {\bibfnamefont {R.~M.}\ \bibnamefont {Fernandes}},\ }\bibfield  {title} {\bibinfo {title} {Crystalline nodal topological superconductivity and bogolyubov fermi surfaces in monolayer nbse2},\ }\bibfield  {journal} {\bibinfo  {journal} {Physical Review B}\ }\textbf {\bibinfo {volume} {101}},\ \href {https://doi.org/10.1103/physrevb.101.224503} {10.1103/physrevb.101.224503} (\bibinfo {year} {2020})\BibitemShut {NoStop}%
\bibitem [{\citenamefont {Flicker}(2015)}]{flicker}%
  \BibitemOpen
  \bibfield  {author} {\bibinfo {author} {\bibfnamefont {F.}~\bibnamefont {Flicker}},\ }\emph {\bibinfo {title} {The Geometry and Topology of Charge-Ordered Quantum Fields in Low-Dimensional Systems}},\ \href@noop {} {Ph.D. thesis} (\bibinfo {year} {2015})\BibitemShut {NoStop}%
\bibitem [{\citenamefont {Ferreira}\ \emph {et~al.}(2021)\citenamefont {Ferreira}, \citenamefont {Enaldiev}, \citenamefont {Fal'ko},\ and\ \citenamefont {Magorrian}}]{Ferreira2021}%
  \BibitemOpen
  \bibfield  {author} {\bibinfo {author} {\bibfnamefont {F.}~\bibnamefont {Ferreira}}, \bibinfo {author} {\bibfnamefont {V.~V.}\ \bibnamefont {Enaldiev}}, \bibinfo {author} {\bibfnamefont {V.~I.}\ \bibnamefont {Fal'ko}},\ and\ \bibinfo {author} {\bibfnamefont {S.~J.}\ \bibnamefont {Magorrian}},\ }\bibfield  {title} {\bibinfo {title} {Weak ferroelectric charge transfer in layer-asymmetric bilayers of 2d semiconductors},\ }\bibfield  {journal} {\bibinfo  {journal} {Scientific Reports}\ }\textbf {\bibinfo {volume} {11}},\ \href {https://doi.org/10.1038/s41598-021-92710-1} {10.1038/s41598-021-92710-1} (\bibinfo {year} {2021})\BibitemShut {NoStop}%
\bibitem [{\citenamefont {Ugeda}\ \emph {et~al.}(2015)\citenamefont {Ugeda}, \citenamefont {Bradley}, \citenamefont {Zhang}, \citenamefont {Onishi}, \citenamefont {Chen}, \citenamefont {Ruan}, \citenamefont {Ojeda-Aristizabal}, \citenamefont {Ryu}, \citenamefont {Edmonds}, \citenamefont {Tsai}, \citenamefont {Riss}, \citenamefont {Mo}, \citenamefont {Lee}, \citenamefont {Zettl}, \citenamefont {Hussain}, \citenamefont {Shen},\ and\ \citenamefont {Crommie}}]{Ugeda2015}%
  \BibitemOpen
  \bibfield  {author} {\bibinfo {author} {\bibfnamefont {M.~M.}\ \bibnamefont {Ugeda}}, \bibinfo {author} {\bibfnamefont {A.~J.}\ \bibnamefont {Bradley}}, \bibinfo {author} {\bibfnamefont {Y.}~\bibnamefont {Zhang}}, \bibinfo {author} {\bibfnamefont {S.}~\bibnamefont {Onishi}}, \bibinfo {author} {\bibfnamefont {Y.}~\bibnamefont {Chen}}, \bibinfo {author} {\bibfnamefont {W.}~\bibnamefont {Ruan}}, \bibinfo {author} {\bibfnamefont {C.}~\bibnamefont {Ojeda-Aristizabal}}, \bibinfo {author} {\bibfnamefont {H.}~\bibnamefont {Ryu}}, \bibinfo {author} {\bibfnamefont {M.~T.}\ \bibnamefont {Edmonds}}, \bibinfo {author} {\bibfnamefont {H.-Z.}\ \bibnamefont {Tsai}}, \bibinfo {author} {\bibfnamefont {A.}~\bibnamefont {Riss}}, \bibinfo {author} {\bibfnamefont {S.-K.}\ \bibnamefont {Mo}}, \bibinfo {author} {\bibfnamefont {D.}~\bibnamefont {Lee}}, \bibinfo {author} {\bibfnamefont {A.}~\bibnamefont {Zettl}}, \bibinfo {author} {\bibfnamefont {Z.}~\bibnamefont {Hussain}}, \bibinfo {author} {\bibfnamefont {Z.-X.}\ \bibnamefont
  {Shen}},\ and\ \bibinfo {author} {\bibfnamefont {M.~F.}\ \bibnamefont {Crommie}},\ }\bibfield  {title} {\bibinfo {title} {Characterization of collective ground states in single-layer {NbSe}2},\ }\href {https://doi.org/10.1038/nphys3527} {\bibfield  {journal} {\bibinfo  {journal} {Nature Physics}\ }\textbf {\bibinfo {volume} {12}},\ \bibinfo {pages} {92} (\bibinfo {year} {2015})}\BibitemShut {NoStop}%
\bibitem [{\citenamefont {Dai}\ \emph {et~al.}(2014)\citenamefont {Dai}, \citenamefont {Calleja}, \citenamefont {Alldredge}, \citenamefont {Zhu}, \citenamefont {Li}, \citenamefont {Lu}, \citenamefont {Sun}, \citenamefont {Wolf}, \citenamefont {Berger},\ and\ \citenamefont {McElroy}}]{Dai2014}%
  \BibitemOpen
  \bibfield  {author} {\bibinfo {author} {\bibfnamefont {J.}~\bibnamefont {Dai}}, \bibinfo {author} {\bibfnamefont {E.}~\bibnamefont {Calleja}}, \bibinfo {author} {\bibfnamefont {J.}~\bibnamefont {Alldredge}}, \bibinfo {author} {\bibfnamefont {X.}~\bibnamefont {Zhu}}, \bibinfo {author} {\bibfnamefont {L.}~\bibnamefont {Li}}, \bibinfo {author} {\bibfnamefont {W.}~\bibnamefont {Lu}}, \bibinfo {author} {\bibfnamefont {Y.}~\bibnamefont {Sun}}, \bibinfo {author} {\bibfnamefont {T.}~\bibnamefont {Wolf}}, \bibinfo {author} {\bibfnamefont {H.}~\bibnamefont {Berger}},\ and\ \bibinfo {author} {\bibfnamefont {K.}~\bibnamefont {McElroy}},\ }\bibfield  {title} {\bibinfo {title} {Microscopic evidence for strong periodic lattice distortion in two-dimensional charge-density wave systems},\ }\bibfield  {journal} {\bibinfo  {journal} {Physical Review B}\ }\textbf {\bibinfo {volume} {89}},\ \href {https://doi.org/10.1103/physrevb.89.165140} {10.1103/physrevb.89.165140} (\bibinfo {year} {2014})\BibitemShut {NoStop}%
\bibitem [{\citenamefont {Arguello}\ \emph {et~al.}(2014)\citenamefont {Arguello}, \citenamefont {Chockalingam}, \citenamefont {Rosenthal}, \citenamefont {Zhao}, \citenamefont {Guti{\'{e}}rrez}, \citenamefont {Kang}, \citenamefont {Chung}, \citenamefont {Fernandes}, \citenamefont {Jia}, \citenamefont {Millis}, \citenamefont {Cava},\ and\ \citenamefont {Pasupathy}}]{Arguello2014}%
  \BibitemOpen
  \bibfield  {author} {\bibinfo {author} {\bibfnamefont {C.~J.}\ \bibnamefont {Arguello}}, \bibinfo {author} {\bibfnamefont {S.~P.}\ \bibnamefont {Chockalingam}}, \bibinfo {author} {\bibfnamefont {E.~P.}\ \bibnamefont {Rosenthal}}, \bibinfo {author} {\bibfnamefont {L.}~\bibnamefont {Zhao}}, \bibinfo {author} {\bibfnamefont {C.}~\bibnamefont {Guti{\'{e}}rrez}}, \bibinfo {author} {\bibfnamefont {J.~H.}\ \bibnamefont {Kang}}, \bibinfo {author} {\bibfnamefont {W.~C.}\ \bibnamefont {Chung}}, \bibinfo {author} {\bibfnamefont {R.~M.}\ \bibnamefont {Fernandes}}, \bibinfo {author} {\bibfnamefont {S.}~\bibnamefont {Jia}}, \bibinfo {author} {\bibfnamefont {A.~J.}\ \bibnamefont {Millis}}, \bibinfo {author} {\bibfnamefont {R.~J.}\ \bibnamefont {Cava}},\ and\ \bibinfo {author} {\bibfnamefont {A.~N.}\ \bibnamefont {Pasupathy}},\ }\bibfield  {title} {\bibinfo {title} {Visualizing the charge density wave transition in 2h-nbse2 in real space},\ }\bibfield  {journal} {\bibinfo  {journal} {Physical Review B}\ }\textbf {\bibinfo
  {volume} {89}},\ \href {https://doi.org/10.1103/physrevb.89.235115} {10.1103/physrevb.89.235115} (\bibinfo {year} {2014})\BibitemShut {NoStop}%
\bibitem [{\citenamefont {Silva-Guill{\'{e}}n}\ \emph {et~al.}(2016)\citenamefont {Silva-Guill{\'{e}}n}, \citenamefont {Ordej{\'{o}}n}, \citenamefont {Guinea},\ and\ \citenamefont {Canadell}}]{SilvaGuilln2016}%
  \BibitemOpen
  \bibfield  {author} {\bibinfo {author} {\bibfnamefont {J.~{\'{A}}.}\ \bibnamefont {Silva-Guill{\'{e}}n}}, \bibinfo {author} {\bibfnamefont {P.}~\bibnamefont {Ordej{\'{o}}n}}, \bibinfo {author} {\bibfnamefont {F.}~\bibnamefont {Guinea}},\ and\ \bibinfo {author} {\bibfnamefont {E.}~\bibnamefont {Canadell}},\ }\bibfield  {title} {\bibinfo {title} {Electronic structure of 2h-nbse2 single-layers in the {CDW} state},\ }\href {https://doi.org/10.1088/2053-1583/3/3/035028} {\bibfield  {journal} {\bibinfo  {journal} {2D Materials}\ }\textbf {\bibinfo {volume} {3}},\ \bibinfo {pages} {035028} (\bibinfo {year} {2016})}\BibitemShut {NoStop}%
\bibitem [{\citenamefont {Guster}\ \emph {et~al.}(2019)\citenamefont {Guster}, \citenamefont {Rubio-Verd{\'{u}}}, \citenamefont {Robles}, \citenamefont {Zald{\'{\i}}var}, \citenamefont {Dreher}, \citenamefont {Pruneda}, \citenamefont {Silva-Guill{\'{e}}n}, \citenamefont {Choi}, \citenamefont {Pascual}, \citenamefont {Ugeda}, \citenamefont {Ordej{\'{o}}n},\ and\ \citenamefont {Canadell}}]{Guster2019}%
  \BibitemOpen
  \bibfield  {author} {\bibinfo {author} {\bibfnamefont {B.}~\bibnamefont {Guster}}, \bibinfo {author} {\bibfnamefont {C.}~\bibnamefont {Rubio-Verd{\'{u}}}}, \bibinfo {author} {\bibfnamefont {R.}~\bibnamefont {Robles}}, \bibinfo {author} {\bibfnamefont {J.}~\bibnamefont {Zald{\'{\i}}var}}, \bibinfo {author} {\bibfnamefont {P.}~\bibnamefont {Dreher}}, \bibinfo {author} {\bibfnamefont {M.}~\bibnamefont {Pruneda}}, \bibinfo {author} {\bibfnamefont {J.~{\'{A}}.}\ \bibnamefont {Silva-Guill{\'{e}}n}}, \bibinfo {author} {\bibfnamefont {D.-J.}\ \bibnamefont {Choi}}, \bibinfo {author} {\bibfnamefont {J.~I.}\ \bibnamefont {Pascual}}, \bibinfo {author} {\bibfnamefont {M.~M.}\ \bibnamefont {Ugeda}}, \bibinfo {author} {\bibfnamefont {P.}~\bibnamefont {Ordej{\'{o}}n}},\ and\ \bibinfo {author} {\bibfnamefont {E.}~\bibnamefont {Canadell}},\ }\bibfield  {title} {\bibinfo {title} {Coexistence of elastic modulations in the charge density wave state of 2h-nbse2},\ }\href {https://doi.org/10.1021/acs.nanolett.9b00268} {\bibfield
  {journal} {\bibinfo  {journal} {Nano Letters}\ }\textbf {\bibinfo {volume} {19}},\ \bibinfo {pages} {3027} (\bibinfo {year} {2019})}\BibitemShut {NoStop}%
\bibitem [{\citenamefont {Lian}\ \emph {et~al.}(2018)\citenamefont {Lian}, \citenamefont {Si},\ and\ \citenamefont {Duan}}]{Lian2018}%
  \BibitemOpen
  \bibfield  {author} {\bibinfo {author} {\bibfnamefont {C.-S.}\ \bibnamefont {Lian}}, \bibinfo {author} {\bibfnamefont {C.}~\bibnamefont {Si}},\ and\ \bibinfo {author} {\bibfnamefont {W.}~\bibnamefont {Duan}},\ }\bibfield  {title} {\bibinfo {title} {Unveiling charge-density wave, superconductivity, and their competitive nature in two-dimensional {NbSe}2},\ }\href {https://doi.org/10.1021/acs.nanolett.8b00237} {\bibfield  {journal} {\bibinfo  {journal} {Nano Letters}\ }\textbf {\bibinfo {volume} {18}},\ \bibinfo {pages} {2924} (\bibinfo {year} {2018})}\BibitemShut {NoStop}%
\bibitem [{\citenamefont {Lim}\ \emph {et~al.}(2020)\citenamefont {Lim}, \citenamefont {Kim}, \citenamefont {Won},\ and\ \citenamefont {Cheong}}]{Lim2020}%
  \BibitemOpen
  \bibfield  {author} {\bibinfo {author} {\bibfnamefont {S.}~\bibnamefont {Lim}}, \bibinfo {author} {\bibfnamefont {J.}~\bibnamefont {Kim}}, \bibinfo {author} {\bibfnamefont {C.}~\bibnamefont {Won}},\ and\ \bibinfo {author} {\bibfnamefont {S.-W.}\ \bibnamefont {Cheong}},\ }\bibfield  {title} {\bibinfo {title} {Atomic-scale observation of topological vortices in the incommensurate charge density wave of 2h-{TaSe2}},\ }\href {https://doi.org/10.1021/acs.nanolett.0c00539} {\bibfield  {journal} {\bibinfo  {journal} {Nano Letters}\ }\textbf {\bibinfo {volume} {20}},\ \bibinfo {pages} {4801} (\bibinfo {year} {2020})}\BibitemShut {NoStop}%
\bibitem [{\citenamefont {McMillan}(1976)}]{McMillan1976}%
  \BibitemOpen
  \bibfield  {author} {\bibinfo {author} {\bibfnamefont {W.~L.}\ \bibnamefont {McMillan}},\ }\bibfield  {title} {\bibinfo {title} {Theory of discommensurations and the commensurate-incommensurate charge-density-wave phase transition},\ }\href {https://doi.org/10.1103/physrevb.14.1496} {\bibfield  {journal} {\bibinfo  {journal} {Physical Review B}\ }\textbf {\bibinfo {volume} {14}},\ \bibinfo {pages} {1496} (\bibinfo {year} {1976})}\BibitemShut {NoStop}%
\bibitem [{\citenamefont {Goodwin}\ and\ \citenamefont {Fal'ko}(2022)}]{Goodwin2022}%
  \BibitemOpen
  \bibfield  {author} {\bibinfo {author} {\bibfnamefont {Z.~A.~H.}\ \bibnamefont {Goodwin}}\ and\ \bibinfo {author} {\bibfnamefont {V.~I.}\ \bibnamefont {Fal'ko}},\ }\bibfield  {title} {\bibinfo {title} {Moir{\'{e}} modulation of charge density waves},\ }\href {https://doi.org/10.1088/1361-648x/ac99ca} {\bibfield  {journal} {\bibinfo  {journal} {Journal of Physics: Condensed Matter}\ }\textbf {\bibinfo {volume} {34}},\ \bibinfo {pages} {494001} (\bibinfo {year} {2022})}\BibitemShut {NoStop}%
\bibitem [{\citenamefont {Chen}\ \emph {et~al.}(2019)\citenamefont {Chen}, \citenamefont {Su}, \citenamefont {Neto},\ and\ \citenamefont {Pereira}}]{Chen2019}%
  \BibitemOpen
  \bibfield  {author} {\bibinfo {author} {\bibfnamefont {C.}~\bibnamefont {Chen}}, \bibinfo {author} {\bibfnamefont {L.}~\bibnamefont {Su}}, \bibinfo {author} {\bibfnamefont {A.~H.~C.}\ \bibnamefont {Neto}},\ and\ \bibinfo {author} {\bibfnamefont {V.~M.}\ \bibnamefont {Pereira}},\ }\bibfield  {title} {\bibinfo {title} {Discommensuration-driven superconductivity in the charge density wave phases of transition-metal dichalcogenides},\ }\bibfield  {journal} {\bibinfo  {journal} {Physical Review B}\ }\textbf {\bibinfo {volume} {99}},\ \href {https://doi.org/10.1103/physrevb.99.121108} {10.1103/physrevb.99.121108} (\bibinfo {year} {2019})\BibitemShut {NoStop}%
\bibitem [{\citenamefont {Joe}\ \emph {et~al.}(2014)\citenamefont {Joe}, \citenamefont {Chen}, \citenamefont {Ghaemi}, \citenamefont {Finkelstein}, \citenamefont {de~la Pe{\~{n}}a}, \citenamefont {Gan}, \citenamefont {Lee}, \citenamefont {Yuan}, \citenamefont {Geck}, \citenamefont {MacDougall}, \citenamefont {Chiang}, \citenamefont {Cooper}, \citenamefont {Fradkin},\ and\ \citenamefont {Abbamonte}}]{Joe2014}%
  \BibitemOpen
  \bibfield  {author} {\bibinfo {author} {\bibfnamefont {Y.~I.}\ \bibnamefont {Joe}}, \bibinfo {author} {\bibfnamefont {X.~M.}\ \bibnamefont {Chen}}, \bibinfo {author} {\bibfnamefont {P.}~\bibnamefont {Ghaemi}}, \bibinfo {author} {\bibfnamefont {K.~D.}\ \bibnamefont {Finkelstein}}, \bibinfo {author} {\bibfnamefont {G.~A.}\ \bibnamefont {de~la Pe{\~{n}}a}}, \bibinfo {author} {\bibfnamefont {Y.}~\bibnamefont {Gan}}, \bibinfo {author} {\bibfnamefont {J.~C.~T.}\ \bibnamefont {Lee}}, \bibinfo {author} {\bibfnamefont {S.}~\bibnamefont {Yuan}}, \bibinfo {author} {\bibfnamefont {J.}~\bibnamefont {Geck}}, \bibinfo {author} {\bibfnamefont {G.~J.}\ \bibnamefont {MacDougall}}, \bibinfo {author} {\bibfnamefont {T.~C.}\ \bibnamefont {Chiang}}, \bibinfo {author} {\bibfnamefont {S.~L.}\ \bibnamefont {Cooper}}, \bibinfo {author} {\bibfnamefont {E.}~\bibnamefont {Fradkin}},\ and\ \bibinfo {author} {\bibfnamefont {P.}~\bibnamefont {Abbamonte}},\ }\bibfield  {title} {\bibinfo {title} {Emergence of charge density wave domain
  walls above the superconducting dome in 1t-{TiSe}2},\ }\href {https://doi.org/10.1038/nphys2935} {\bibfield  {journal} {\bibinfo  {journal} {Nature Physics}\ }\textbf {\bibinfo {volume} {10}},\ \bibinfo {pages} {421} (\bibinfo {year} {2014})}\BibitemShut {NoStop}%
\bibitem [{\citenamefont {Leridon}\ \emph {et~al.}(2020)\citenamefont {Leridon}, \citenamefont {Caprara}, \citenamefont {Vanacken}, \citenamefont {Moshchalkov}, \citenamefont {Vignolle}, \citenamefont {Porwal}, \citenamefont {Budhani}, \citenamefont {Attanasi}, \citenamefont {Grilli},\ and\ \citenamefont {Lorenzana}}]{Leridon2020}%
  \BibitemOpen
  \bibfield  {author} {\bibinfo {author} {\bibfnamefont {B.}~\bibnamefont {Leridon}}, \bibinfo {author} {\bibfnamefont {S.}~\bibnamefont {Caprara}}, \bibinfo {author} {\bibfnamefont {J.}~\bibnamefont {Vanacken}}, \bibinfo {author} {\bibfnamefont {V.~V.}\ \bibnamefont {Moshchalkov}}, \bibinfo {author} {\bibfnamefont {B.}~\bibnamefont {Vignolle}}, \bibinfo {author} {\bibfnamefont {R.}~\bibnamefont {Porwal}}, \bibinfo {author} {\bibfnamefont {R.~C.}\ \bibnamefont {Budhani}}, \bibinfo {author} {\bibfnamefont {A.}~\bibnamefont {Attanasi}}, \bibinfo {author} {\bibfnamefont {M.}~\bibnamefont {Grilli}},\ and\ \bibinfo {author} {\bibfnamefont {J.}~\bibnamefont {Lorenzana}},\ }\bibfield  {title} {\bibinfo {title} {Protected superconductivity at the boundaries of charge-density-wave domains},\ }\href {https://doi.org/10.1088/1367-2630/ab976e} {\bibfield  {journal} {\bibinfo  {journal} {New Journal of Physics}\ }\textbf {\bibinfo {volume} {22}},\ \bibinfo {pages} {073025} (\bibinfo {year} {2020})}\BibitemShut {NoStop}%
\end{thebibliography}%

\begin{widetext}
\newpage
\section{Supplementary Information}
%\subsection{Elastic constants}
%Elastic parameters are extracted from DFT calculations of strained bulk material in the lowest energy (AP) stacking. Calculated bulk lattice constants, a = 3.XXX and c = 12.XXX [cite, cite], and elastic moduli are in good agreement with experiment \cite{Lv2017, Gardos1990}. $C_{11}$ \; is about 11\% of that of graphite, and about half that of the group-VI TMDs, implying a relatively soft material which will be more amenable to interlayer effects. 

%A uniaxial strain is applied, following which the positions of atoms are relaxed (subject to a fixed uniaxial pressure) and elastic constants are then calculated from the linear stress-strain relationship, $u_{ii} = C_{11} \sigma_{ii}$ and $u_{ii} = C_{12} \sigma_{jj}$, where $i,j$ are lattice directions [cite, cite]. 

\subsection{Stacking configurations.}
\begin{figure*}[ht]
  % \centering
    \includegraphics[width=0.75\columnwidth]{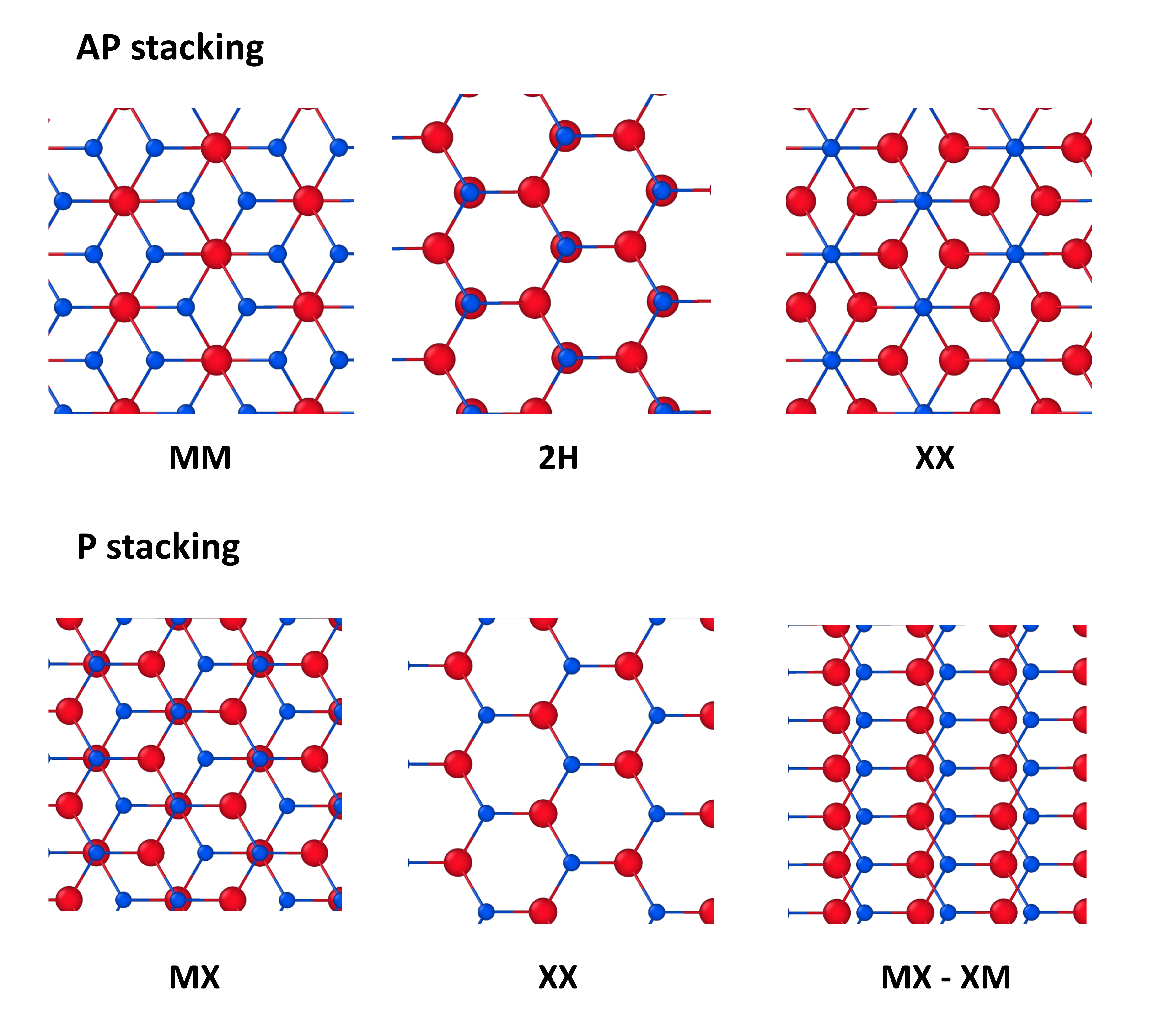}
    \caption{Plan view of the main stacking orders occurring in twisted NbSe$_2$ domains.}
    \label{fig:SI_stacking}
\end{figure*}

\subsection{DFT-calculated Adhesion fit \& Elastic  parameters.}
\begin{table*}[h]
\begin{ruledtabular}
\begin{tabular}{lllllllll}
        C$_1$ (eV nm$^2$) & C$_2$ (eV nm$^6$) & C$_3$ (eV nm$^{10}$) & A$_1$ (eV/nm$^2$) & A$_2$ (eV/nm$^2$) & $\rho_1$ (nm) & $\rho_2$ (nm) & $d_0$ (nm) & $\varepsilon$ (eV/nm$^4$) \\
       \hline
 0.136                             & 0.208                             & 0.029                             & 0.178       & -0.016      & 0.051   & 0.042   & 0.66 & 214
\end{tabular}
\end{ruledtabular}
\caption{\label{table2} DFT-calculated adhesion fit parameters.}
\end{table*}

\begin{table}[h]
\begin{ruledtabular}
\begin{tabular}{llll}
        \color{black} a ({\AA}) & \color{black}c ({\AA}) & Young's modulus (N/m) & Poisson's ratio \\
        \hline
        \color{black} 3.45  & \color{black}12.74 & 77.92  & 0.334                      
\end{tabular}
\end{ruledtabular}
\caption{\label{elasticity} DFT-calculated \color{black} lattice parameters and \color{black} elastic constants.}
\end{table}

\subsection{DFT calculation of ferroelectric charge transfer.}
%%%%%%%%%%%%%%%%%%%%%%%%%%%%%%%%%%%%%%%%%%%%%%%%%%%%%%%%%%
\begin{figure*}[ht]
  % \centering
    \includegraphics[width=0.95\columnwidth]{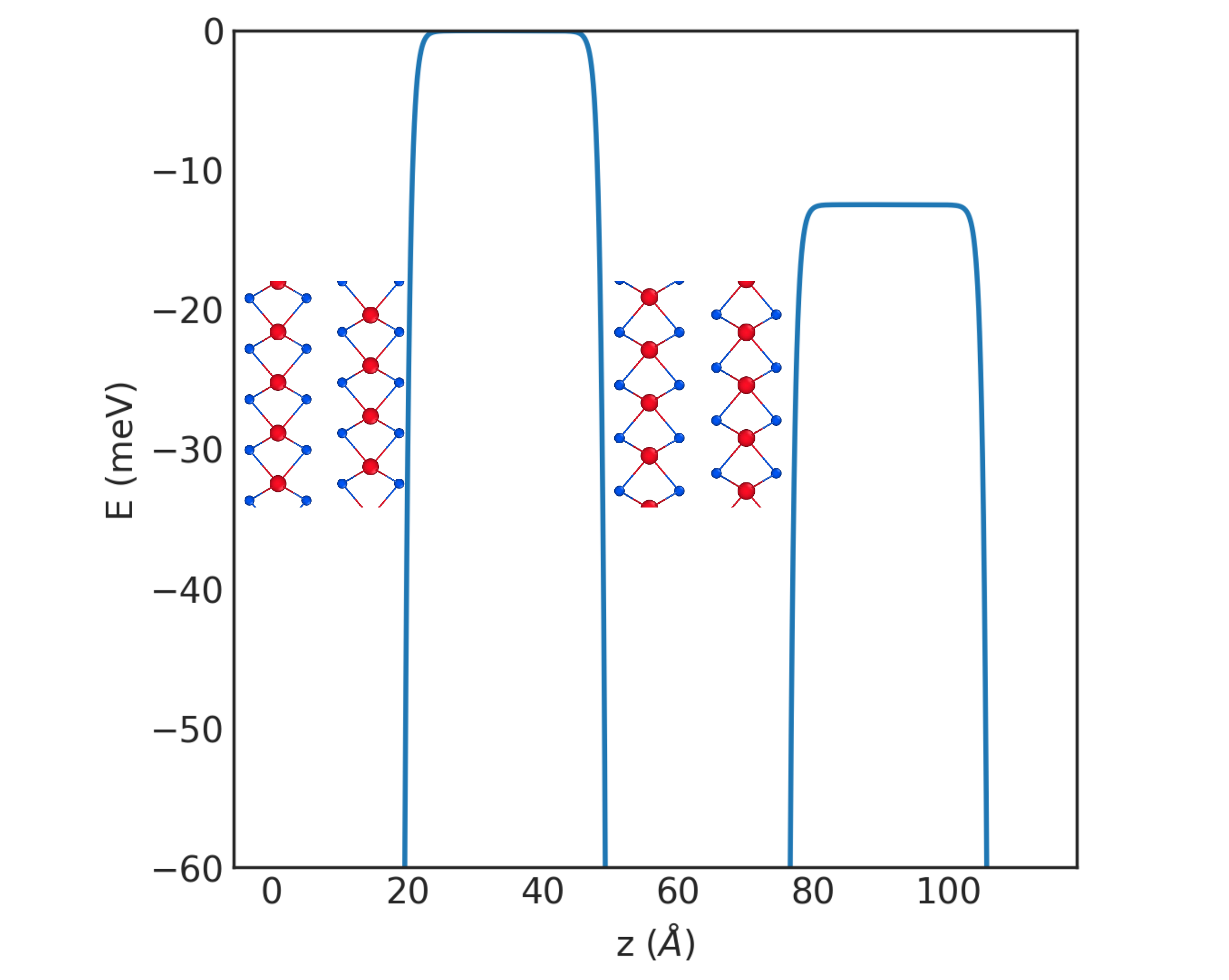}
    \caption{Potential drop across a mirrored P-bilayer shows ferroelectricity in P domains. There is a smaller ferroelectric charge transfer in comparison to semiconducting TMDs, $\Delta^P \approx$ 11 meV.}
    \label{fig:SI_FE}
\end{figure*}
%%%%%%%%%%%%%%%%%%%%%%%%%%%%%%%%%%%%%%%%%%%%%%%%%%%%%%%%%%

\newpage
\color{black}
\newpage
\subsection{Layer polarization }
\begin{figure}[H]
   \centering
    \includegraphics[width=1.0\columnwidth]{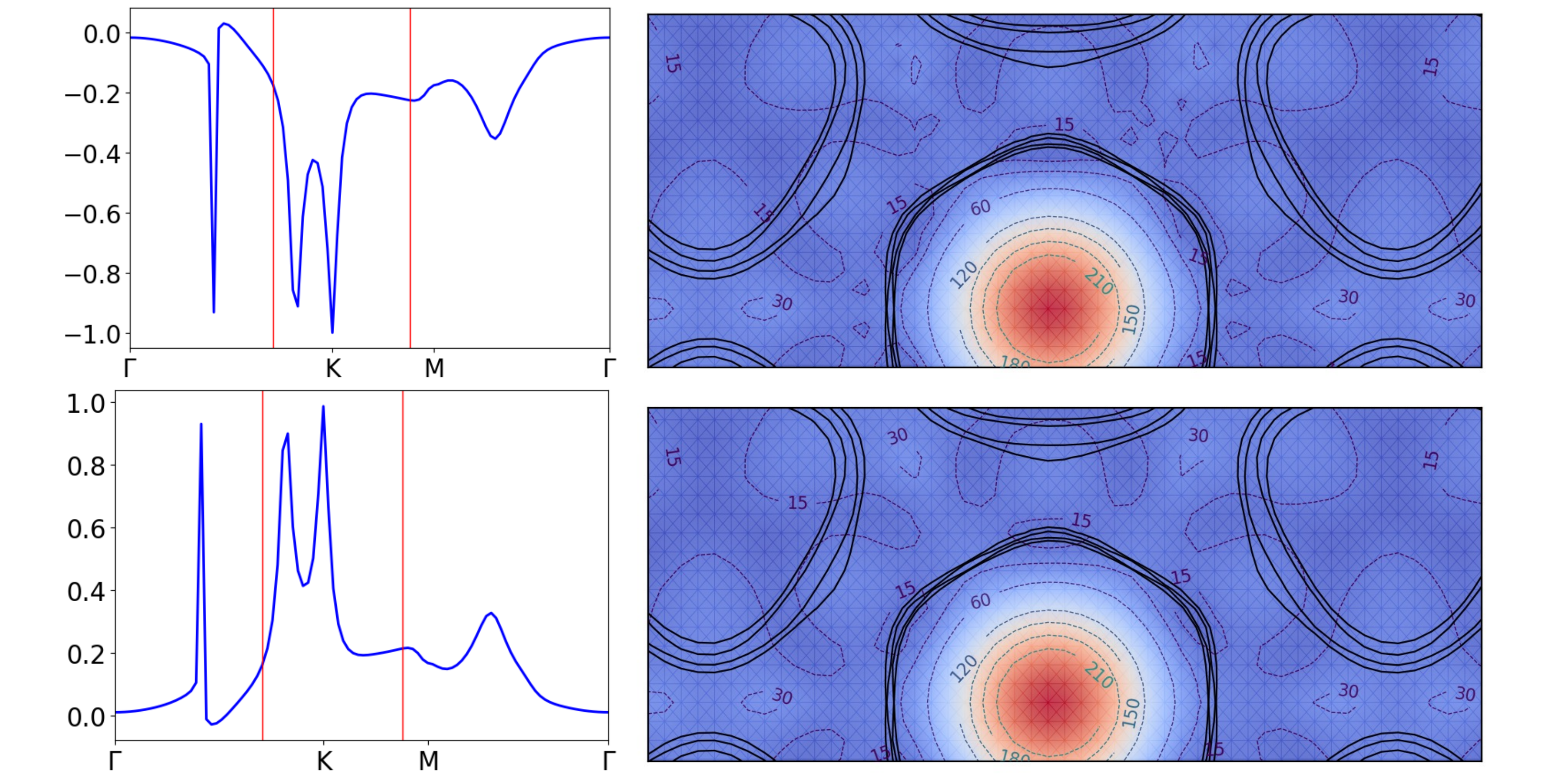}
    \caption{\color{black} Difference in the projection, onto the top and bottom layers, of DFT wavefunctions of the first (top left) and second metallic band (bottom left), of an XM bilayer. Top right: Calculated interlayer hybridization with an additional layer-dependent term $\Delta_z \sigma_0 \eta_z$, for $\Delta_z = 0$ meV (same as main text) and  $\Delta_z$ = 20 meV (bottom right). }
    \label{fig:Layer_pol}
\end{figure}

\color{black}

\newpage
\color{black}
\subsection{Band Structure, Orbital Projections}
\color{black}

\begin{figure*}[ht]
  % \centering
    \includegraphics[width=0.7\columnwidth]{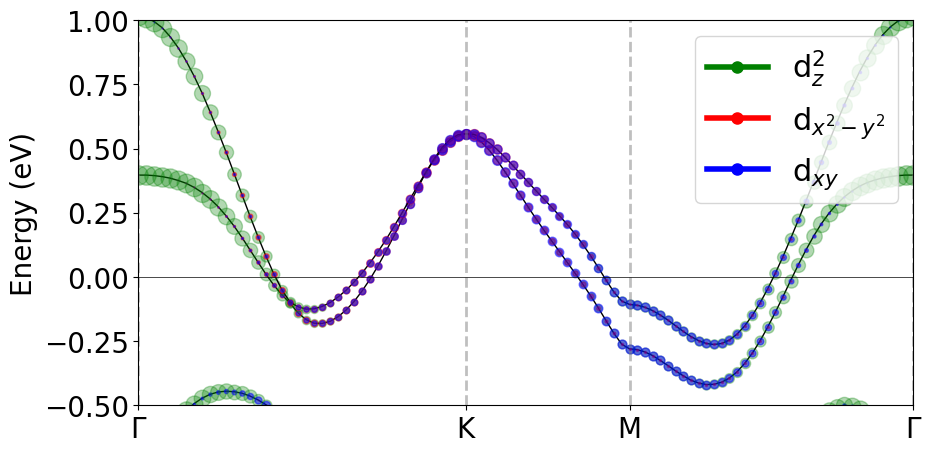}
    \caption{\color{black}Electronic structure and orbital projection of the MM-AP bilayer, where there is stronger interlayer splitting along the $\Gamma$-K and K-M directions, due to a higher proportion of $d_z^2$-orbitals.}
    \label{fig:SI_Fermi}
\end{figure*}

\color{black}

\newpage
\subsection{Full expression for hybridization around $\Gamma$ and K pockets.}

We model spin and interlayer splitting at each pocket with a Hamiltonian
\begin{equation}
    \mathcal{H}_i (\mathbf{p}) = \epsilon_i(\mathbf{p}) \sigma_0 \eta_0 d^\dagger_{\mathbf{p}, i} d_{\mathbf{p}, i}  + \alpha_i (\mathbf{p}) \sigma_0 \eta_x + \beta_i (\mathbf{p}) \sigma_z^a \eta_0.
\end{equation}

The interlayer term $\alpha$ varies with interlayer displacement between the layers, which also has periodic dependence on crystal momentum around a pocket,
\begin{equation}
    \alpha_\Gamma (\mathbf{p}) = \bar{t}_\Gamma + t_{\Gamma, 6} cos(6 \phi) + t_{\Gamma, 2} cos(2 \phi),
\end{equation}
and
\begin{equation}
    \alpha_K (\mathbf{p}) = \bar{t}_K + t_{K, 3} cos(3 \phi) + t_{K, 1} cos(\phi),
\end{equation}
the 3/6 cosine terms capture periodic dependence inside domains, while the 2/1 terms are due to nematic distortion of pockets in the vicinity of dislocations. The parameters $\bar{t}_{\Gamma,K}$ are the average interlayer hybridization around a pocket. Spin-orbit terms are also included,
\begin{equation}
    \beta_\Gamma (\mathbf{p}) = \lambda_\Gamma |\mathbf{p}|^3 cos(3 \phi),
\end{equation}
such that spin splitting vanishes along $\Gamma$-M directions, and  with fixed splitting in the K$_\pm$-pockets,
\begin{equation}
    \beta_K (\mathbf{p}) = \lambda_K.
\end{equation}

\subsection{Full hybridization parameters around $\Gamma$ and K pockets in domains and at domain boundaries.}

\begin{table}[h]
\begin{ruledtabular}
\begin{tabular}{llll}
           & $\bar{t}_\Gamma$ (meV)   & $t_{\Gamma,6}$ (meV)   & $t_{\Gamma,2}$ (meV)  \\
           \hline
2H         & 18.22 & 15.00 & 0.00   \\
MM         & 52.73 & 21.00 & 0.00   \\
XX         & 57.50 & 16.40 & 0.00   \\
DW (2H-XX) & 56.27 & -4.33 & 24.54  \\
DW (XX-MM) & 47.05 & 3.78  & -22.57
\end{tabular}
\end{ruledtabular}
\caption{\label{SI_AP_Gamma} AP-bilayer: Fit of $\alpha_\Gamma$ at selected high-symmetry stackings.}
\end{table}

\begin{table}[h]
\begin{ruledtabular}
\begin{tabular}{llll}
           & $\bar{t}_K$ (meV)    & $t_{K,6}$ (meV)   & $t_{K,2}$  (meV) \\
           \hline
2H         & 15.00 & 0.00  & 0.00 \\
MM         & 21.00 & 0.00  & 0.00 \\
XX         & 16.40 & 0.00  & 0.00 \\
DW (2H-XX) & 23.16 & 16.29 & 0.00 \\
DW (XX-MM) & 15.11 & 12.21 & 0.00
\end{tabular}
\end{ruledtabular}
\caption{\label{SI_AP_K} AP-bilayer: Fit of $\alpha_K$ at selected high-symmetry stackings.}
\end{table}

\begin{table}[h]
\begin{ruledtabular}
\begin{tabular}{llll}
   & $\bar{t}_\Gamma$ (meV)   & $t_{\Gamma,3}$ (meV)   & $t_{\Gamma,1}$ (meV)   \\
   \hline
MX & 19.55 & 24.00 & 0.00   \\
XX & 66.46 & 52.95 & 0.00   \\
DW (MX-XM)& 34.94 & 0.00  & -27.67
\end{tabular}
\end{ruledtabular}
\caption{\label{SI_P_Gamma} P-bilayer: Fit of $\alpha_\Gamma$ at selected high-symmetry stackings.}
\end{table}

\begin{table}[h]
\begin{ruledtabular}
\begin{tabular}{llll}
   & $\bar{t}_K$ (meV)   & $t_{K,3}$ (meV)   & $t_{K,1}$ (meV) \\
   \hline
MX & 24.00 & 4.16  & 0.00  \\
XX & 52.95 & 10.49 & 0.00  \\
DW (MX-XM)& 11.62 & 13.63 & 21.75
\end{tabular}
\end{ruledtabular}
\caption{\label{SI_P_K} P-bilayer: Fit of $\alpha_K$ at selected high-symmetry stackings.}
\end{table}

\subsection{Fourier expansion of hopping parameter across moir\'e supercell.}
We have evaluated the average interlayer hybridization term, $\bar{t}$, at individual pockets using DFT. We do this by performing band structure calculations along each of six individual high-symmetry lines in momentum space around each pocket, using fully-relaxed cells (at fixed disregistry - i.e. fixed planar coordinates). We then average the value of $\alpha(\mathbf{p})$ at the Fermi level along these directions. The obtained DFT data (as a function of disregistry) is fitted to a Fourier expansion using the first star of reciprocal lattice vectors around the lattice site $r'$,
\begin{equation}
    \bar{t}_i (r_0) = V_0 + \sum_g V(g) e^{i g \cdot (r_0 - r')},
\end{equation}
where the vectors $g$ are the six TMD reciprocal lattice vectors in the first star, $g_n = \pm g_{1,2,3}$, and
\begin{equation}
    V(g) = V^*(-g) = V_1 e^{i\psi}.
\end{equation}

\begin{table}[h]
\begin{ruledtabular}
\begin{tabular}{lllll}
          & V$_0$ (meV)  & V$_1$ (meV) & $\psi (^o)$ & r' \\
          \hline
AP, $\Gamma$ & 36.55 & 3.88  & 7.44          & XX         \\
AP, K     & 14.90 & 1.20  & 112.00        & XX         \\
P, $\Gamma$  & 41.53 & -4.97 & 59.80         & MX         \\
P, K      & 35.22 & -3.84 & 59.70         & MX     
\end{tabular}
\end{ruledtabular}
\caption{\label{Fourier_expansion} Parameters of Fourier expansion of the interlayer coupling as a function of disregistry.}
\end{table}

\newpage
\subsection{Fourier expansion of average interlayer hybridization parameter }
\begin{figure}[H]
   \centering
    \includegraphics[width=0.7\columnwidth]{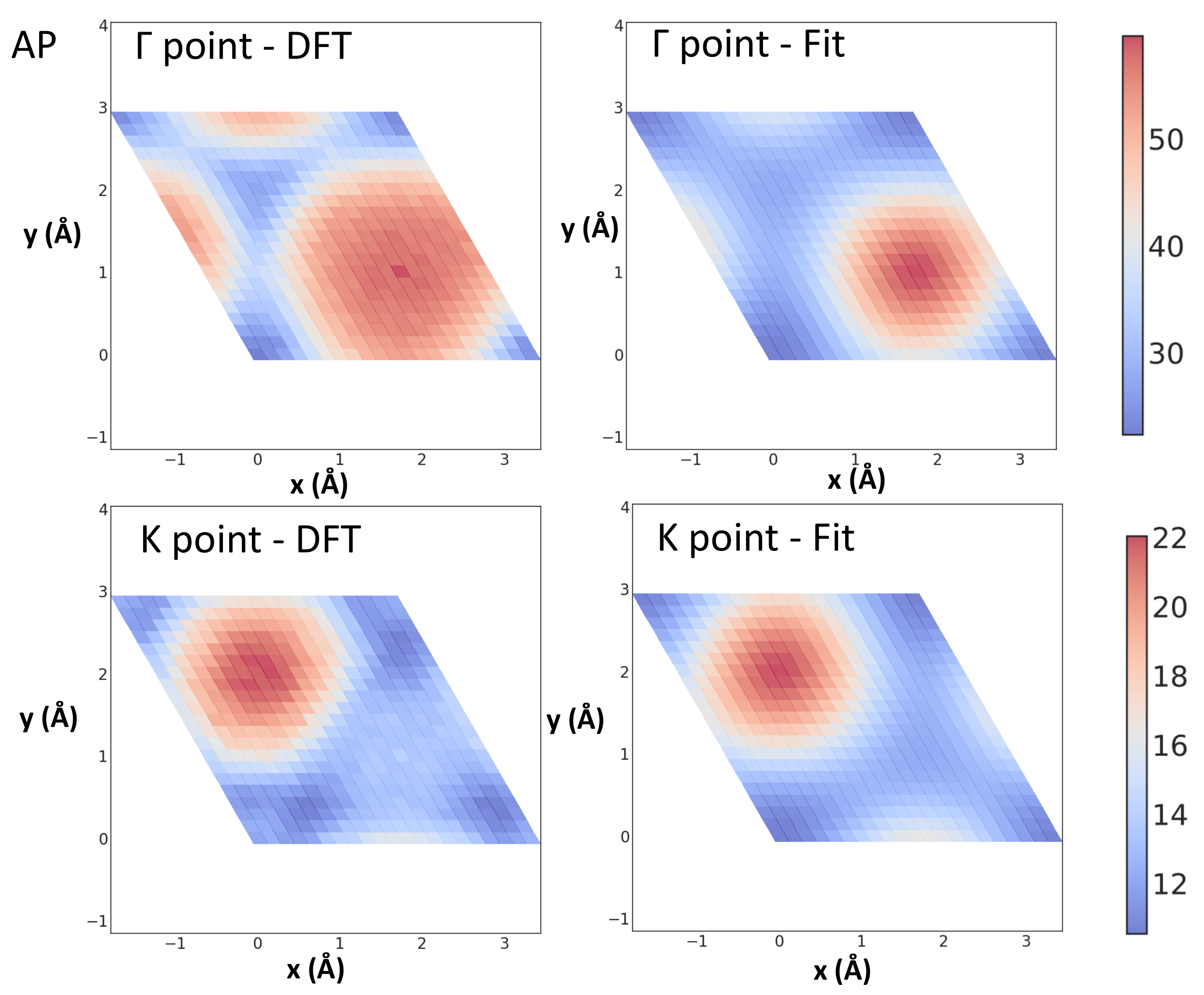}
    \includegraphics[width=0.7\columnwidth]{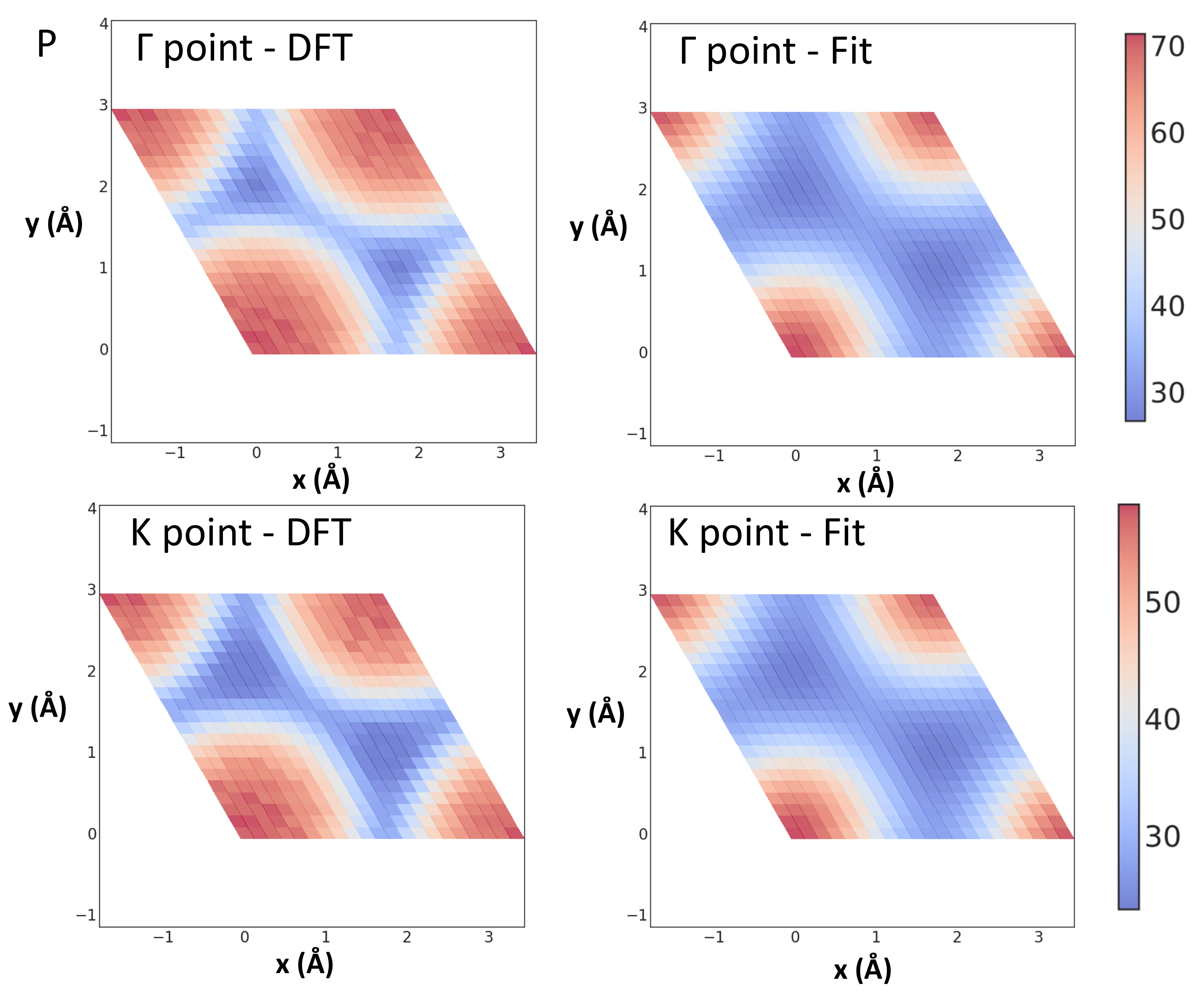}
    \caption{DFT-calculated vs Fourier expansion of $\bar{t}_i$. \color{black}All units are in meV.}
    \label{fig:Hybrid_fits}
\end{figure}
\color{black}

\subsection{CDW structures}

\begin{figure}[H]
   \centering
    \includegraphics[width=0.9\columnwidth]{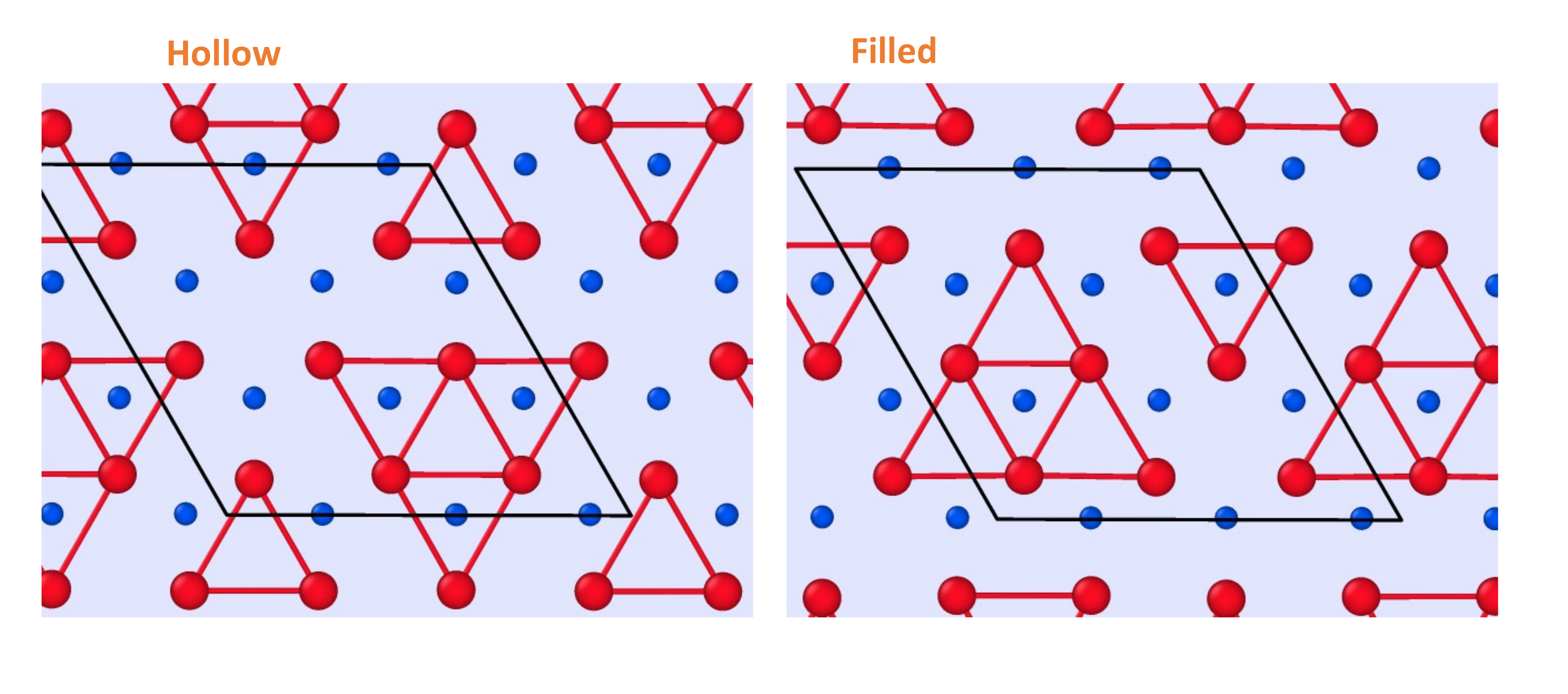}
    \caption{Hollow \& filled charge density wave structures with both Nb and Se sublattices shown. Nb distances are shown to demonstrate the degree of reconstruction due to the CDW phase.}
    \label{fig:SI_CDW_ML}
\end{figure}
\color{black}

\begin{figure}[H]
   \centering
    \includegraphics[width=0.9\columnwidth]{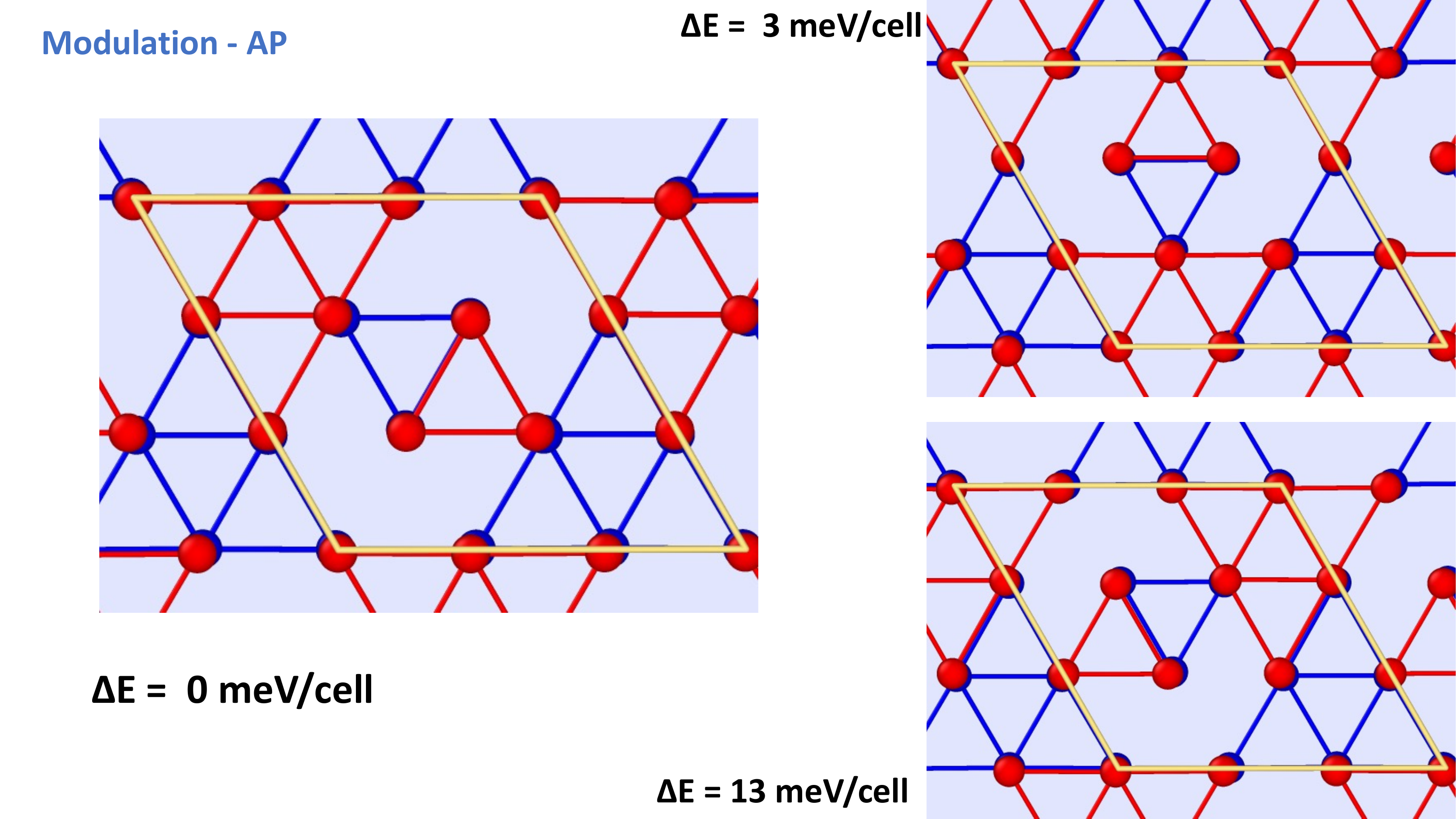}
    \caption{Three distinct interlayer structures for AP-stacked, hollow CDW structures. The other nine stackings are 120-degree rotations of these structures. }
    \label{fig:SI_CDW_AP}
\end{figure}
\color{black}

\begin{figure}[H]
   \centering
    \includegraphics[width=0.9\columnwidth]{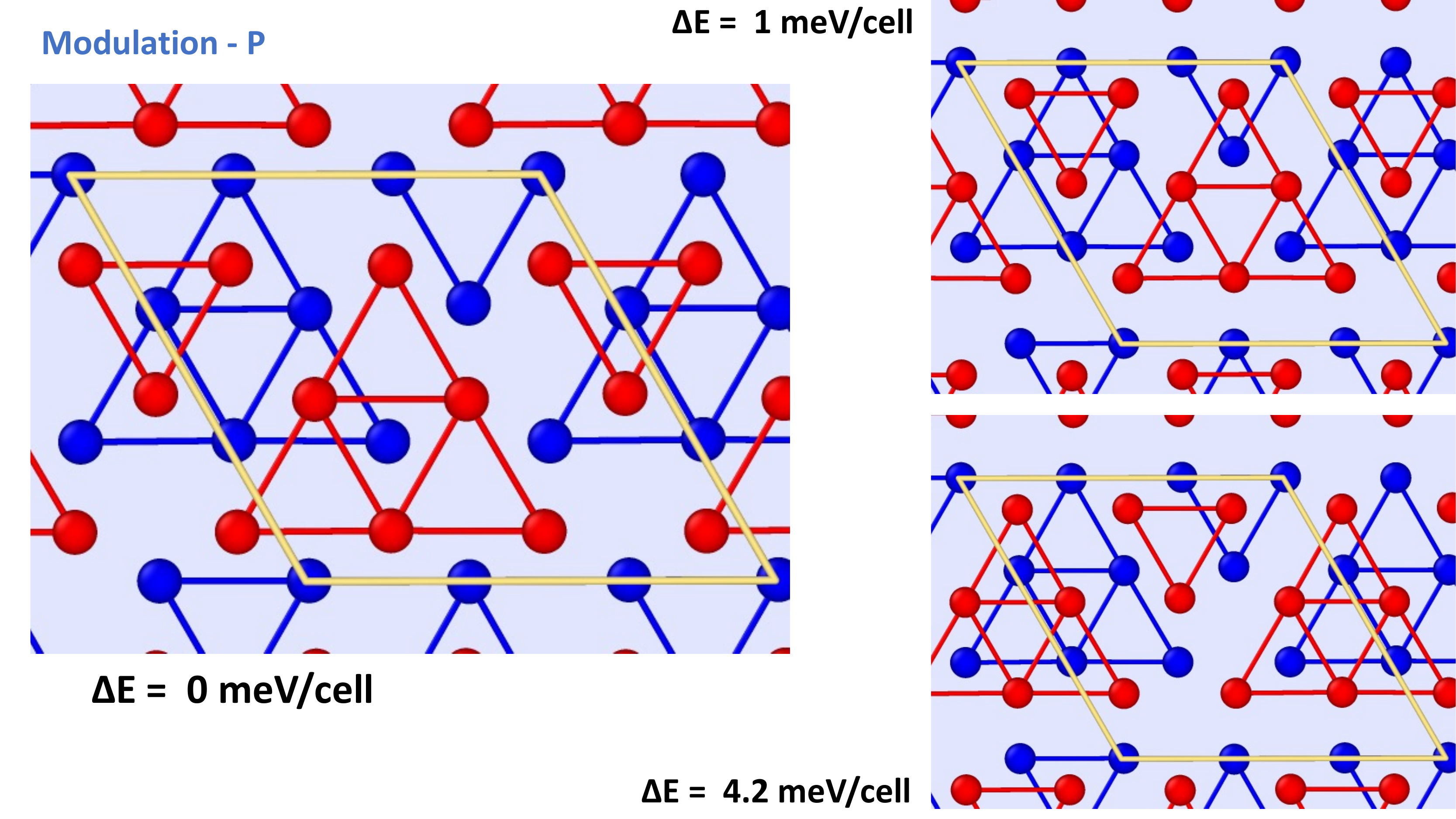}
    \caption{Three distinct interlayer structures for P-stacked, hollow CDW structures. The other eighteen stackings are 120-degree rotations of these structures.}
    \label{fig:SI_CDW_P}
\end{figure}
\color{black}

\end{widetext}

%%%%%%%%%%%%%%%%%%%%%%%%%%%%

\end{document}